\documentclass[onecolumn,amsmath,amssymb,nofootinbib,letterpaper,11pt]{article}

\pdfoutput=1
\usepackage{jheppub}
\usepackage{ifpdf}
\usepackage[utf8]{inputenc}
\usepackage[T1]{fontenc}

\input{epsf}
\usepackage{epsfig}
\usepackage{epstopdf}
\usepackage{arcs}

\usepackage{graphicx}
\usepackage{subfigure}

\usepackage{tensor}
\usepackage{braket}
\usepackage{float}
\usepackage[dvipsnames]{xcolor}
\usepackage{amsmath}
\usepackage{morefloats}
\newcommand{\be}{\begin{equation}}
\newcommand{\ee}{\end{equation}}

\usepackage{tikz}
\usetikzlibrary{calc}   
\usetikzlibrary{topaths}
\usepackage{hyperref}
\hypersetup{
	colorlinks =WildStrawberry,
	linkcolor =red,
	pdftitle={Mimetic Inflation},
	citecolor=NavyBlue,
	filecolor=WildStrawberry,
	urlcolor=WildStrawberry,
}

\usepackage{color}

\usepackage{array}
\usepackage{bm}

\usepackage[toc]{appendix}
\usepackage{color,soul}
\usepackage{datetime}
\newcommand{\bea}{\begin{eqnarray}}
\newcommand{\eea}{\end{eqnarray}}
\newcommand{\ba}{\begin{eqnarray}}
\newcommand{\ea}{\end{eqnarray}}

\newcommand{\beq}{\begin{equation}}
\newcommand{\eeq}{\end{equation}}
\newcommand{\beqa}{\begin{eqnarray}}
\newcommand{\eeqa}{\end{eqnarray}}
\newcommand{\beqar}{\begin{eqnarray*}}
\newcommand{\eeqar}{\end{eqnarray*}}

\renewcommand{\c}{$c$}

\newcommand{\calR}{{\cal{R}}}

\def\bfk{{\bf k}}

\newcommand{\bfx}{{\bf{x}}}

\renewcommand{\href}[2]{#2}

\title{Mimetic Inflation}
\author{Seyed Ali Hosseini Mansoori $^{a}$\footnote{shosseini@shahroodut.ac.ir},}
\author{Alireza Talebian $^{b}$\footnote{talebian@ipm.ir}, }
\author{Hassan Firouzjahi $^{b}$\footnote{firouz@ipm.ir}}

\affiliation[a]{Faculty of Physics, Shahrood University of Technology, P.O. Box 3619995161 Shahrood, Iran}
\affiliation[b]{School of Astronomy, Institute for Research in Fundamental Sciences (IPM), P.O. Box 19395-5531, Tehran, Iran\vspace{0.1cm}}

\vspace{0.1cm}

\abstract{ We study inflationary solution in  an extension of  mimetic gravity with the higher derivative interactions coupled to gravity. Because of the higher derivative interactions  the setup is free from the ghost and gradient instabilities while it hosts a number of novel properties. The dispersion relation of 
scalar perturbations develop quartic momentum correction similar to the setup of ghost inflation. Furthermore, the tilt of tensor perturbations can take either signs with a modified consistency relation between the tilt and the amplitude of tensor perturbations. Despite the presence of higher derivative interactions coupled to gravity the tensor perturbations propagate with the speed equal to the speed of light as required by the LIGO observations.  Furthermore, the higher derivative interactions induce non-trivial interactions in cubic Hamiltonian, generating non-Gaussianities in various shapes such as the equilateral, orthogonal and squeezed configurations with observable amplitudes. 
}

\preprint{
}

\begin{document}
\maketitle

\section{Introduction}

Mimetic gravity is a novel scalar-tensor theory proposed by Chamseddine and Mukhanov  \cite{Chamseddine:2013kea} as a modification of General Relativity (GR). The idea is to express the physical metric $g_{\mu\nu}$ in the Einstein-Hilbert action by performing a conformal transformation $g_{\mu\nu}=-(\tilde{g}^{\alpha \beta} \partial_{\alpha}\phi\partial_{\beta}\phi) \tilde{g}_{\mu \nu}$ from an auxiliary metric $\tilde{g}_{\mu\nu}$ in which $\phi$ is a scalar field. As a result,   the longitudinal mode of gravity becomes dynamical even in the absence of any matter source. The above transformation  can also be considered as a singular limit of the general disformal transformation where the transformation is not invertible \cite{Deruelle:2014zza,Yuan:2015tta}.  With the physical metric, the scalar field is subject to the constraint\footnote{We use the mostly positive signature for the metric.},
\begin{equation}\label{mimetric con}
 g^{\mu\nu} \partial_{\mu}\phi \partial_{\nu} \phi=-1.
\end{equation}

As a consequence  of this constraint, the theory mimics the roles of cold dark matter in cosmic expansion, hence the theory is dubbed as the mimetic dark matter.    The original mimetic model was then extended to inflation, dark energy and also theories with non-singular cosmological and black hole solutions \cite{Chamseddine:2014vna,Chamseddine:2016uef,Chamseddine:2016ktu}. See also Refs. \cite{Mirzagholi:2014ifa,Myrzakulov:2015kda, Arroja:2015yvd, Sebastiani:2016ras, Dutta:2017fjw, 
Saadi:2014jfa,Firouzjahi:2018xob,Gorji:2019rlm,Matsumoto:2015wja,Momeni:2015aea,Astashenok:2015qzw,Sadeghnezhad:2017hmr, Nozari:2019esz, 
Solomon:2019qgf,Shen:2019nyp,Ganz:2019vre,deCesare:2019pqj,Nozari:2019shm,
deCesare:2018cts,Ganz:2018mqi,Ganz:2018vzg, Sheykhi:2019gvk, Sheykhi:2020dkm, Sheykhi:2020fqf, Nojiri:2014zqa,Astashenok:2015haa,Nojiri:2016ppu,Nojiri:2017ygt,Nojiri:2016vhu,Odintsov:2018ggm, Casalino:2018wnc}  for further theoretical developments in mimetic gravity.

The original version of the mimetic theory is free from instabilities \cite{Barvinsky:2013mea, Chaichian:2014qba}, but there is no nontrivial dynamics for scalar-type fluctuations. In order to circumvent this problem, the higher derivative term  $(\Box \phi)^2$ is added to the original action which generates a dynamical scalar degree of freedom  propagating with a nonzero sound speed \cite{Chamseddine:2014vna,Mirzagholi:2014ifa}. In addition, the mimetic model with a general higher derivative function in the form $f(\Box \phi)$ has been considered in Refs. \cite{Chamseddine:2016uef,Chamseddine:2016ktu}.  However, these extended  mimetic setups with a propagating scalar degree of freedom are plagued with the ghost and the gradient instabilities \cite{Ijjas:2016pad,Firouzjahi:2017txv,Ramazanov:2016xhp}\footnote{The mimetic dark matter scenario also suffers from caustics \cite{Capela:2014xta}, see also 
\cite{DeFelice:2015moy, Gumrukcuoglu:2016jbh, Babichev:2016jzg, Babichev:2017lrx}.  It should be noted that the gauge field extensions of the mimetic scenario potentially avoid caustics formations \cite{Gorji:2018okn,Gorji:2019ttx}.}. To remedy these issues, it was suggested in \cite{Zheng:2017qfs,Hirano:2017zox,Gorji:2017cai} to extend the mimetic model further by considering direct couplings of the higher derivative terms to the curvature tensor of the spacetime such as $\Box \phi R$, $\nabla_{\mu} \nabla_{\nu}\phi R^{\mu \nu}$, $\Box \phi\nabla_{\mu} \phi \nabla_{\nu}\phi R^{\mu \nu}$ and so on. By appropriate choices of these higher derivative couplings one can bypass the problems of the  gradient and the ghost instabilities. However,  now the background dynamics is more complicated and a simple dark mater solution is not a direct outcome of the analysis. 

In this work our goal is to construct inflationary solutions in the extended mimetic setup with the effects of higher derivative couplings taking into account. As we will see the presence of higher derivative couplings to gravity generate new interactions and the analysis of cosmological perturbations become non-trivial. For example, because of these higher derivative interactions the dispersion relation of scalar perturbations receive higher order corrections resembling non-relativistic dispersion relation as in ghost inflation  setup \cite{ArkaniHamed:2003uz}.  In addition, the predictions for the tensor perturbations are modified with a new consistency condition between the scalar spectral index $n_s$, the sound speed of scalar perturbations $c_s$ and the tensor to scalar ratio $r_t$.

Because of the  higher derivative  interactions, the model predicts novel non-Gaussianity features. The situation here is somewhat similar to the EFT studies of  higher derivative corrections to the  single field model \cite{Cheung:2007st} where large non-Gaussianity of various shapes such as equilateral and orthogonal types can be generated. In addition, similar to models with a non-standard kinetic energy such as DBI model \cite{Alishahiha:2004eh}, the sound speed of scalar perturbations play non-trivial roles in generating large non-Gaussianities.  The strong observational bounds on primordial non-Gaussianities can be used to constrain the model parameters. More specifically,  the amplitude of non-Gaussianity parameter $f_{_{\rm NL}}$ in  the squeezed, equilateral and orthogonal configurations from the Planck observations \citep{Akrami:2018odb,Akrami:2019izv} are constrained  to be 
\begin{equation}
\label{fnl-Observ}
f_{_{\rm NL}}^{\rm sq} = -0.9\pm 5.1, \hspace{1cm}
f_{_{\rm NL}}^{\rm equi} = -26 \pm 47, \hspace{1cm}
f_{_{\rm NL}}^{\rm ortho} = -38 \pm 24 \hspace{0.25cm} (68\% \rm CL) \, .
\end{equation}
We shall use these bounds to constrain model parameters and various couplings.

The organization of the paper is as follows. In next Section, we present our setup and construct   the background solutions  which mimic cold dark matter even in the absence of normal matter. Then we extend these analysis to obtain an inflationary solution.  In Section \ref{power} we obtain the power spectrum of the curvature and 
tensor perturbations and calculate various cosmological observables. In Section \ref{Bispec}, we study bispectrum  and calculate the non-Gaussianity parameter $f_{_{\rm NL}}$ for local, equilateral and orthogonal configurations numerically, followed by discussions and summaries in Section \ref{Conclusion}. Many technical analysis of cosmological perturbations   associated to power spectra and scalar bispectrum  are relegated to Appendices \ref{Ap2} and \ref{AppC}. 

\section{Inflationary Solution}
\label{setup-sec}

In this section we study the background dynamics to obtain a period of inflation in early universe.
 
As summarized in Introduction, to remedy the ghost and gradient instabilities various higher derivative terms are added to the mimetic setup. Besides the higher derivative terms such as   $(\Box \phi)^2$ and $\nabla_{\mu}\nabla_{\nu}\phi \nabla^{\mu}\nabla^{\nu}\phi$,  we also require the  higher derivative couplings of  of the mimetic field  to the curvature of the spacetime such as $\Box \phi R$, $\nabla_{\mu} \nabla_{\nu}\phi R^{\mu \nu}$, $\Box \phi\nabla_{\mu} \phi \nabla_{\nu}\phi R^{\mu \nu}$ and so on
\cite{Zheng:2017qfs,Hirano:2017zox,Gorji:2017cai}. Here we restrict ourselves to 
the simplest case where  there is only a direct coupling of  the higher derivative term  $\Box \phi \equiv \chi$ to the Ricci scalar as follows,  
\ba
\label{action0}
S= \int {\rm d}^4 x  \sqrt{-g} \left[ \frac{{ M}_{ \rm P}^2}{2} F(\chi)R + \lambda \left( g^{\alpha \beta } \partial_\alpha \phi \partial_\beta \phi + 1 \right) + P (\chi) - V(\phi)
\right] \,,
\ea
in which $M_{\rm P} $ is the reduced Planck mass \footnote{The Hamiltonian analysis of this model have also been investigated in Ref \cite{Zheng:2018cuc} in both Einstein frame and Jordan frame.}. The Lagrangian multiplier $\lambda$ enforces the constraint Eq. (\ref{mimetric con}) \cite{Golovnev:2013jxa}. In addition, we have allowed a potential term for the mimetic field which 
will drive inflation. In this setup $P$ and $F$ are arbitrary smooth function of $\chi$.
The former is added to make the scalar perturbation propagating (i.e. inducing a non-zero $c_s$ ) while the latter is required to remedy the gradient and the ghost instabilities \cite{Zheng:2017qfs,Hirano:2017zox,Gorji:2017cai}. As mentioned before, more complicated function of  derivatives of $\phi$ and  $\chi$ such as $F_2(\nabla_{\mu} \nabla_{\nu}\phi R^{\mu \nu})$ and  $F_3(\chi \nabla_{\mu} \phi \nabla_{\nu}\phi R^{\mu \nu})$
can also be added along with the simple function $F(\chi)$. However, the analysis even in the simplest setup of action (\ref{action0}) is complicated enough so we do not consider models with other higher derivative couplings. 

Before presenting the fields equation one important comment is in order. In the action  (\ref{action0}) the effective gravitational coupling (effective reduced Planck mass) is actually $M_{\rm P} F(\chi)^{1/2}$.  We can perform the calculations in the given ``Jordan frame'' but with a proper interpretation of the physical gravitational coupling. Alternatively, we may perform a metric field redefinition  and go to the ``Einstein frame'' where the  gravitational coupling is simply $M_{\rm P}$. The latter is rather complicated as the model presented in  action (\ref{action0}) contains various higher derivative terms. Instead, we follow the first approach and work in the original Jordan frame. However, we make a further assumption that at the end of inflation  the fields $\phi$ becomes trivial with $F(\chi_e)=1$ 
and one recovers the standard GR afterwards. This is a simplification made based on the intuitive ground though  we do not have a dynamical mechanism to enforce it. We leave it as an open question as how or whether this transition from a mimetic setup to a  standard GR setup can be achieved at the end of inflation.  With these discussions in mind, we set  $M_{\rm P}=1$ in the rest of the analysis.

By taking the variation of the action \eqref{action0} with respect to the inverse metric $g^{\mu \nu}$, one obtains the Einstein field equations as 
$F(\chi) G_{\mu \nu} = T_{\mu \nu} $ where $G_{\mu \nu}$ is the Einstein tensor and $T_{\mu \nu}$ is the effective energy momentum tensor,  given by
\begin{eqnarray}
\label{Tmu-nu}
\nonumber T^{\mu }_{\nu }&=& -2\lambda {{\partial }^{\mu }}\phi {{\partial }_{v}}
\phi +\left[ {{\partial }^{\mu }}P_{\chi}{{\partial }_{\nu }}\phi +
{{\partial }^{\mu }}\phi {{\partial }_{\nu }}P_{\chi} +\frac{1}{2} \Big(\partial^{\mu} \phi \partial_{\nu}(R F_{\chi})+\partial_{\nu} \phi \partial^{\mu}(R F_{\chi})\Big)+\nabla^{\mu}\partial_{\nu} F\right]\\
&+&
\delta _{\nu }^{\mu }\left[ P-\chi P_{\chi }-V-{{g}^{\alpha \beta }}{{
		\partial }_{\alpha }}P_{\chi}{{\partial }_{\beta }}\phi  -\frac{1}{2}\nabla^{\alpha} \Big( R \partial_{\alpha} \phi F_{\chi}\Big)+ \Box F\right ] \, ,
\end{eqnarray}
in which ${ P}_{\chi} \equiv \partial { P}/\partial \chi$ and so on. In obtaining the above expression we have implemented  the mimetic constraint \eqref{mimetric con}. 
Clearly the energy momentum tensor given above can not be cast into the form of the energy momentum tensor of a perfect fluid.

Moreover, varying the action \eqref{action0} with respect to the scalar filed $\phi$ gives the following modified Klein-Gordon equation, 
\begin{equation}
\label{KG-eq}
\frac{1}{\sqrt{-g}}{{\partial }_{\mu }}\Big[ \sqrt{-g}\Big( 2\lambda {{\partial }^{\mu }}\phi -{{\partial }^{\mu }}\big(P_{\chi}+\frac{1}{2} R F_{\chi}\big)\Big) \Big]+\frac{\partial V\left( \phi  \right)}{\partial \phi }=0 \, .
\end{equation}

The background cosmological solution is in the form of FRLW universe  with the metric
\ba
{\rm d}s^2 = - {\rm d}t^2 + a(t)^2 {\rm d} \bfx^2 \,,
\ea
in which $t$ and $a$ are the cosmic time and the scale factor respectively. 
One can check that the  background field equations take the following forms
\ba
\nonumber 3 F H^2 &=& V- 2 \lambda - \big( P + 3 H P_{\chi } + 3 \dot H P_{\chi \chi} \big) -3 \Big(6 H^3 F_{\chi}-4H \dot{H} F_{\chi}+6 H^2 \dot{H}F_{\chi \chi}\\
&+&3 \dot{H}^2 F_{\chi \chi}- \ddot{H} F_{\chi} \Big) \label{Eqq8}\, ,
\ea
and
\ba
\label{Ein2}
F (2 \dot H + 3 H^2 ) = V- ( P + 3 H P_{\chi } - 3 \dot H P_{\chi \chi})-3H \Big(F_{\chi} (6 H^2+5\dot{H})-6H \dot{H} F_{\chi \chi} \Big) \, ,
\ea
where $H = \dot a(t)/a(t)$ is the Hubble expansion rate. Note that at the background level, the  mimetic constraint \eqref{mimetric con} enforces $\dot \phi=1$ and correspondingly $\chi  = -3 H$. 

Before constructing the inflationary solution, let us for the moment set $V=0$ to see how the mimetic setup can yield the dark matte solution.  Using  Eq. (\ref{KG-eq})  the Lagrangian multiplier is obtained as follows
\begin{equation}
\lambda=\frac{\mathcal{C}}{2 a^3}+ 6 H \dot{H} F_\chi -9H^2 \dot{H} F_{\chi\chi} - \dfrac{9}{2}\dot{H}^2 F_{\chi\chi} - \dfrac{3}{2}\dot{H} P_{\chi\chi} + \dfrac{3}{2}\ddot{H} F_\chi \quad \quad (V=0) \, ,
\end{equation}   
in which $\mathcal{C}$ is an integration constant. Plugging this into the Friedmann equation
\eqref{Eqq8}, one obtains
\begin{equation}
\label{Fried1}
\frac{3 H^2}{2} \Big[F + 6 H F_\chi + \dfrac{1}{3H^2}(P+3H P_\chi)\Big]=-\frac{\mathcal{C}}{a^3} \quad \quad (V=0) \, .
\end{equation}
The constant $\mathcal{C}$ above indicates that a dark matter type solution can exist. However, in order for this conclusion to be valid we require 
the combination in the big bracket in Eq. (\ref{Fried1}) to be a constant. 
Since  this combination  plays important roles in our analysis below let us define 
\begin{align}
F+ 6 H F_\chi + \dfrac{1}{3H^2}(P+3H P_\chi) \equiv - \mathcal{K}  \,.
\label{A}
\end{align}
Using this definition of the function $\mathcal{K} $ and taking the time derivative of Eq. (\ref{Fried1}) once more, we obtain 
\ba
\label{Fried2}
2 \dot H + 3 H^2 = -\frac{H \dot {\mathcal{K}}}{ \mathcal{K}} \, , \quad \quad (V=0) \, .
\ea
For a dark matter solution with $a \sim t^{2/3}$, the left hand side 
of Eq. (\ref{Fried2}) vanishes so indeed to have a dark matter solution we require the 
function $\mathcal{K}$ to be constant. In this case the Friedmann equation simplifies to $3 H^2 = \tilde{ \mathcal{C}}/a^3 = \rho $ where $\rho$ is the effective energy density and $\tilde {\mathcal{C}} \equiv  2 \mathcal{C}/\mathcal{K}$. 
To have a consistent solution one requires that $\tilde {\mathcal{C}} >0$ while there 
is no restriction on the signs of  $\mathcal{C}$ and $\mathcal{K}$ separately at this level.   For example, in the original mimetic setup with $F=1$ and $P=0$, one obtains $\mathcal{K}=-1$ and the dark matter solution is a direct outcome of the analysis.  In conclusion, while the functions $P(\chi)$ and $F(\chi)$ are arbitrary, but in order to obtain  a dark matter candidate in this setup 
one requires the combination $\mathcal{K}$ defined in Eq. (\ref{A}) to be a  constant. 

Now we consider the case when $V(\phi)=V(t) \neq 0$ in order to obtain inflationary solution.   Starting with the second Einstein equation (\ref{Ein2}) and using the definition of $\mathcal{K}$ to eliminate the combinations containing $F_{\chi \chi}$ and $P_{\chi \chi}$ in favours of $\dot {\mathcal{K}}$ we obtain
\begin{equation}
\label{equ}
2\dot{H} + 3 H^2 = -\frac{H \dot {\mathcal{K}}}{ \mathcal{K}}+ 
\dfrac{V}{\mathcal{K}}  \,.
\end{equation}
In particular, if we set $V=0$, we recover Eq. (\ref{Fried2}) as expected. 
Now noting that 
\ba
2\dot{H} + 3 H^2 + \frac{H \dot {\mathcal{K}}}{ \mathcal{K}} = \frac{1}{a^3 H^2  \mathcal{K}} \frac{d}{d t}\Big( a^3 H^2  \mathcal{K} \Big) \, ,
\ea
we can integrate Eq. (\ref{equ}) to obtain
\begin{align}
3 H^2 = \frac{1}{\mathcal{K}\,a^3} \Big(\mathcal{C} -3 \int {\rm d}a\, a^2 V\Big) \, ,
\label{3H2}
\end{align}
in which as in previous case $\mathcal{C}$ is a constant. 

In an inflationary background one can neglect the constant term $\mathcal{C}$ in 
Eq. (\ref{3H2}) as it is diluted rapidly. In addition, if $V$ is nearly flat as in conventional slow-roll scenarios then we can obtain a phase of near dS spacetime with $ H^2 \simeq -V/\mathcal{K}$.  Now we see the  curious effect that in order to obtain an inflationary background we require the signs of $V$ and 
$\mathcal{K}$ to be opposite. However, as we shall see in nest section, in order to have healthy scalar and tensor perturbations we require $\mathcal{K} >0$.
As a result, to obtain an inflationary solution in this setup we need a negative potential. This should be compared with the analysis in \cite{Chamseddine:2014vna} where $F=1$, $P(\chi)=\frac{\gamma}{2} \chi^2 $ and $V>0$. For these values of $F$ and $P$ we obtain $\mathcal{K} = -1+ 3\gamma/2$.  On the other hand,  the sound speed of scalar perturbations in the model of \cite{Chamseddine:2014vna} is $c_s^2 = \gamma/(2- 3 \gamma)$ so  $\mathcal{K}$ has opposite sign compared to $c_s^2$ (with $\gamma>0)$. Now in order to avoid the gradient instability one requires $c_s^2 >0$ so in their inflationary solution they need $\mathcal{K}<0$. But as we shall see in next section the sign of the quadratic action for the scalar perturbations is proportional to the sign of $\mathcal{K}$ so a 
negative $\mathcal{K}$ indicates the propagation of ghost  as pointed out in details in  \cite{Firouzjahi:2017txv}.   As we mentioned above, this problem arises because in the analysis of  \cite{Chamseddine:2014vna} they considered $V>0$ to construct inflationary solution. 

Although the mimetic field $\phi$ is not an ordinary ``rolling" scalar field in the sense that it appears with a constraint in the setup, but the requirement of a negative potential looks  unexpected. However, we remind that negative potentials have been employed in the past in other contexts such as in contracting universes  \cite{Khoury:2001wf,Kallosh:2001ai,Finelli:2001sr,Buchbinder:2007ad,Buchbinder:2007tw}, see also \cite{Linde:2001ae,Hartle:2012qb}. 

To construct a specific inflationary setup, we consider the inverted quadratic potential
as follows 
\ba 
\label{pot-V}
V(\phi)= \Bigg\{
\begin{array}{ll}
-\frac{1}{2}m^2 \phi^2 & ~~~~~~~~ t<0 \\
0 & ~~~~~~~~ t>0 \\
\end{array} 
,~~~~ \hspace{0.5cm} \phi=t
\ea
Inflation occurs when $t<0$ while the hot big bang phase follows  inflation for $t>0$.  We obtain the inflationary phase  as the field rolls up the negative potential toward the origin. In addition,  we assume that the potential vanishes for $t>0$, so as discussed below Eq. (\ref{Fried2}),  the seeds of observed dark matter can be obtained in this setup when inflation ends. 

So far our analysis were general and we did not specify the forms of the functions 
$F(\chi)$ and $P(\chi)$, unless we require the combination $\mathcal{K}$ to be positive. From now one, we further demand that $\mathcal{K}$ is a constant which simplifies the construction of the inflationary solution greatly. Now by introducing the new variable $y\equiv a^{\frac{3}{2}}$,  Eq.  \eqref{equ} becomes a linear differential equation,
\begin{equation}\label{dif1}
\ddot{y} - \frac{3 m^2 t^2}{8  \mathcal{K}}  y=0 \,, \quad \quad (t<0)
\end{equation}
which is similar to Eq. (31) in Ref. \cite{Chamseddine:2014vna}. 

The inflationary branch of the solution is given by  
\begin{align}
y(t) = \sqrt{-t}~ K_{\frac{1}{4}} \Big(\sqrt{\dfrac{3}{32 \mathcal{K}}}~ m \, t^2 \Big) \, ,
\end{align}
in which $K_\nu (x)$ is the modified Bessel function of the second kind.  At large negative $t$, one finds that the scale factor grows as
\begin{equation}
a \propto (-t)^{-\frac{1}{3}} \exp \big(-\frac{m t^2}{2\sqrt{6 \mathcal{K}}} \big) \,,
\label{scale factor}
\end{equation}
whereas it is proportional to $(t)^{2/3}$ for positive $t$ after inflation which is the indication of a dark matter dominated universe. Of course, we have to include  reheating  where radiation should be generated after 
inflation.   

\begin{figure}[t!]
	\centering
	\includegraphics[width=0.7\linewidth]{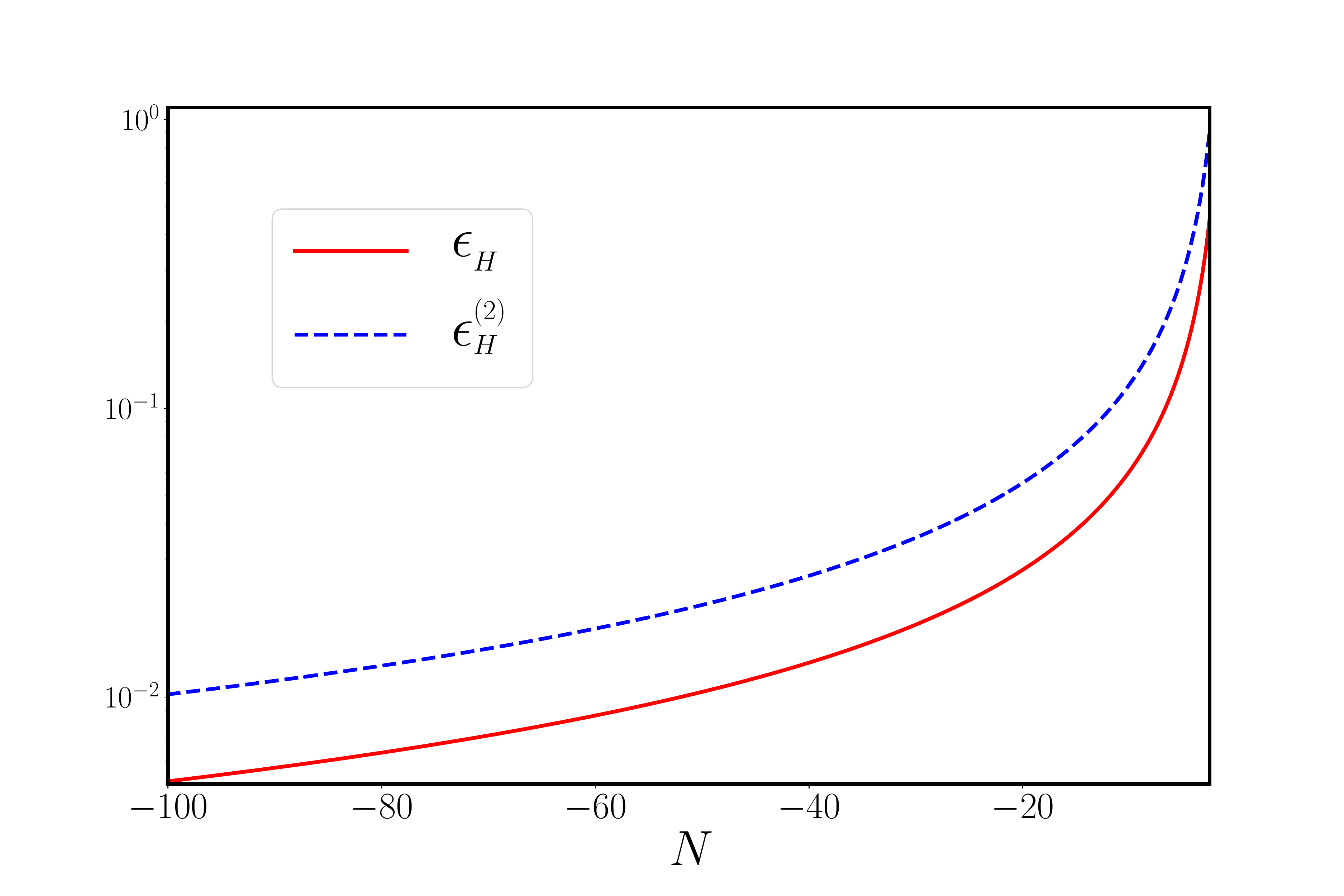}
	\caption{The first and second slow-roll parameters associated  to the scale factor \eqref{scale factor} in terms of the number of e-folds $N$. These behaviours are independent from the values of $m$ and ${\cal K}$.
	}
	\label{fig:slow-roll}
\end{figure}

The behaviours of the corresponding slow-roll parameters, 
\begin{align}
\epsilon_{_H} \equiv -\dfrac{\dot{H}}{H^2} \,, \hspace{1cm} \epsilon_{_H}^{_{(2)}} \equiv \dfrac{\dot{\epsilon}_{_H}}{H\,\epsilon_{_H}} \,,
\label{epsilonH}
\end{align}
are shown in Fig.~\ref{fig:slow-roll}. One can easily satisfy the conditions $\epsilon_{_H}, \ \epsilon_{_H}^{_{(2)}} \ll 1$ to obtain $50- 60$\, number of e-folds 
to solve the flatness and the horizon problems. As usual, inflation ends when either of the slow-roll parameters approach order of unity.  It should be noted that these slow-roll conditions are independent of the values of  $m$ and ${\cal K}$.

For the future uses,  let us define the slow-roll parameters associated with the background functions such as $H$,  $F$ etc as follows 
\begin{align}
\epsilon_{_X}^{_{(1)}} \equiv \epsilon_{_X} &\equiv \pm \dfrac{\dot{X}}{H \ X} \hspace{0.25cm} \text{and} \hspace{0.25cm} \epsilon_{_X}^{_{(n)}} \equiv \dfrac{\dot{\epsilon}_{_X}^{_{(n-1)}}}{H \ \epsilon_{_X}^{_{(n-1)}}} \hspace{0.25cm} \text{for} \hspace{0.25cm} n\geq2 \, .
\label{epX}
\end{align}
Note that the minus sign is chosen when $X=H$ while for other background functions we chose the plus sign. 
Using the scale factor \eqref{scale factor} in the inflationary phase, we find the following relations among the Hubble slow-roll parameters: 
\begin{align}
\label{eh}
\epsilon_{_H}^{_{(4)}} \approx \epsilon_{_H}^{_{(3)}} \approx \epsilon_{_H}^{_{(2)}} \approx 2 \epsilon_{_H} \approx \biggr| \frac{1}{ N} \biggr| \ll 1 \, ,
\end{align}
which are satisfied  for $ N \sim 50-60$ number of e-folds before the end of inflation.

As mentioned before, our functions $F(\chi)$ and $P(\chi)$ are arbitrary except that we have imposed that the combination $\mathcal{K}$ defined in  Eq. (\ref{A}) to be constant.   For example, the following pair of the polynomial functions 
\begin{equation*}
	\label{F-P}
	F(\chi)=1+\alpha \chi +\beta \chi^2 \,,\hspace{0.5cm} P(\chi)=\frac{\gamma}{2} \chi^2-\frac{\alpha}{6} \chi^3-\frac{\beta}{3} \chi^4 \, ,
	\end{equation*}
satisfy the constraint \eqref{A} with $\mathcal{K}= -1+ 3\gamma/2$. 
	
In general case, the polynomial functions $F(\chi) = \sum_{n=0}^{} f_n \, \chi^n \,,$ and $P(\chi) = \sum_{n=0}^{} p_n \, \chi^n$
can be considered with $p_0 = 0$ and $p_1$ an arbitrary constant. If the 
rest of coefficients $f_n$ and $p_n$ satisfy  the relations $p_2 \neq \dfrac{f_0}{3} $ and 
\begin{align}
p_{n+2} = \dfrac{1-2n}{3n+3}\,f_{n} \,~~~~~ n\geq 1 \, ,
\end{align}
then  we obtain  ${\cal K}= 3p_2- f_0$.

One open question in this setup is the issue of reheating after inflation. To be consistent with the big bang cosmology, the inflationary phase has to be followed by a hot radiation dominated background. In conventional slow-roll models this is achieved via the (p)reheating mechanism in which the inflaton field transfers its energy to the Standard Model (SM) particles and fields while oscillating in its global minimum. In our mimetic scenario the field $\phi$ is not a rolling field in the usual sense  but instead it is a space-filling field with the profile $\phi=t$.
So in order to achieve reheating one has to modify the current setup and couple the mimetic field to the SM fields one way or another. This is an open question which deserves a separate study elsewhere. We also comment that in the current setup 
with the potential (\ref{pot-V})  a dark matter solution is inherited in the solution for 
$t>0$ so one may only need reheating to generate the host radiation while the dark matter can come from the mimetic source.


\section{Primordial Power Spectra}
\label{power}

In this section we calculate the power spectra of the curvature  and  tensor perturbations. For this purpose we calculate the quadratic actions associated to these perturbations.

The details of the analysis of the quadratic actions are presented in Appendix \ref{Ap2}.  The quadratic action for the comoving curvature perturbation $\calR$ and the tensor perturbations $\gamma_{ij}$ is obtained to be 
\ba
\label{action-R2b}
S_2& = &  \int dt d^3{\bf  x} \, 
\vartheta \frac{a^3}{2} \left[  \dot{\calR}^2  - \frac{c_s^2}{a^2} {{\big( \partial \calR \big)}^2} - \sigma^2 \big(\dfrac{\partial^2 \calR}{a^2}\big)^2
+\frac{F}{4\vartheta} \Big( (\dot{\gamma}_{ij})^2 - \dfrac{(\partial \gamma_{ij})^2}{a^2} \Big) 
\right ]
\,.
\ea
in which we have defined the parameters $\vartheta$ and $\sigma$ as 
\begin{align}
\vartheta \equiv \dfrac{3{\mathcal{K}}\,F}{{\mathcal{K}}+F}, \hspace{1cm}
\sigma^2 \equiv \dfrac{F^2_\chi}{{\mathcal{K}}\,F} \, ,
\end{align}
and during inflation the sound speed of scalar perturbations $c_{s}^2$ as
\ba\label{cs2}
c_s^2 &\equiv  -\dfrac{1}{3{\mathcal{K}}} \left({\mathcal{K}}+F+3HF_\chi\right)  \, .
\ea
Note that $\vartheta$ is dimensionless while $\sigma$ has the dimension of length square.

In order for the perturbations to be free from the ghost and  gradient instabilities we require that all three parameters $ \vartheta $, $c_s^2$ and  $ \sigma^2$ to be positive.  Correspondingly we require  
\begin{align}
{\mathcal{K}}>0 \,, ~~~~~~~~~~ F>0 \,, ~~~~~~~~~~ F_\chi<0\, .
\end{align}
In particular note that if $ \mathcal{K} <0$ then the scalar perturbations develop ghost instability. This is  the reason why we needed to couple the higher derivative terms  to gravity to cure the ghost and gradient instabilities in the original setup of mimetic gravity \cite{Zheng:2017qfs,Hirano:2017zox,Gorji:2017cai}.

\subsection{Scalar power spectrum}

The quadratic action for  the scalar perturbations from the quadratic action \eqref{action-R2b} in Fourier space can be written as
\begin{align}
\label{action-u2}
S^{(2)}_{\rm Scalar} = \dfrac{1}{2}\int{\rm d}\tau\, {\rm d}^3k~ \Big[ (u_k')^2 - \left( c_s^2k^2+\dfrac{\sigma^2}{a^2}k^4-\dfrac{\tilde{z}''}{\tilde{z}} \right) u_k^2 \Big] \,.
\end{align}
in which the prime indicates the derivative with respect to the conformal time $d \tau =dt/a(t) $ and we have defined the canonically normalized field $u \equiv z \mathcal{R}$ with $z\equiv \sqrt{\vartheta} a$. 

In a near de Sitter background  where $\vartheta$, $\c_s$ and $\sigma$ are approximately constant\footnote{The mode equation \eqref{w} was extensively analyzed without any approximations in Ref. \cite{Fujita:2015ymn}.} we have  $z''/z \simeq 2/\tau^2$ and  the corresponding mode function equation is given by 
\begin{align}
\label{w}
u''_{k} + \Big( \omega(\tau)^2-\dfrac{2}{\tau^2} \Big) u_k = 0 \, ; \hspace{0.5 cm} \omega(\tau)^2 \equiv c_s^2 k^2 + \sigma^2 H^2 k^4 \tau^2 \, .
\end{align}
The above equation indicates that we are dealing with a modified dispersion relation.
With $\mathcal{K} >0$ we have $\sigma^2 >0$ and  the dispersion relation \eqref{w} is known as the {Corley-Jacobson} dispersion relation which was studied for investigating the black holes physics \cite{Corley:1996ar,Corley:1997pr} and for the effects of trans-Planckian physics on cosmological perturbations \cite{Martin:2000xs,Martin:2002kt}. In addition, this type of dispersion relation occurs in ghost inflation  \cite{ArkaniHamed:2003uz} where a timelike scalar field  fills the entire spacetime  with the profile $\phi=t$ as in our mimetic setup.

Such modified dispersion relations indicate the violation of Lorentz invariance in the UV limit. However, for low physical momentum when
\begin{align}
\label{k-sigma}
\dfrac{k}{a} \ll \dfrac{c_s}{\sigma} \,,
\end{align}
the linear dispersion relation is recovered.  We can define the scale at which the modification to the linear dispersion relation becomes important as $\Lambda \equiv c_s/\sigma$. For the physical momentum larger than this scale, $k_{\rm phy} \gtrsim \Lambda$, the quartic contribution to the dispersion relation becomes important. For future purpose we introduce the  parameter $\nu$ via
\begin{align}
\nu \equiv \dfrac{c_s^{-1}H}{\Lambda}=\dfrac{\sigma H}{c_s^2} \,,
\label{nu}
\end{align}
which quantifies the ratio of the sound-Hubble horizon parameter over the momentum scale $\Lambda$ around when the behaviour of the dispersion relation changes. For the models in which $\nu \ll 1$ the dispersion relation is a linear relation, $\omega \propto k$,  as in standard slow-roll models, while for $\nu \gg 1$ the mode function $u_k$ is described by the non-relativistic  dispersion relation $\omega \propto k^2$.

Imposing the adiabatic vacuum initial conditions, the mode function of the comoving curvature perturbation is obtained to be \cite{Ashoorioon:2011eg,Ashoorioon:2018uey,Ashoorioon:2018ocr} 
\begin{equation}\label{sol-u}
\mathcal{R}(\textbf{k},\tau)=w_{k}(\tau) {a}_{\textbf{k}}+w_{k}^{*}(\tau) {a}^{\dagger}_{-\textbf{k}};\hspace{0.5cm} w_{k}(\tau)=\frac{i H \sqrt{\tau} e^{-\frac{\pi}{8\nu}}}{c_s k\sqrt{2\nu \vartheta}}  ~ {\rm W}_{ \frac{i}{4\nu}, \frac{3}{4}}\left(-i \nu c_s^2 k^2 \tau^2\right) \, ,
\end{equation}
where ${a}_{\textbf{k}}$ and ${a}^{\dagger}_{\textbf{k}}$ are the creation and the annihilation operators as usual and  ${\rm W}_{ \frac{i}{4\nu}, \frac{3}{4}}$ is the Whittaker function.

With the help of the above mode function, it is easy to calculate the super horizon ($c_{s}k\tau \rightarrow 0$) limit of the power-spectrum for comoving curvature perturbation. Taking into account the asymptotic behaviour of Whittaker function, i.e. $W_{a, b}(z) \approx z^{1/2-b}\Gamma(2 b)/\Gamma(b-a+1/2)$ for $z \to 0$ \cite{abramowitz1948handbook},  the curvature perturbations power spectrum on superhorizon scales  is given by
\begin{align}
\label{PR-2}
{\cal P}_{\cal R}=
\dfrac{1}{16\pi c_s^3}\dfrac{H^2}{\vartheta} g(\nu); \hspace{0.5cm} g(\nu) \equiv\dfrac{\nu^{-3/2}~e^{-\frac{\pi}{4\nu}}}{|\Gamma(\frac{5}{4}+ \frac{i}{4\nu})|^2} .
\end{align}

Let us now discuss about the asymptotic behaviour of the 
the power spectrum in small and large $\nu$ limits.   In the limit $\nu \ll 1$~\footnote{We use the relation $|\Gamma(x+iy)| \sim \sqrt{2\pi} \ |y|^{x-1/2} \ e^{-\pi|y|/2}$ as $y \rightarrow \infty$.}, we find out 
\begin{align}
\label{PRvLarge}
{\cal P}_{\cal R} &\simeq \dfrac{1}{4\pi^2c_s^3}\dfrac{H^2}{\vartheta}
\left( 1-\dfrac{5}{4}\nu^2 \right) \,.
\end{align}
Since in the limit $\nu \ll 1$ we have a relativistic dispersion relation, one expects that the power spectrum in this limit resembles that of standard slow-roll inflation. Indeed, if we formally identify the coefficient $\vartheta$ in the quadratic action (\ref{action-R2b}) 
with the corresponding factor in the action of slow-roll models \cite{Chen:2006nt}, $\vartheta \leftrightarrow 2\epsilon_H/c_s^2$, then the power spectrum in Eq. (\ref{PRvLarge}) reduces to the standard result ${\cal P}_{\cal R}  = H^2/8 \pi^2 \epsilon_H c_s$ in slow-roll  models.

On the other hand, in the limit $\nu \gg 1$ the quartic term  $k^4$ dominates in  
$\omega^2$  and the dispersion relation becomes non-relativistic as in the model of 
ghost inflation \cite{ArkaniHamed:2003uz}.  In this limit the  power spectrum (\ref{PR-2}) reduces to
\begin{align}
{\cal P}_{\cal R} = \dfrac{H^{1/2} \sigma^{-3/2}}{\pi \vartheta \, \Gamma(\frac{1}{4})^2}. 
\end{align}
Identifying a suitable choice of the parameters $\vartheta$ and $\sigma$ with the corresponding parameters in \cite{ArkaniHamed:2003uz} we reproduce the 
 power spectrum for ghost inflation as well.

Having calculated the curvature perturbation power spectrum, we can also calculate the spectral index $n_{\rm s}$ as
\begin{align}
n_{\rm s} -1 = \dfrac{\mathrm{d} \ln {\cal P}_{\cal R}}{\mathrm{d}\ln k}\bigg|_{*}
\simeq 
-2\epsilon_{_H} - 3\epsilon_{_{c_s}}-\epsilon_{_\vartheta} + \epsilon_{_g} ,
\end{align}
where the subscript $*$ shows the time of horizon crossing for the mode of interest $k$ and  we have used our slow-roll notation \eqref{epX} for the background variables $X= c_s, \vartheta, g(\nu)$. In order to have an almost scale invariant power spectrum, one requires the four parameters $\epsilon_{H}$, $\epsilon_{\vartheta}$, $\epsilon_{c_{s}}$, and $\epsilon_{g}$ to be very small.

\subsection{Tensor power spectrum}
\label{tensor}

To calculate the power spectrum of tensor perturbations,  let us first expand the tensor modes of the quadratic action \eqref{action-R2b} in terms of their polarization tensors $e_{ij}^{+}$ and $e_{ij}^{\times}$ as $\gamma_{ij}=\sum_{+,\times} \gamma^{\lambda} e_{ij}^{\lambda}$ where $\lambda=+,\times$ and 
$e_{ij}^{\lambda}$ are symmetric,  transverse and traceless tensors. Moreover, using the normalization condition, $e_{ij}^{\lambda} e_{ij}^{\lambda'}=2 \delta_{\lambda \lambda'}$, we obtain the second-order action for the tensor modes in Fourier space as follows
\ba
\label{action-v}
S^{(2)}_{\rm Tensor} & = &\frac{1}{2} \sum_{\lambda}\int{\rm d}\tau\, {\rm d}^3k~\tilde{z}^2 \Big( \gamma_{\lambda}'^2 - k^2 \gamma_{\lambda}^2 \Big) \, ,
\ea
where  $\tilde{z}^2\equiv  F(\chi)\,a^2/2$.  In order for the perturbation to be stable, we require that $F>0$.

Interestingly, from the above action we see that the tensor modes propagate with the speed equal to unity, $c_T =1$, i.e. the tensor perturbations propagate with the speed of light. This is because we considered  the special case of higher derivative coupling to gravity in the form of $F (\chi)$. However, it is well-known that for  general  higher derivative interactions with gravity,  $c_T$ is not equal to speed of light. These types of  modified gravity theories are under strong constraints from the LIGO observations which require that  $|c_{\rm T}-1|<5 \times 10^{-15} $ \cite{Monitor:2017mdv, PhysRevLett.119.251303,PhysRevLett.119.251301}.  
For example, in our setup if we allow more general higher derivative  interactions 
such as  the curvature independent quadratic higher derivative terms $\nabla_{\mu}\nabla_{\nu}\phi \nabla^{\mu}\nabla^{\nu} \phi$ and the curvature dependent cubic higher derivative terms  $\Box \phi \nabla^{\mu} \phi \nabla^{\nu} \phi R_{\mu \nu}$ and $ \nabla^{\mu}  \nabla^{\nu} \phi R_{\mu \nu}$ then $c_{T} \neq 1$ \cite{Gorji:2018okn}.

Upon defining the canonically normalized field associated with $\gamma_{\lambda}$ by $v_\lambda \equiv\tilde{z} \gamma_{\lambda}$ and imposing the the Minkowski (Bunch-Davies) initial condition,  the mode function is obtained to be 
\begin{equation}
\gamma_{\lambda}(\tau, \textbf{k})=\frac{i H e^{-i k \tau}}{k^{3/2} \sqrt{2F}}(1+i k \tau).
\end{equation}
Defining the power spectrum of the gravitational tensor modes  
via 
\begin{eqnarray}\label{CF-hh}
\sum_{\lambda} \big\langle \gamma^\lambda  {\gamma}^{\lambda} 
\big\rangle \equiv \frac{2 \pi^2}{k^{3}} 
\mathcal{P}_{\gamma} \, (2\pi)^3 \delta^{(3)}({\bf k}-{\bf k}') \,,
\end{eqnarray}
we obtain 
\begin{align}
{\cal P}_{\cal \gamma} &= 
\dfrac{2H^2}{\pi^2 \,F} \,,
\label{PowerTensor}
\end{align}
where both $H$ and $F$ are evaluated at the time of horizon crossing. 
Compared to conventional models of inflation, we see the additional factor $1/F$ in tensor power spectrum. This is understandable if one notes that, naively speaking, we  have rescaled the gravitational coupling $M_{\rm P}^2 \rightarrow M_{\rm P}^2 F$ in the starting action Eq. (\ref{action0}).

The spectral index of $\mathcal{P}_{\gamma}$ is also given by
\begin{align}\label{nt}
n_{\rm t} \equiv  \dfrac{{\rm d}\ln {\cal P}_\gamma}{{\rm d}\ln k}|_{*}
= -2\epsilon_{_H} - \epsilon_{_F}
\end{align}
where 
$\epsilon_{_F}$ is the slow-roll parameter associated with $F$ as defined in Eq. (\ref{epX}) which is given by  
\begin{align}
\label{eF}
 \epsilon_{_F}  \equiv  \dfrac{\dot{F}}{H \ F}
= -\dfrac{3  H F_\chi}{F}\ \epsilon_{_H} \,=-\epsilon_{_H}\Big[1+\tilde{\nu}^2(1+3c_s^2)\Big] \, ,
\end{align}
where $\tilde{\nu} \equiv  \sqrt{{\mathcal{K}}/{F}}$ and we have used Eq. \eqref{cs2} in the last step. In particular, we see that $n_{\rm t}$ depends  on $c_{s}$ and $\tilde{\nu}$ after plugging the slow parameter $\epsilon_{F}$ from Eq. \eqref{eF} into Eq. \eqref{nt}. 

\begin{figure}[t!]
	\centering
	\includegraphics[width=0.7\linewidth]{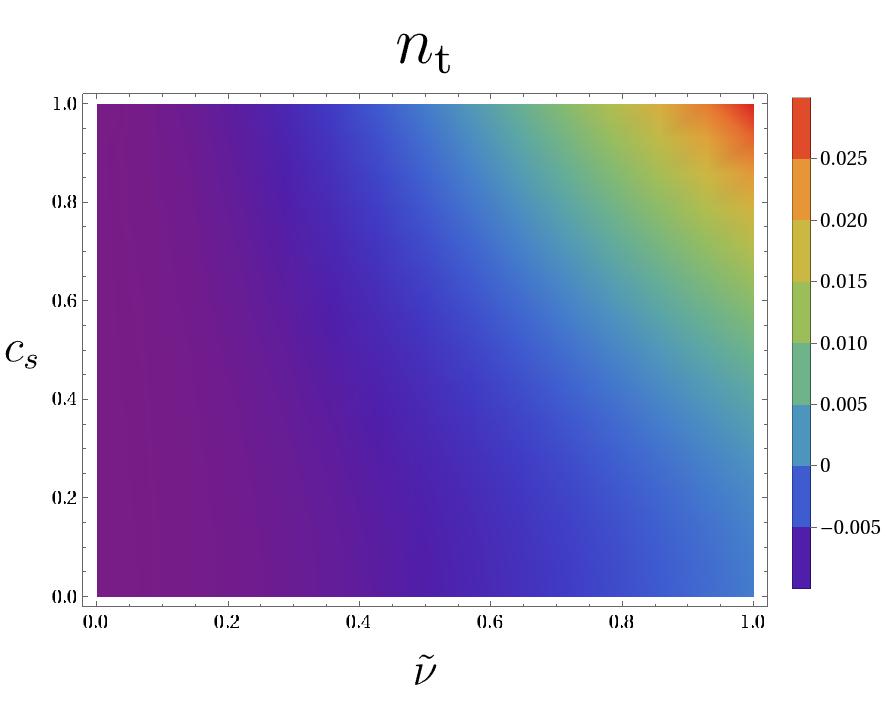}
	\caption{The density plot of $n_{\rm t}$ versus $c_{s}$ and $\tilde{\nu} \equiv  \sqrt{{\mathcal{K}}/{F}}$. We have taken $\epsilon_{H}=0.01$. }
	\label{fig:nt}
\end{figure}

In Fig. \ref{fig:nt} we have presented the predictions for $n_{\rm t}$ for some values of $(c_{s},\tilde{\nu})$ in the parameter space. Interestingly, we see that in some regions of parameter space $n_{\rm t} >0$, i.e. the tensor power spectrum is blue-tilted.  This is unlike the conventional slow-roll models which generally predict a red-tilted tensor power spectrum. As is the case in our model, the detection of a blue-tilted tensor perturbations  cannot rule out inflation automatically \cite{Khoury:2001wf, Khoury:2006fg,Koshelev:2020foq}.  

As long as we assume $\tilde{\nu} \lesssim \mathcal{O}(\sqrt{\epsilon_{_H}})$, then
Eq. \eqref{eF} guarantees that $\epsilon_{_F} \simeq -\epsilon_{_H}$ in the subluminal regime with $0<c_{s}<1$. It means that the function $F(\chi)$ changes very slowly during slow-roll inflation. Therefore, we can consider it approximately 
as  a constant during inflation.  To estimate this value, let us first define the tensor to scalar ratio as follows,
\begin{align}
\label{rt}
r_{\rm t} &\equiv \dfrac{{\cal P}_\gamma}{{\cal P}_{\cal R}}=
 \dfrac{32 c_s^3 \vartheta}{\pi F g(\nu)} \, .
\end{align}
Then, by restoring ${ M}_{\rm P}$ in the scalar and tensor power spectra and using the current observational constraint on inflationary parameters \cite{Akrami:2018odb}, i.e.  $r_{\rm t} \lesssim 0.056$ and
${\cal P}_{\cal R} \simeq 2.1 \times 10^{-9}$, the value of  $F$ at horizon crossing can be estimated as
\begin{align}
\label{Fchi0}
F_{*}
 = 1.72 \times 10^{-3} \left(\dfrac{r_{\rm t}}{0.056}\right)^{-1} \Big(\dfrac{H}{10^{-6} \ {\rm M}_{_{\rm P}}}\Big)^2 \, ,
\end{align}
which implies that we need to choose $\mathcal{K}\lesssim 10^{-5}$ to satisfy the condition $\tilde{\nu} \lesssim {\cal O}(\sqrt{\epsilon_{_H}})$.
As mentioned before, the slow roll approximation $\epsilon_{_F} \approx -\epsilon_{_H}$ is valid only in the region where the comoving curvature perturbation propagates with $c_s <1$.  The superluminal propagation speed with $c_s >1$  is not a problem per se as it does not directly violate causality  on the background \cite{ Babichev:2007dw, PhysRev.182.1400}. However, we restrict ourselves to  scalar perturbations with subluminal speeds.

\begin{figure}[t!]
	\centering
	\includegraphics[width=0.7\linewidth]{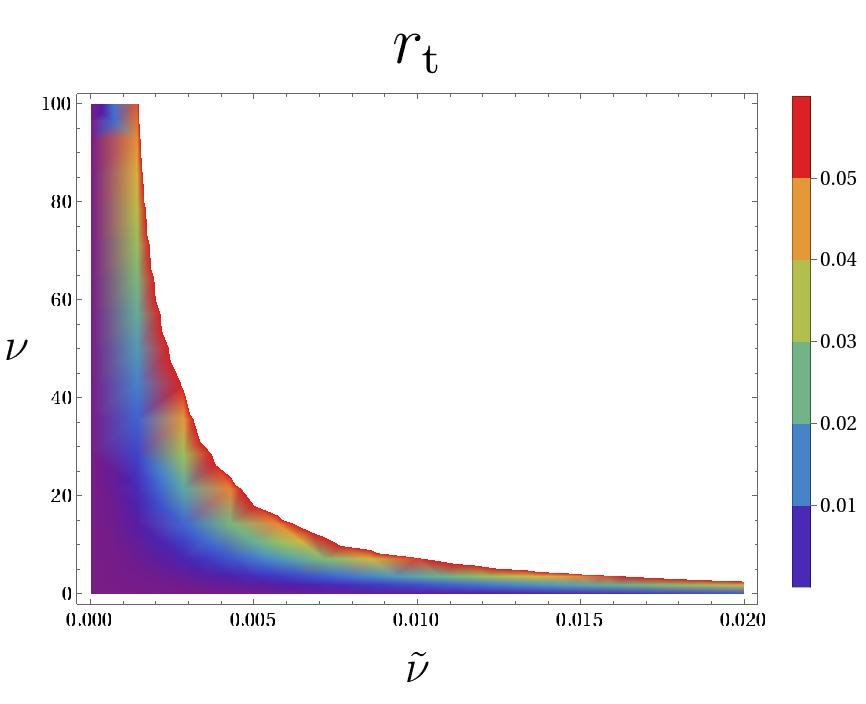}
	\caption{The density plot of $r_{\rm t}$ versus $\nu$ and $\tilde{\nu}$ for $c_s=1$. The white regions are excluded by the observational bound $r_t\leq0.056$ \cite{Akrami:2018odb}.}
	\label{fig:rt}
\end{figure}

Using Eqs. \eqref{eF} and \eqref{nt}, we can also obtain the following generalized 
consistency relation between $r_{\rm t}$ and $n_{\rm t}$,
\begin{equation}
r_{\rm t}=\frac{32 c_{s}}{\pi g(v)(\tilde{\nu}^2+1)} \Big(\frac{n_{\rm t}}{\epsilon_{H}}+1-\tilde{\nu}^2\Big) \, .
\end{equation}
We see that the  consistency relation in conventional models of inflation \cite{GARRIGA1999219}, $r_{\rm t}=-8 c_{s} n_{\rm t}$, is modified  in our model due to the mimetic constraint. In Fig. \ref{fig:rt} we have presented the predictions for $r_{\rm t}(\nu,\tilde{\nu})$ for $c_s=1$. The white areas correspond to the regions of 
 parameter space which are not allowed   due to the observational bound $r_t\leq0.056$ \cite{Akrami:2018odb}. By choosing smaller values of $c_s$ the allowed regions become more extended.

Before closing this section, here we compare our results for the inflationary background
 with those of  \cite{Zheng:2017qfs}. We have shown that in order for a consistent inflationary solution to exist in this  setup, the potential has to be negative. Then imposing the additional condition that the parameter ${\cal K}$ defined in Eq. (\ref{A}) be a constant we have verified the existence of a period of slow-roll inflation as demonstrated in Fig. \ref{fig:slow-roll}. Then calculating the quadratic actions and performing the perturbation analysis  we have shown that the 
 spectral tilt of tensor perturbations can take either signs. On the other hand, Ref. \cite{Zheng:2017qfs} claimed the existence of slow-roll solution with a potential which is (implicitly) positive\footnote{There is a discrepancy with the signature of the action used in  \cite{Zheng:2017qfs}. While they use the $(+, -, -, -)$ signature as in Chamseddine-Mukhanov \cite{Chamseddine:2013kea}, but their action has an opposite sign for the Einstein-Hilbert term. Fortunately, this sign discrepancy does not affect their perturbation analysis about the ghost/gradient instabilities but it has important effects when writing the background equation. More specifically, their potential should be replaced by $-V$. }. As for the predictions of the scalar and tensor power spectra they have borrowed the analysis of \cite{Fujita:2015ymn} which was in a different context. As a result, they have obtained the standard result $n_{\rm t}= -2 \epsilon_H$, so the tilt of tensor perturbations is always negative.

\section{Primordial Bispectra}
\label{Bispec}

In this section, we calculate the three-point correlation of the scalar perturbations 
 $\langle\mathcal R \mathcal R\mathcal R \rangle$ and look at the 
 amplitudes and shapes of non-Gaussianity in various limits.

Utilizing the standard methods, the expectation value of the three point correlation is given by  \cite{Maldacena:2002vr}
\begin{equation}
\Big\langle \mathcal R \left(\bfk_{1}\right)\mathcal R\left(\bfk_{2}\right)\mathcal R\left(\bfk_{3}\right) \Big\rangle  = -i \int_{\tau_{\rm i}}^{\tau_{\rm e}}  a \ {\rm d}\tau \Big\langle 0 \Big\vert \Big[ \mathcal R(\tau_{\rm e},\,\bfk_1)\mathcal R(\tau_{\rm e},\,\bfk_2)\mathcal R(\tau_{\rm e},\,\bfk_3)~,\, H_{\rm int} \Big] \Big\vert 0 \Big\rangle \,,
\label{3-point-f}
\end{equation}
where $H_{\rm int}$ is the interaction Hamiltonian which is calculated from expanding the  Lagrangian \eqref{action0}  up to 3rd orders in curvature perturbations, given in  \eqref{cubic action2},  with $H_{\rm int}=-\mathcal{L}_{3}$. Moreover, $\bfk_{\rm i}$ are the wave vectors and $\tau_{\rm i}$ is the initial time when the inflationary perturbations are deep inside the Hubble radius. Since during a quasi-de Sitter expansion $\tau \simeq -1/(a H)$, it is a good 
approximation to calculate the integral in the limit $\tau_{\rm i} \to -\infty$ and $\tau_{\rm e} \to 0$. 

In the Fourier space, we can write the three-point correlation function of curvature perturbations as
\begin{eqnarray}
\Big \langle \mathcal R\left(\bfk_{1}\right) \mathcal R\left(\bfk_{2}\right)\mathcal R\left(\bfk_{3}\right) \Big \rangle  \equiv  \left(2\pi\right)^{3}\delta^{3}\left(\bfk_{1}+\bfk_{2}+\bfk_{3}\right) \ \mathcal{B}_{\mathcal R}\left(k_{1},k_{2},k_{3}\right)\,,
\end{eqnarray}
in which $k_i=\vert\bfk_i\vert$ and $B_{\mathcal R}\left(k_{1},k_{2},k_{3}\right)$ is called  the bispectrum \footnote{ Because of the translational invariance, the total momentum ${\bf K}\equiv \bfk_{1}+\bfk_{2}+\bfk_{3}$ is conserved.} which can be parameterized as 
\begin{equation}\label{fnl1}
B_{\mathcal R}\left(k_{1},k_{2},k_{3}\right) \equiv \frac{(2 \pi)^4 \ \mathcal{P}_{\cal R}^2}{
\prod_{n=1}^{3} k_{i}^3} \mathcal{A}(k_{1},k_{2},k_{3})
\end{equation}
where $\mathcal{A}$ is called the amplitude of bispectrum.

Finally,  the non-linearity  parameter $f_{_{\rm NL}}$ associated with the amplitude of bispectrum  is defined by the following relation
\begin{equation}\label{fnl2}
f_{_{\rm NL}}  \equiv \frac{10}{3}\frac{1}{
\sum_{i=1}^{3} k_{i}^3} \mathcal A \left(k_{1},k_{2},k_{3}\right) \, .
\end{equation}

As we see from Eq. \eqref{cubic action2},  our interaction Hamiltonian contains 22 independent terms (interactions). These complicated interactions originate from the higher derivative terms in $F(\chi)$ and $P(\chi)$. 
Each of them induce different shapes and amplitudes of non-Gaussianities.  

As examples, let us calculate the bispectrum   for the following two  terms of the cubic action \eqref{cubic action2}, 
\ba
\label{HInt}
{\cal L}_{\rm int} \supset f_2 ~ \dot{\calR}\dfrac{{\left( \partial \calR \right)}^2}{a^2} + f_8~\dot{\calR}^3 \, ,
\ea
which also exist in the model of ghost inflation  with the modified dispersion relation $\omega^2 \propto k^4$. 

The bispectrum for each term in Eq. \eqref{HInt} is evaluated using the mode function  of $\mathcal{R}$ given in Eq. \eqref{sol-u}  as follows 
\begin{align}
\label{eq:integral2} 
\mathcal B_{\mathcal R}\left(k_{1},k_{2},k_{3}\right)_{2} & =
i f_{2} \
w^{*}_{k_1}(0) w^{*}_{k_2}(0) w^{*}_{k_3}(0) 
\int_{-\infty}^0 {\rm d}\tau \ \frac1{H\tau} w_{k_1}(\tau)
w_{k_2}(\tau) w^{\prime}_{k_3}(\tau) \ (\bfk_1 \cdot \bfk_2)
\\ \nonumber & + {\rm symm.} + {\rm {c.c.}}
\end{align}
and
\begin{align}
\label{eq:integral8} 
\mathcal B_{\mathcal R}\left(k_{1},k_{2},k_{3}\right)_{8} & =
6 i f_{8} \
w^{*}_{k_1}(0) w^{*}_{k_2}(0) w^{*}_{k_3}(0) 
\int_{-\infty}^0 {\rm d}\tau \frac1{H\tau} w^{\prime }_{k_1}(\tau)
w^{\prime }_{k_2 }(\tau) w^{\prime }_{k_3}(\tau) 
 +{\rm {c.c.}}\,  .
\end{align}
Using the explicit expression for the wave function \eqref{sol-u} and substituting the above results into Eq. \eqref{fnl1} for $\mathcal{A}$, we obtain the following expressions for the amplitudes $\mathcal{A}^{(2)}$ and $\mathcal{A}^{(8)}$ associated with each interaction:
\begin{align}
\label{eq:NGfinal} \mathcal{A}^{(2)} & = \frac{ f_{2} H^5}{c_{s}^8 \mathcal{P}_{\cal R}^2 \vartheta^3 }
 \sum_{i,j,l=1}^{3} |\epsilon_{ijl}| \ \mathcal{I}^{^{p_{i},p_{j},p_{l}}}_{_{1,-1}}(v) \ k_{i} (\bfk_{j}.\bfk_{l}) \, ,
\end{align}
with $p_{1}=1$ and $p_{2}=p_{3}=0$, and
\begin{align}
\label{eq:NGfinal2} \mathcal{A}^{(8)} & = \frac{6 f_{8} H^5}{c_{s}^6 \mathcal{P}_{\cal R}^2 \vartheta^3 }
k_{1}k_{2}k_{3} \  \mathcal{I}^{^{1,1,1}}_{_{0,-1}}(\nu) \, .
\end{align}
Here the  function $\mathcal{I}_{_{n_{_1},n_{_2}}}^{^{p_{_1},p_{_2},p_{_3}}}$ is defined via 
\begin{equation}
\label{I-def}
 \mathcal{I}^{^{{p_{_1},p_{_2},p_{_3}}}}_{_{n_{_1},n_{_2}}}(\nu,k_{1},k_{2},k_{3})  \equiv  \mathrm{Re} \Big[ \alpha(\nu) \nu^{-(n_{1}+\frac{3}{4})} \int_{-\infty}^0 dx \;x^{n_{2}} h^{(p_{1})}_{\nu}(x)
h^{(p_{2})}_{\nu}\big( \frac{k_2}{k_1} x\big) h^{(p_{3})}_{\nu}\big( \frac{k_3}{k_1} x\big) \Big] \, ,
\end{equation}
where 
\begin{equation}
\alpha(\nu) \equiv \frac{15 e^{\frac{(i-3)\pi}{8 \nu}}}{7680\pi^{5/2}} \Gamma
   \Big(\frac{5}{4}+\frac{i}{4 \nu}\Big)^{-3} \hspace{0.5cm} \text{and} \hspace{0.5cm} h_{\nu}(x)  \equiv  \frac{e^{-\frac{\pi }{8 \nu}} \sqrt{x} \
   W_{\frac{i}{4 \nu},\frac{3}{4}}\left(-i
   x^2\right)}{\nu^{3/4}} \; ,
\end{equation}
and the upper index $p_{_i}$ denotes the order of derivative with respect to the function variables. For example $h^{(0)}_{\nu}(x) = h_\nu(x), h^{(1)}_{\nu}(x) = \frac{d\, h_\nu(x)}{d x}  $ and so on.  The amplitudes for all other interactions are listed in Appendix \ref{AppC}. 

\begin{figure}[h!]
	\centering
	\includegraphics[width=0.32\linewidth]{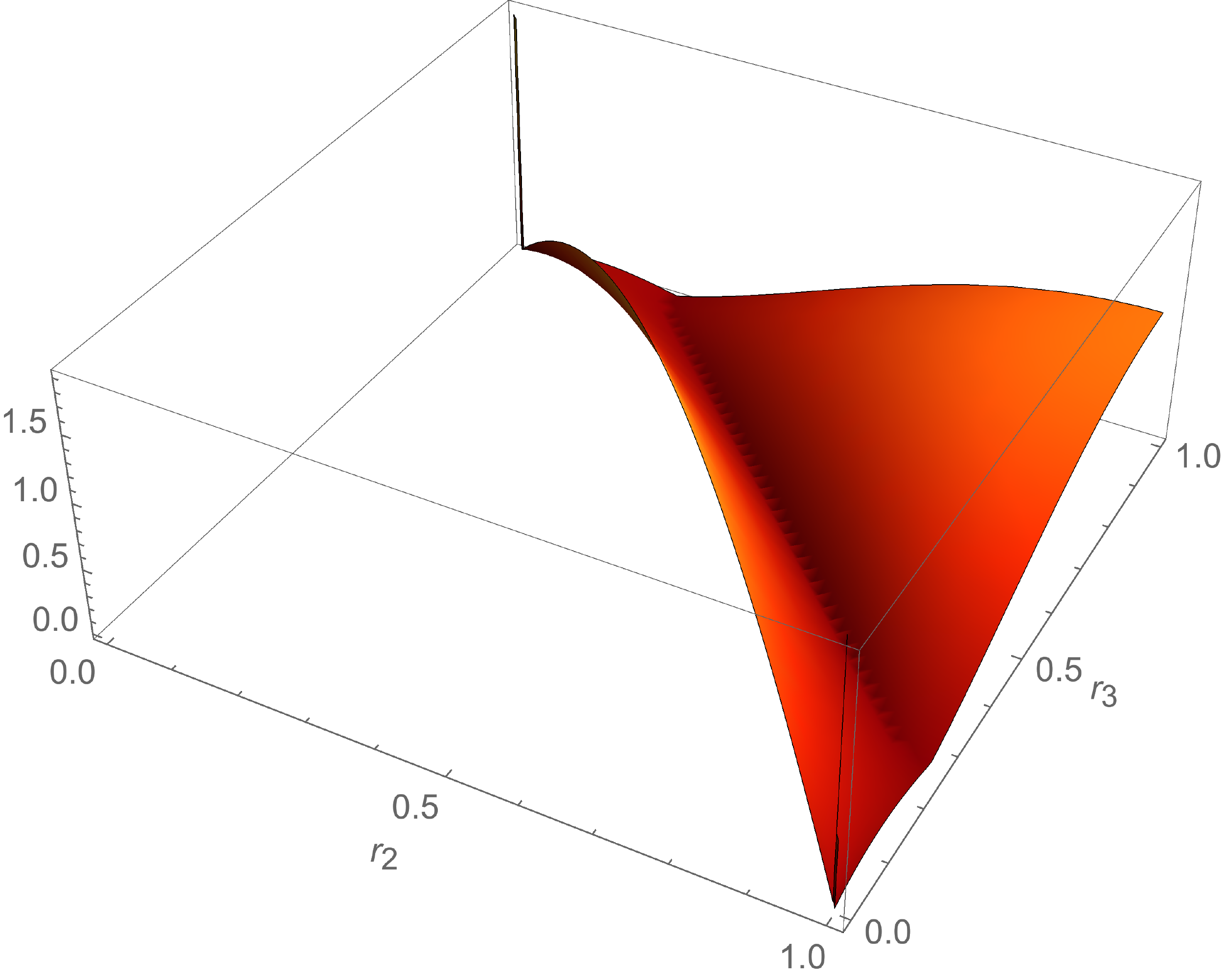}
	\includegraphics[width=0.32\linewidth]{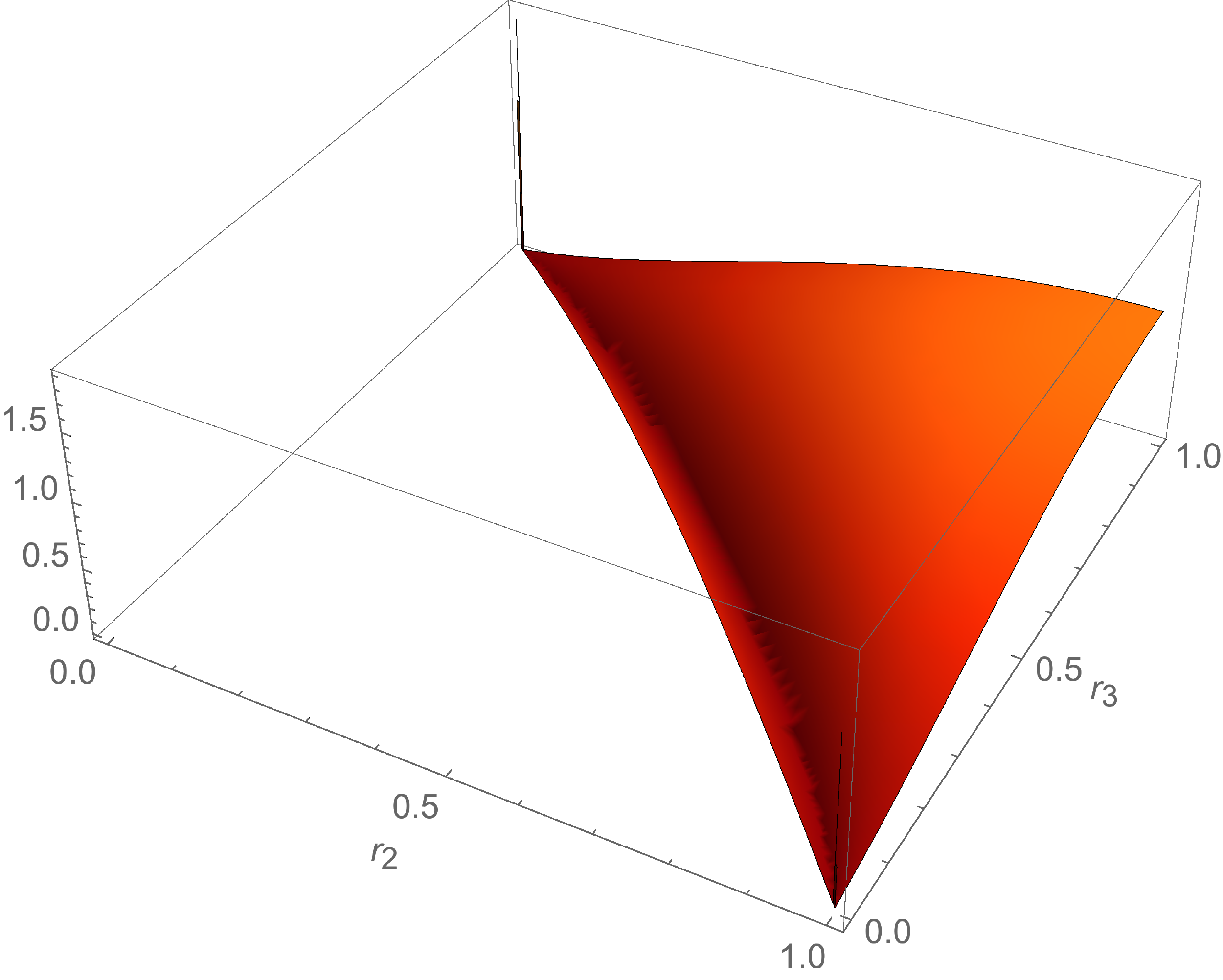}
	\includegraphics[width=0.32\linewidth]{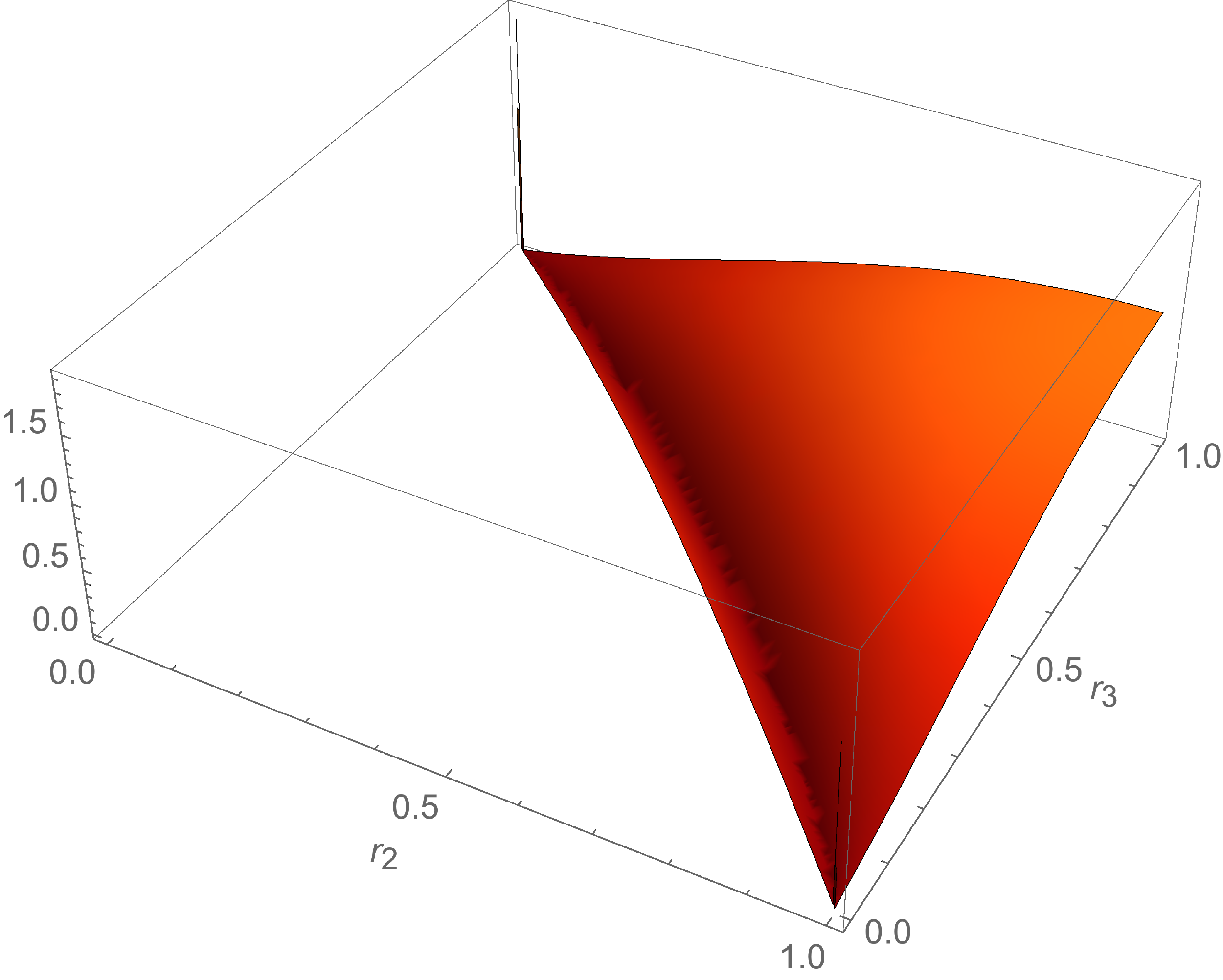}
	\caption{Numerical results for the shapes of  $|\mathcal{A}^{(2)}|/k_{1}k_{2}k_{3}$ for $\nu=1,10,50$ from left to right. The amplitude is  normalized by the value obtained in the equilateral limit.}
	\label{shape2}
\end{figure}

\begin{figure}[h]
	\centering
	\includegraphics[width=0.32\linewidth]{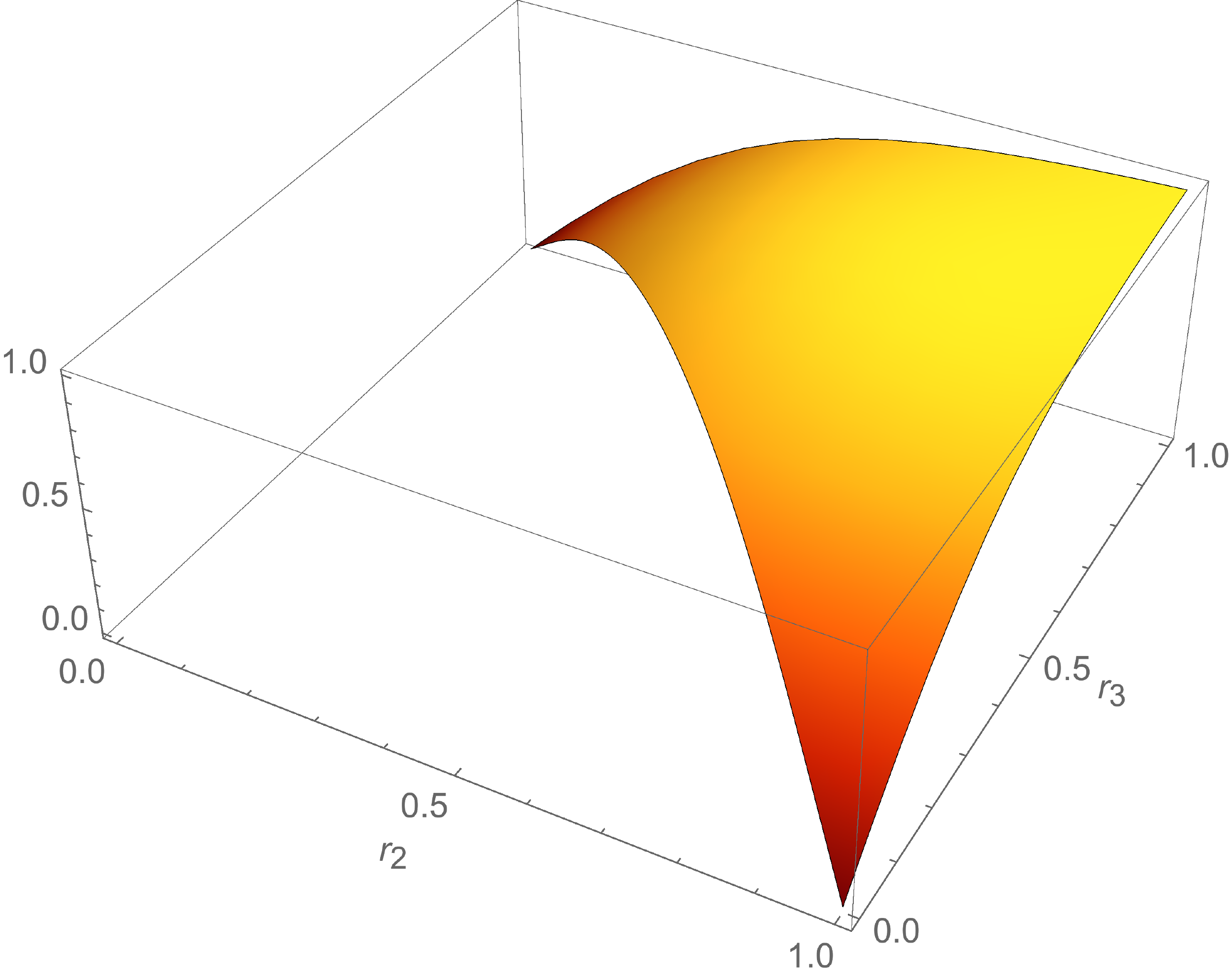}
	\includegraphics[width=0.32\linewidth]{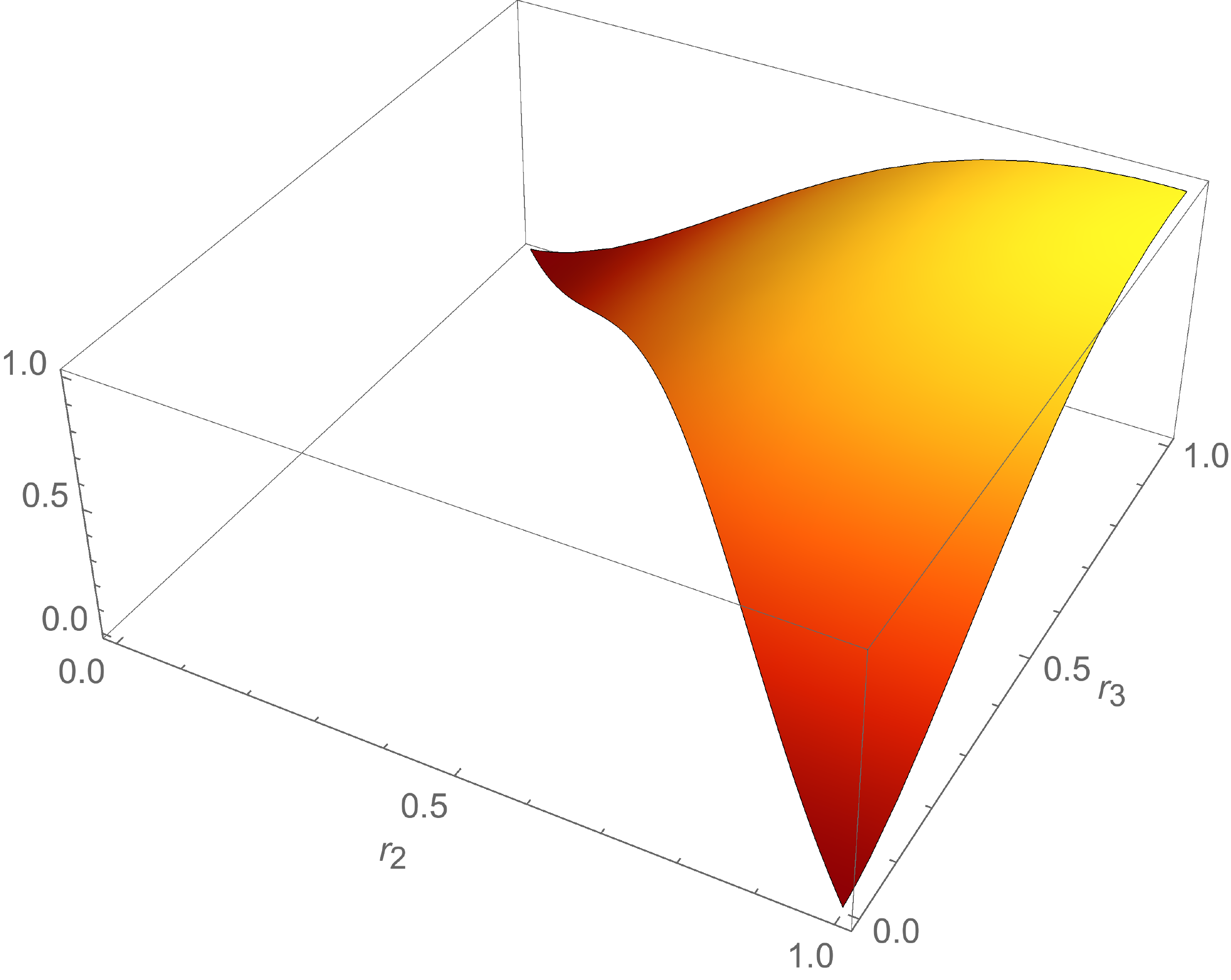}
	\includegraphics[width=0.32\linewidth]{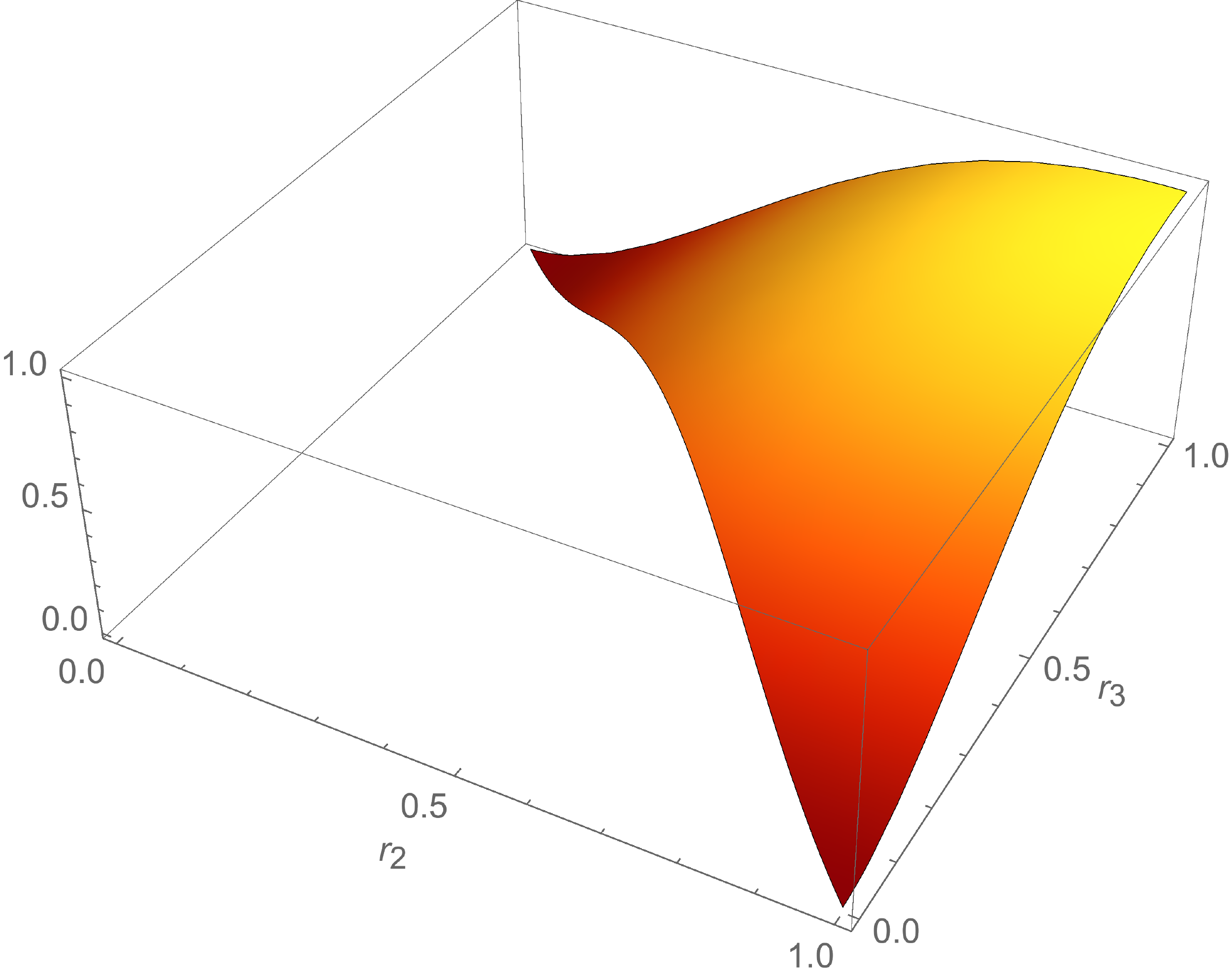}
	\caption{Numerical results for the shapes of  $|\mathcal{A}^{(8)}|/k_{1}k_{2}k_{3}$ for $\nu=1,10,50$ from left to right with the description as in Fig. \ref{shape2}.  }
	\label{shape8}
\end{figure}

To study the shape function of the above amplitudes,  in Figs. \ref{shape2} and \ref{shape8} we have presented the 3D plot of   $r_{2}^{-1} r_{3}^{-1} \mathcal{A}(1,r_{2},r_{3})$ as a function of $r_{2}\equiv k_{2}/k_{1}$ and $r_{3}\equiv k_{3}/k_{2}$ for $\nu=\{1,10,50\}$. The plots are produced numerically, after rotating the contour of integration over $\tau$ along the direction
$\propto -(1+i)$ so that they converge exponentially. We  see that $\mathcal{A}^{(2)}$ and $\mathcal{A}^{(8)}$ roughly have similar shapes and amplitudes and both roughly peak at the equilateral limit $k_{1}=k_{2}=k_{3}=k$. In addition,  the variation of $\nu$ has no significant effects on the shapes. 

\begin{figure}[h!]
	\centering
	\includegraphics[width=0.4\linewidth]{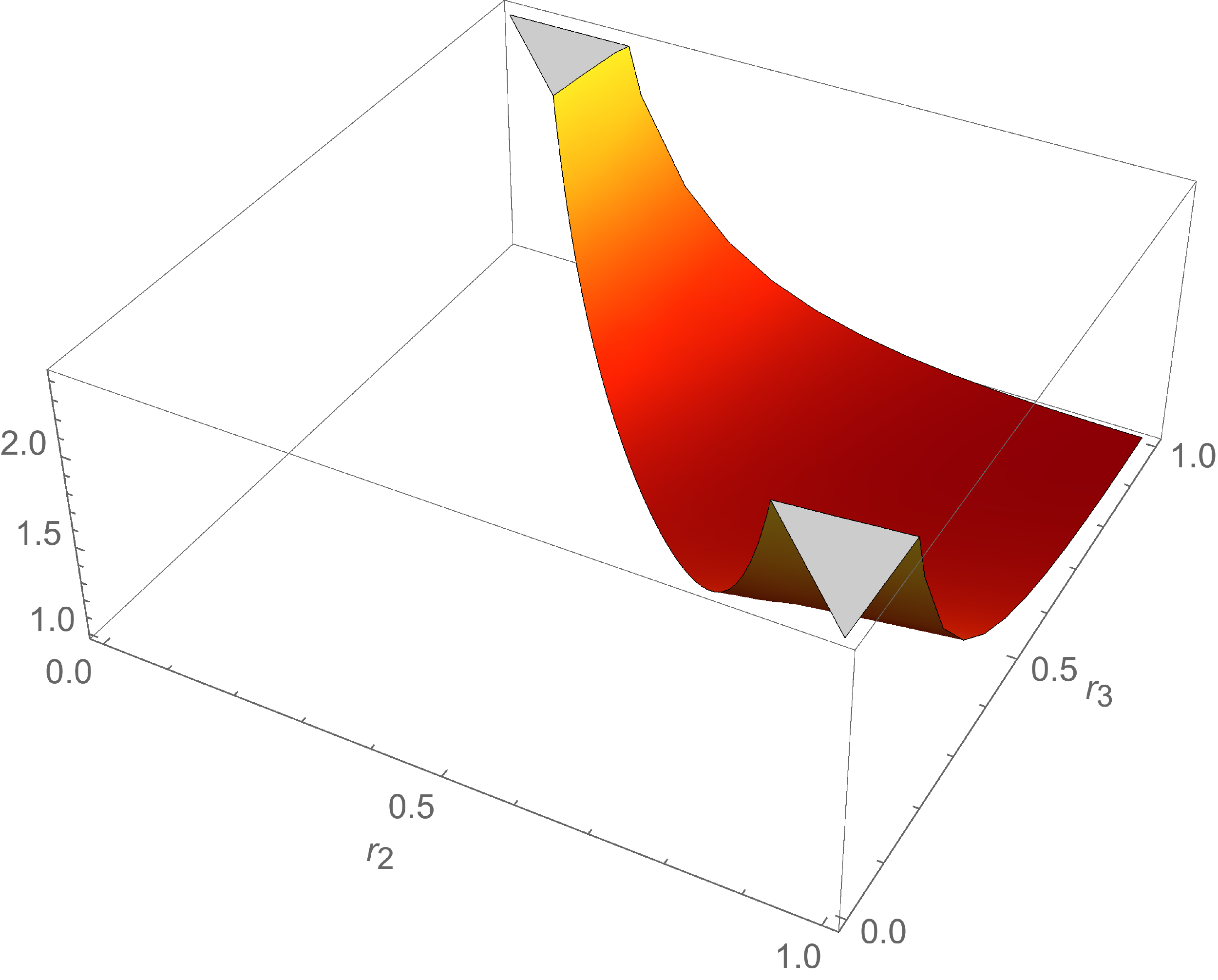}
	\includegraphics[width=0.4\linewidth]{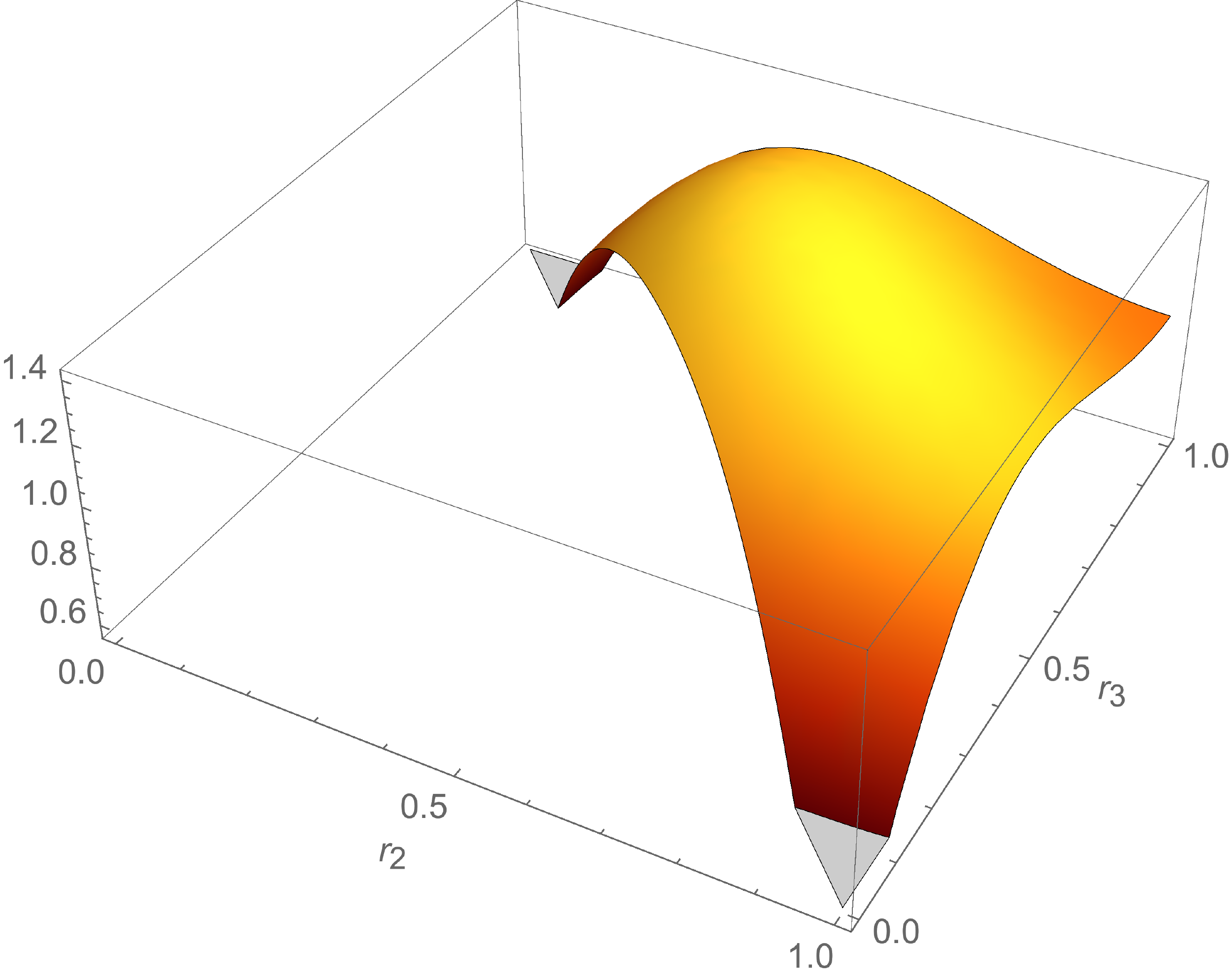}
	\caption{Numerical results for the shapes of  $|\mathcal{A}^{(4)}|/k_{1}k_{2}k_{3}$ (left panel) and $|\mathcal{A}^{(10)}|/k_{1}k_{2}k_{3}$ (right panel) for $\nu=1$. The amplitude is  normalized by the value obtained in the equilateral limit.\label{shape410}}
	\end{figure}

\begin{table}[]
\begin{center}
\begin{tabular}{c ||cccccccccc}
 \hline\hline
 Amplitude & $\mathcal{A}^{(1)}$ & $\mathcal{A}^{(3)}$ & $\mathcal{A}^{(4)}$ &$\mathcal{A}^{(5)}$&$\mathcal{A}^{(6)}$ &$\mathcal{A}^{(7)}$ &$\mathcal{A}^{(9)}$ &$\mathcal{A}^{(10)}$ &$\mathcal{A}^{(11)}$ &$\mathcal{A}^{(12)}$ 
 \\  \hline\hline 
  Shape & Local & Equi & Local & Equi &Equi & Local & Equi & Ortho & Equi &Equi\\ 
   \hline\hline  \hline \hline
 Amplitude & $\mathcal{A}^{(13)}$ & $\mathcal{A}^{(14)}$ & $\mathcal{A}^{(15)}$ &$\mathcal{A}^{(16)}$&$\mathcal{A}^{(17)}$ &$\mathcal{A}^{(18)}$ &$\mathcal{A}^{(19)}$ &$\mathcal{A}^{(20)}$ &$\mathcal{A}^{(21)}$ &$\mathcal{A}^{(22)}$ 
  \\ \hline\hline
  Shape & Equi & Local & Equi & Equi &Equi & Equi & Equi & Equi & Ortho &Equi  \\\hline \hline
\end{tabular}
\caption{The shape of bispectrum for each interaction listed in Appendix. \ref{AppC}.   }\label{tab3}
\end{center}
\end{table}	

In Table \ref{tab3} we list the shape of each contribution presented in  Appendix \ref{AppC}. One can see that most of the non-Gaussianity shapes peak at the equilateral limit where all three  modes have comparable wavelengths. However, some  shapes are close to the orthogonal shape and the local shape which has a peak in the squeezed limit. For example, as shown in Fig. \ref{shape410}, $\mathcal{A}^{(4)}$ and $\mathcal{A}^{(10)}$ peak in  the squeezed triangle limit ($k_{3}\ll k_{1} \simeq k_{2}$) and in orthogonal triangle limit ($k_{3}=k_{2}=k_{1}/2$), respectively.

Combining the contributions from all interactions listed  in Appendix \ref{AppC}, the total non-Gaussianity parameter $f_{_{\rm NL}}$ is  given by
\begin{equation}
f_{_{\rm NL}} = \frac{10}{3}\frac{1}{\sum_{i=1}^{3} k_{i}^3} \sum_{j=1}^{22} \mathcal A^{(j)} \, .
\end{equation}   
Correspondingly, we can calculate $f_{_{\rm NL}}$ numerically for squeezed $(k_{1} =k_{2}=k, k_{3} \to 0)$, equilateral ($k_{1} = k_{2} = k_{3} = k$) and 
orthogonal ($k_{1} =k, k_{2}= k_{3} = k/2)$ shapes.

\begin{figure}[t!]
	\centering
	\includegraphics[width=0.49\linewidth]{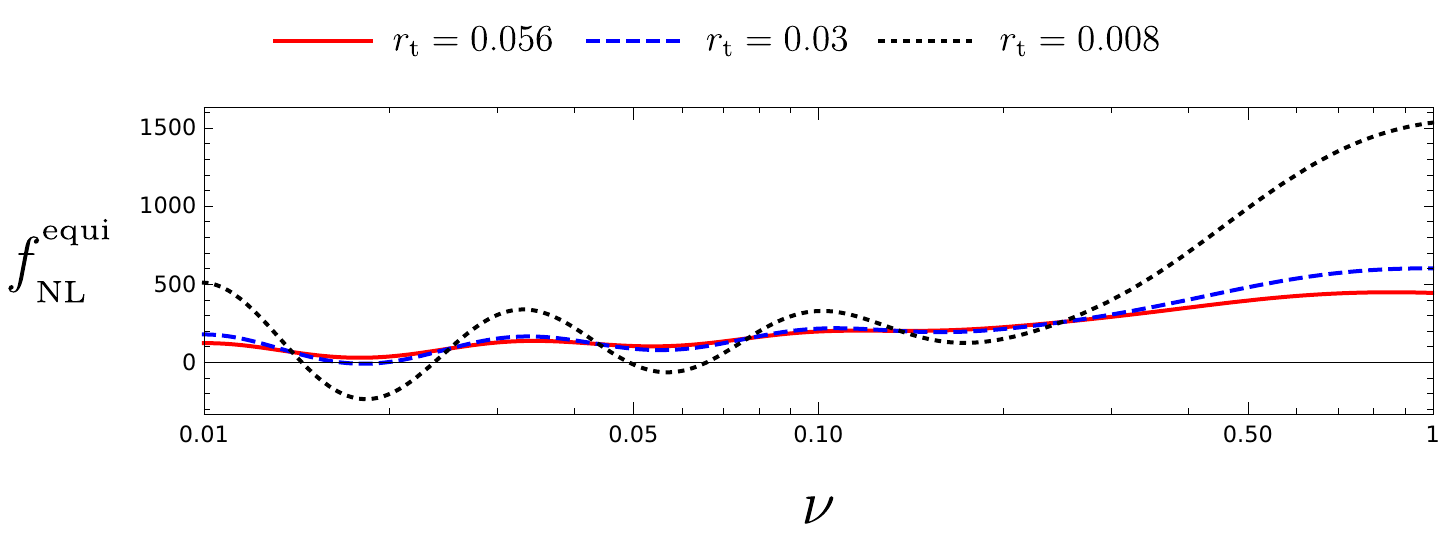}
	\includegraphics[width=0.49\linewidth]{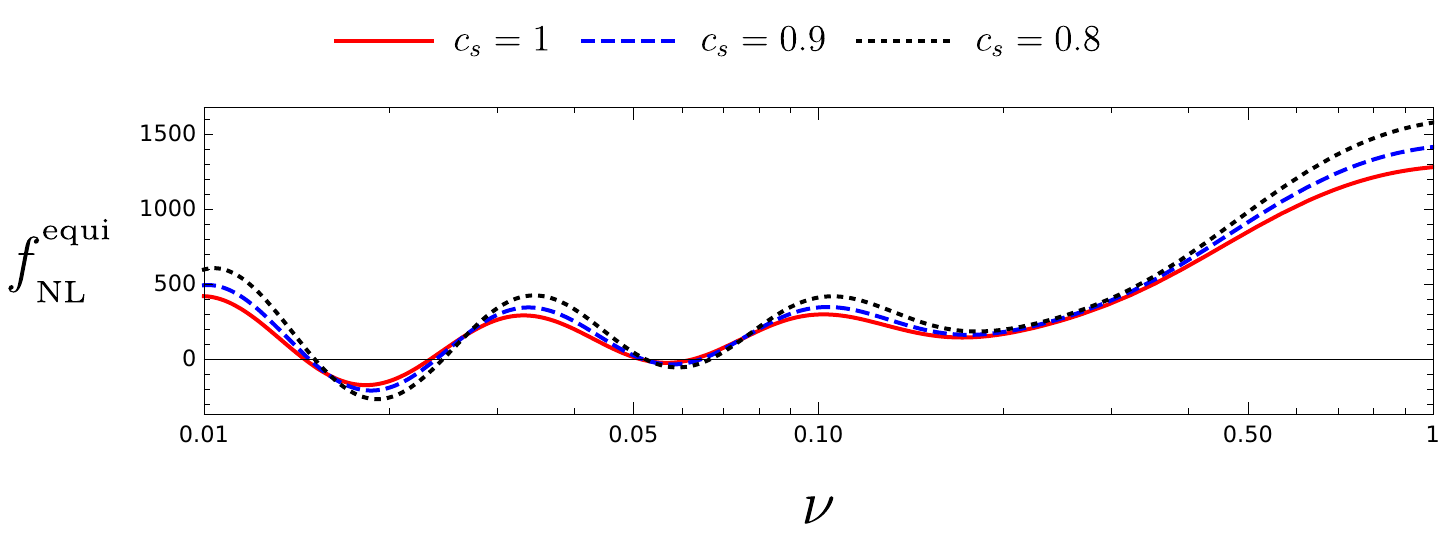}
	\\
	\includegraphics[width=0.49\linewidth]{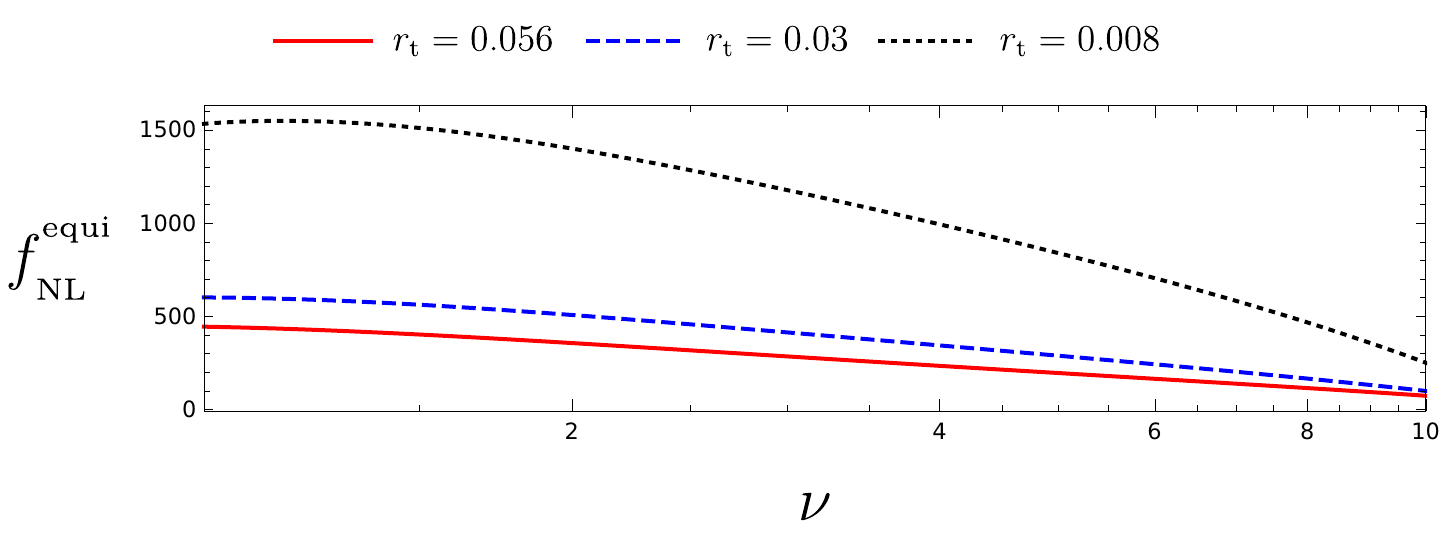}
	\includegraphics[width=0.49\linewidth]{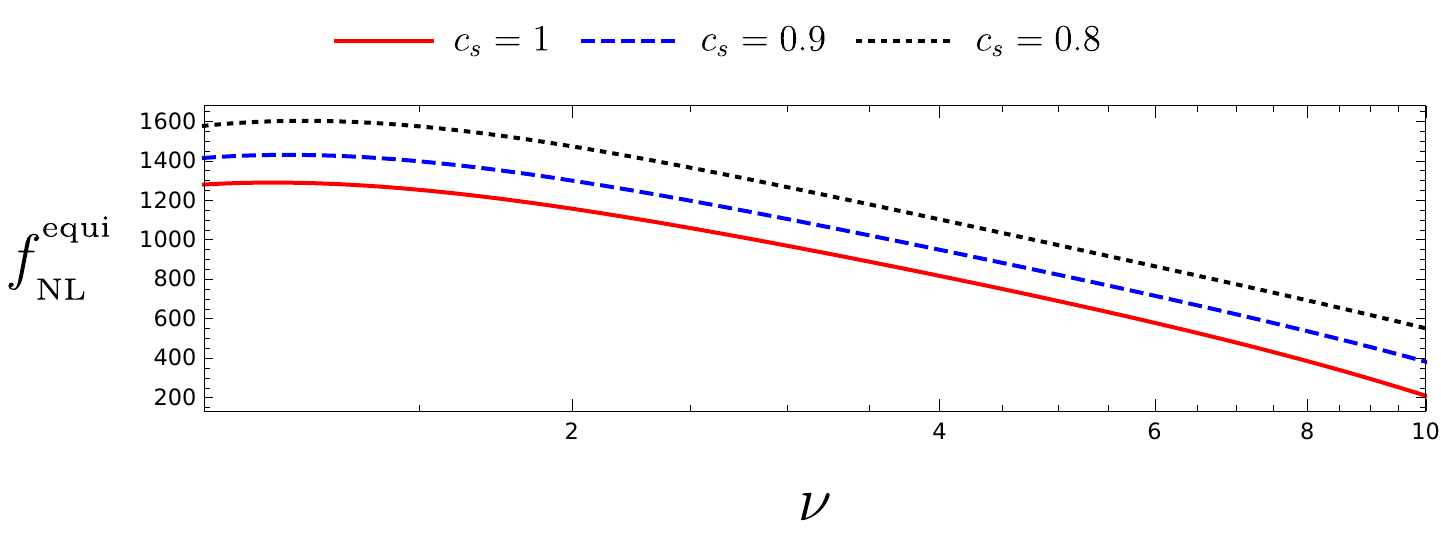}
	\\
	\includegraphics[width=0.49\linewidth]{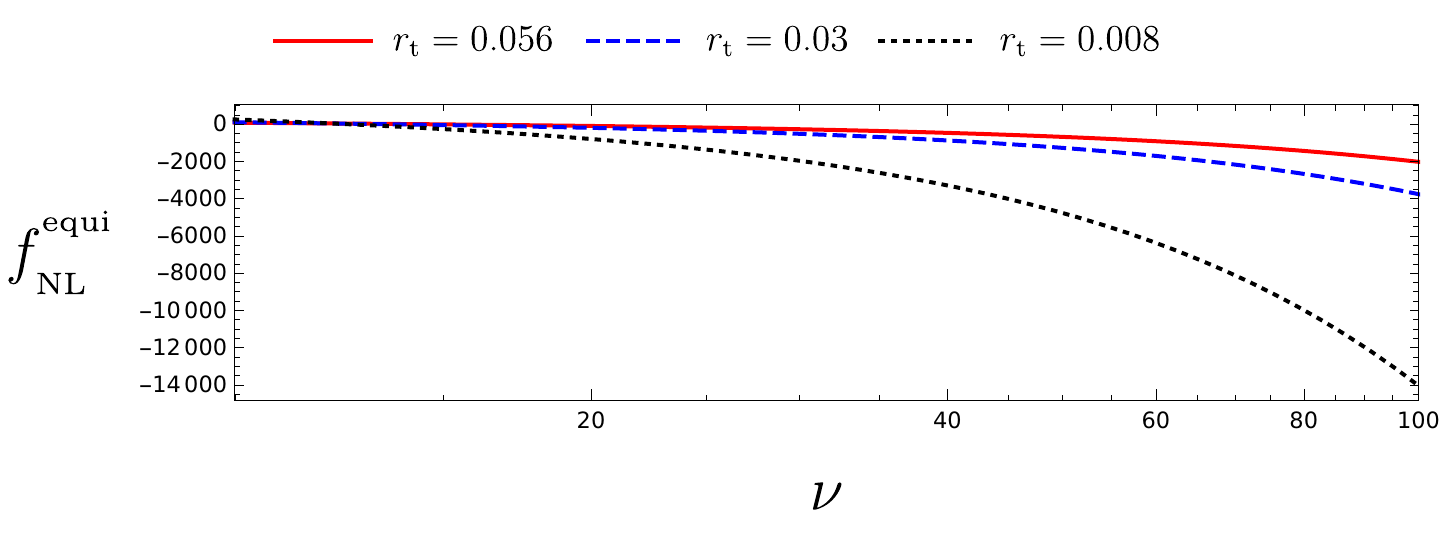}
	\includegraphics[width=0.49\linewidth]{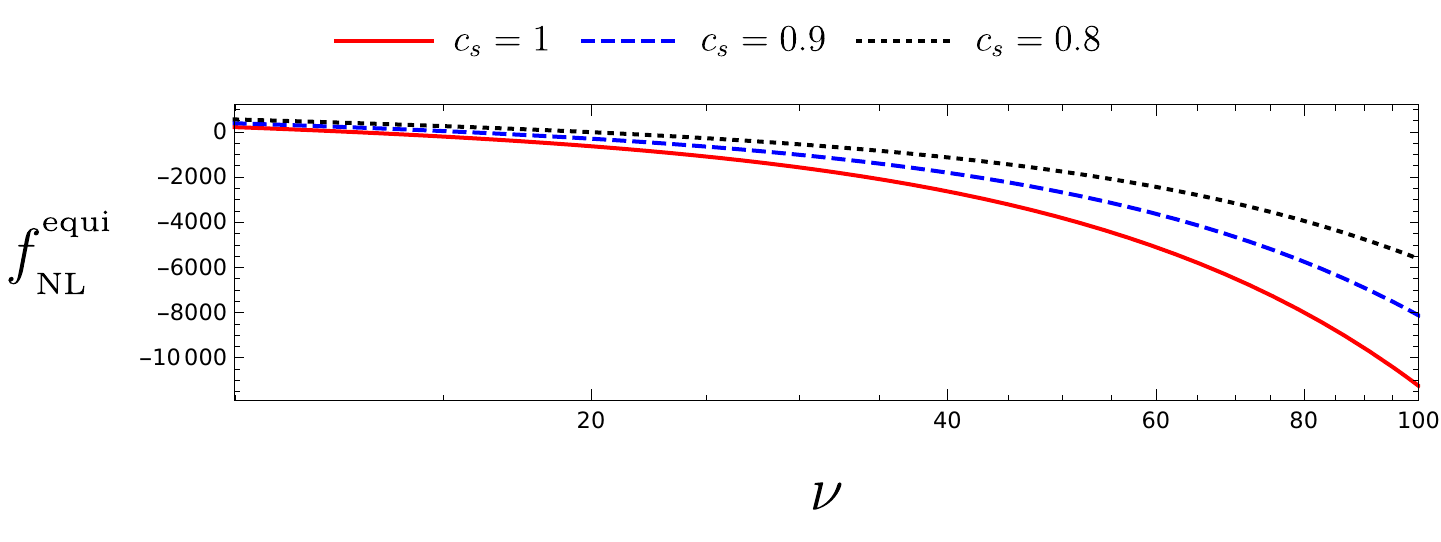}
	\caption{Numerical results for $f_{_{\rm NL}}$ in the equilateral configuration
	with  $\nu $ in the range $[0.01,100]$. In the left panels we have fixed $c_{s}=1$ while  $r_{\rm t}$ is varied whereas in the right panels we have set $r_{\rm t}=0.01$ and $c_s$ is varied.}
	\label{fnv1}
\end{figure}

\begin{figure}[h!]
	\centering
	\includegraphics[width=0.49\linewidth]{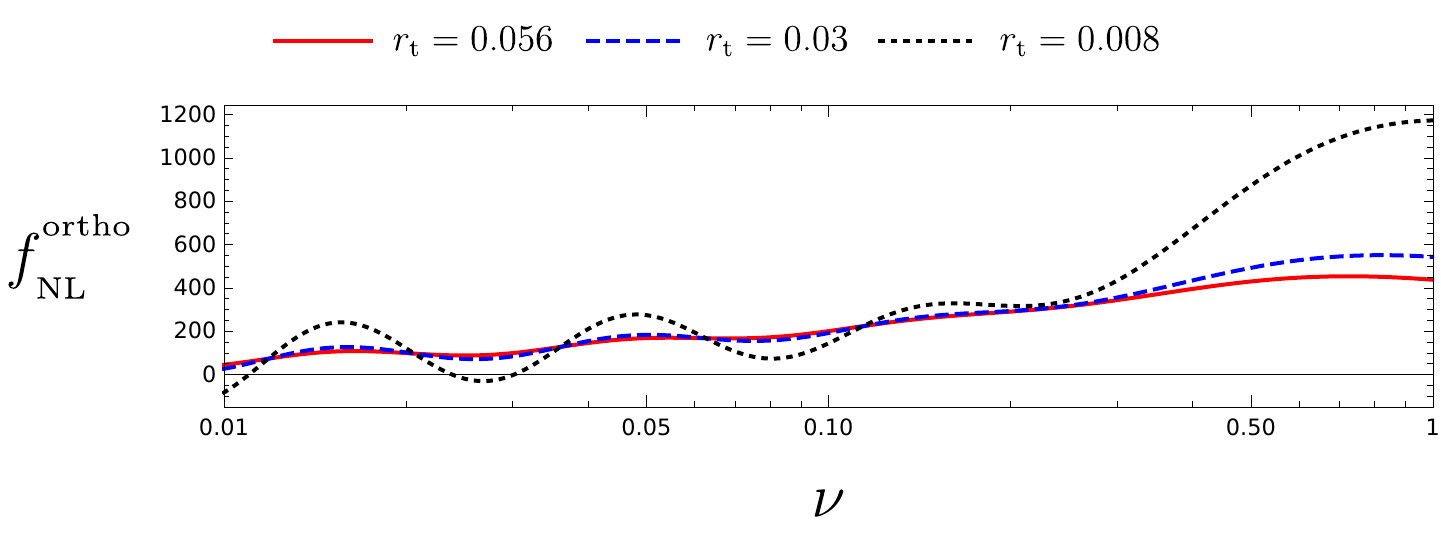}
	\includegraphics[width=0.49\linewidth]{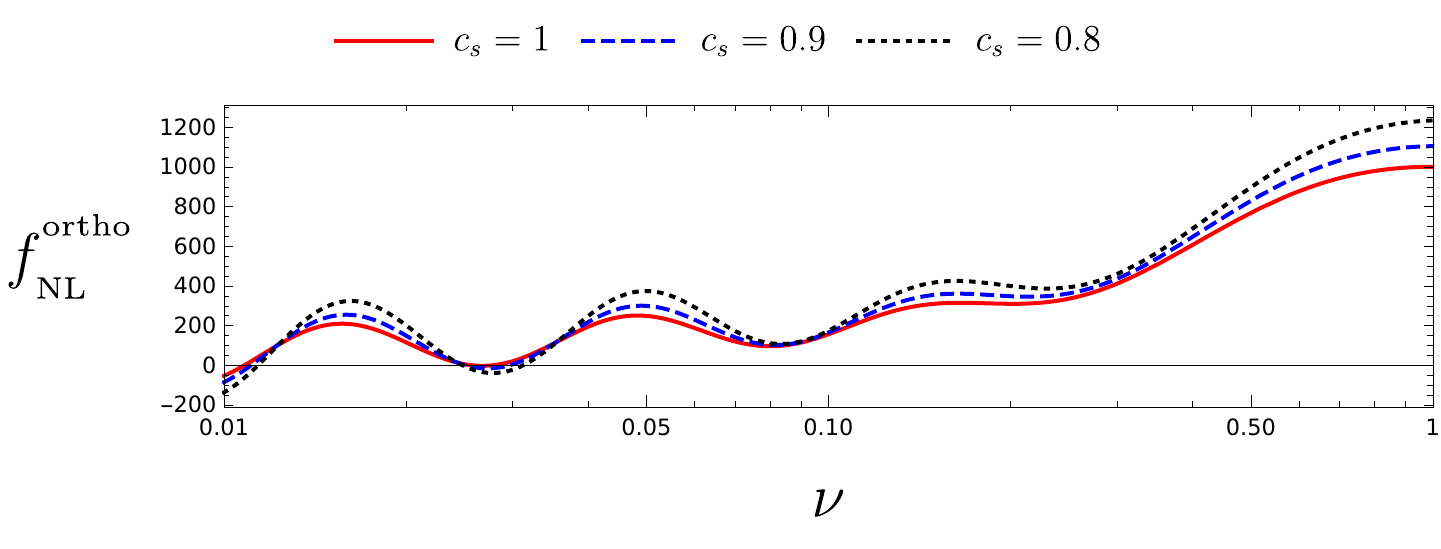}
	\\
	\includegraphics[width=0.49\linewidth]{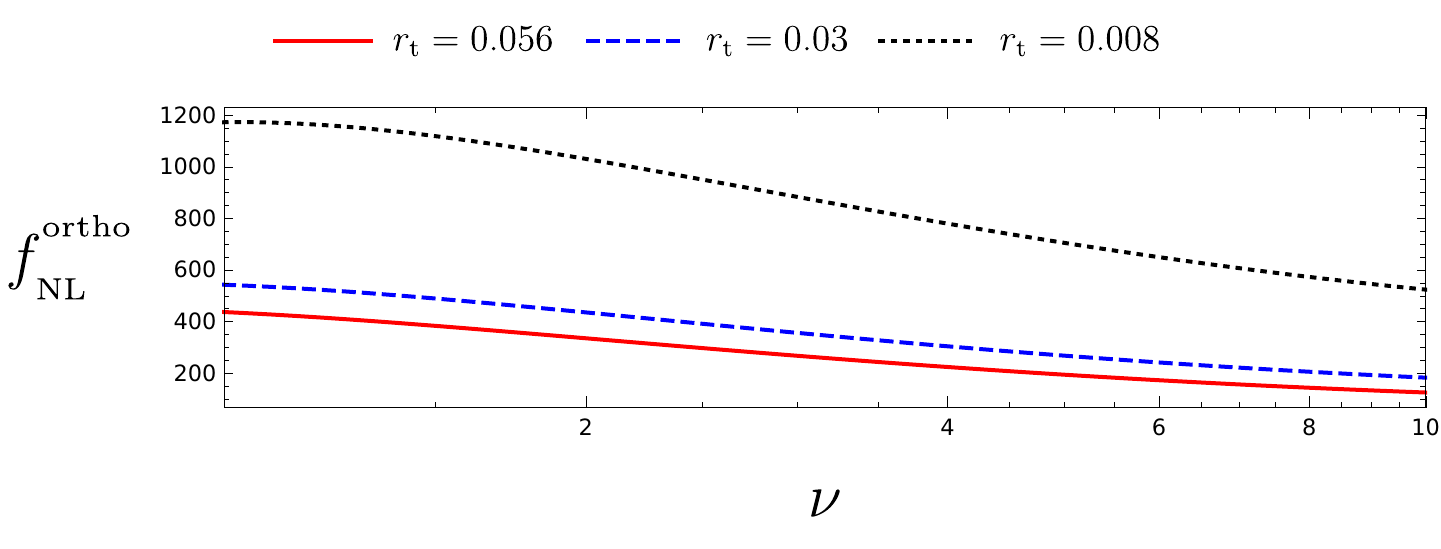}
	\includegraphics[width=0.49\linewidth]{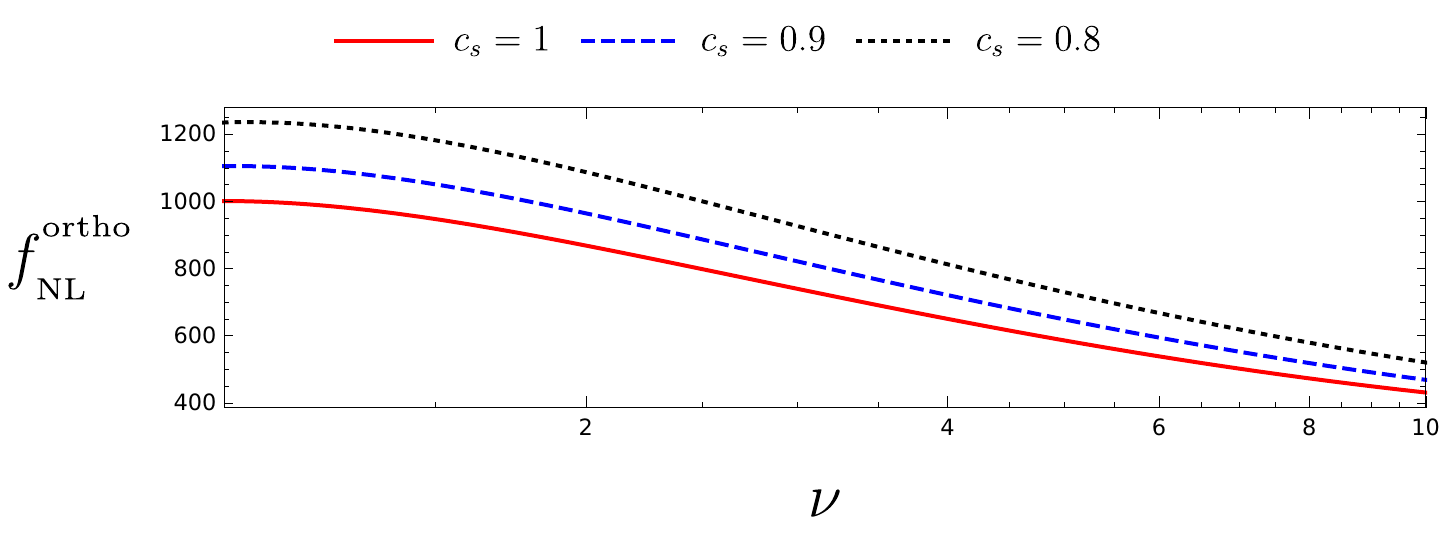}
	\\
	\includegraphics[width=0.49\linewidth]{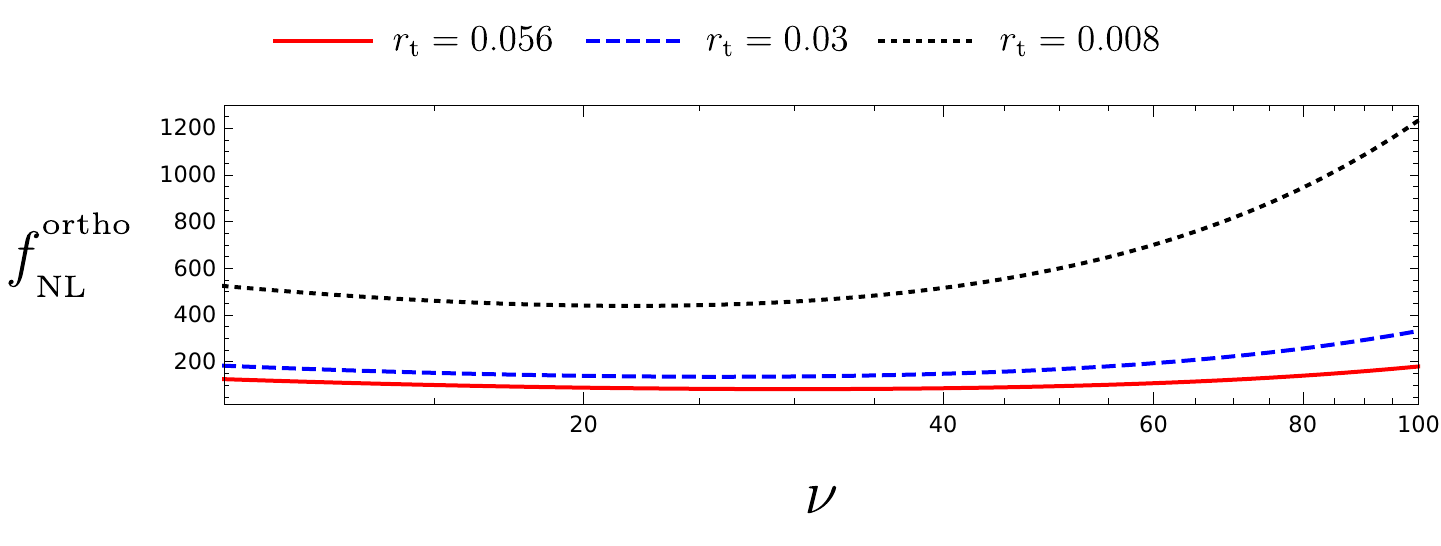}
	\includegraphics[width=0.49\linewidth]{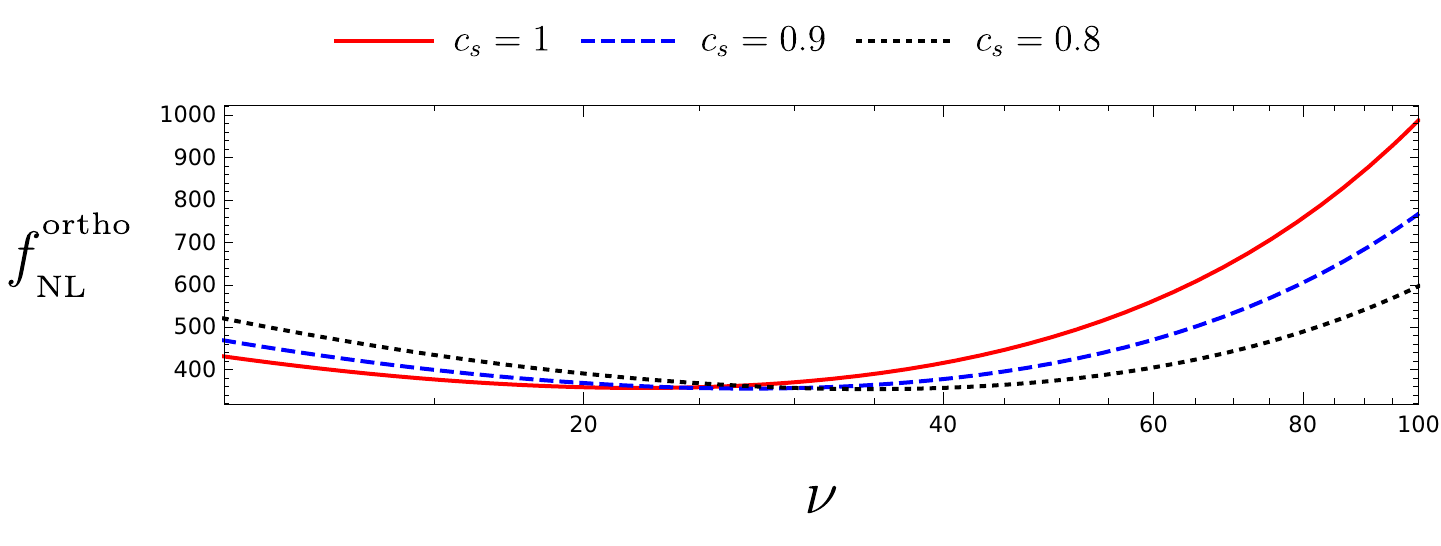}
	\caption{Numerical results for $f_{_{\rm NL}}$ in the range of $\nu=[0.01,100]$ for orthogonal configuration. In the left panels we have considered $c_{s}=1$, whereas in the right panels we have set $r_{\rm t}=0.01$.}
	\label{fnv2}
\end{figure}

\begin{figure}[h!]
	\centering
	\includegraphics[width=0.49\linewidth]{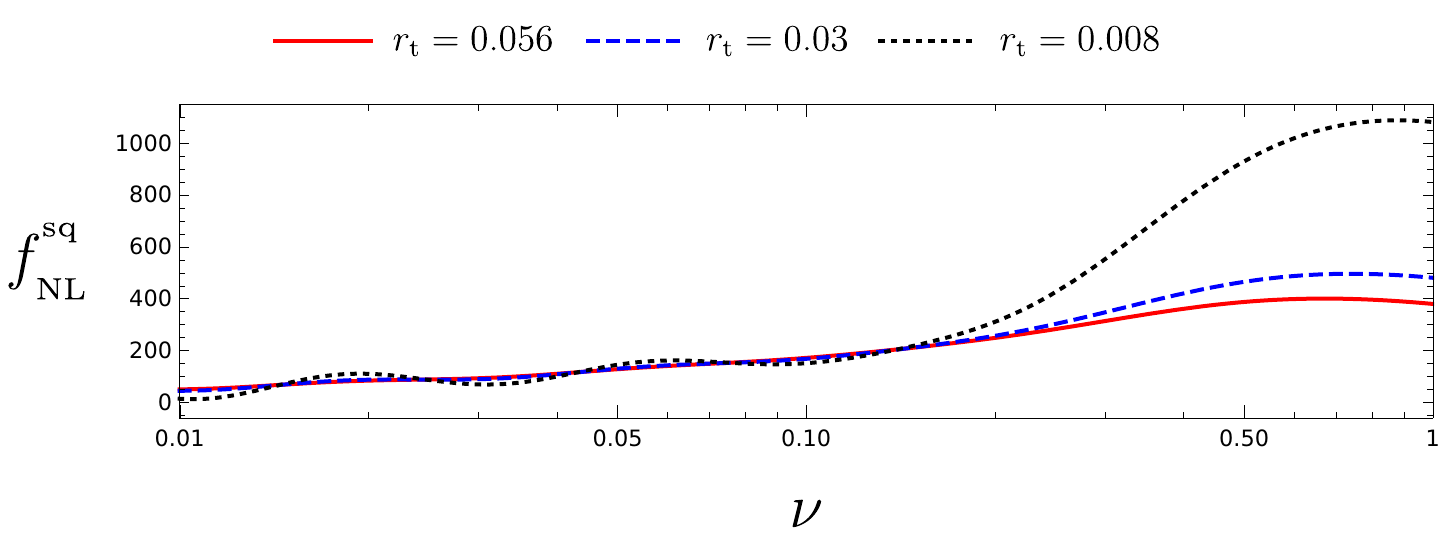}
	\includegraphics[width=0.49\linewidth]{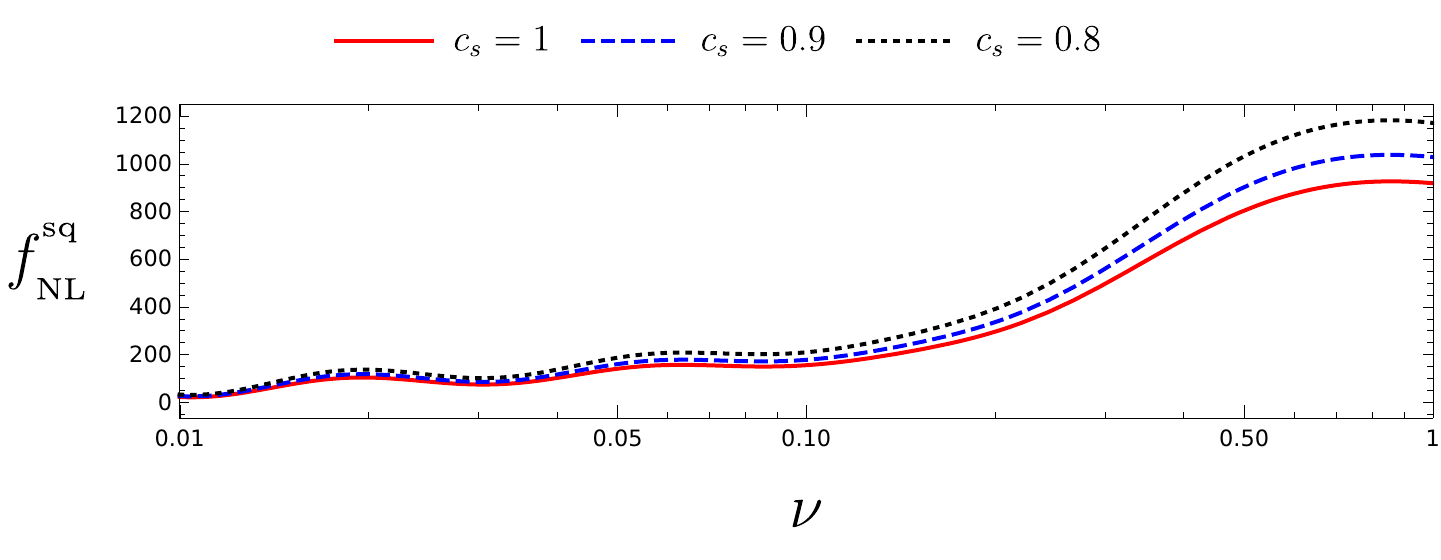}
	\\
	\includegraphics[width=0.49\linewidth]{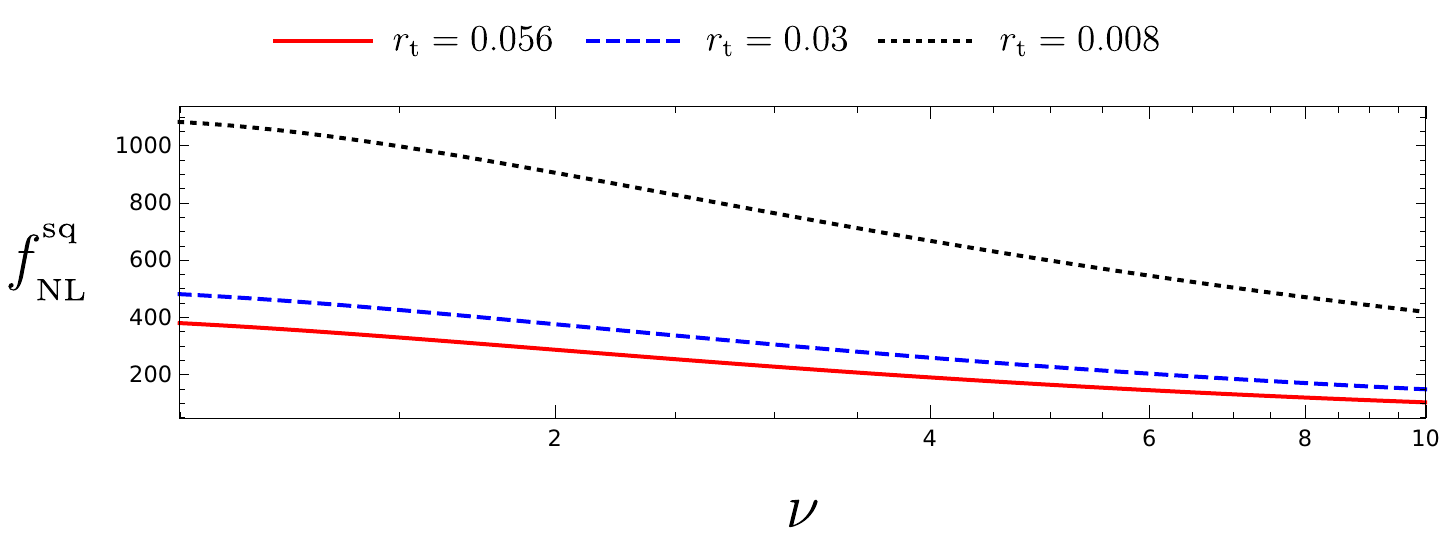}
	\includegraphics[width=0.49\linewidth]{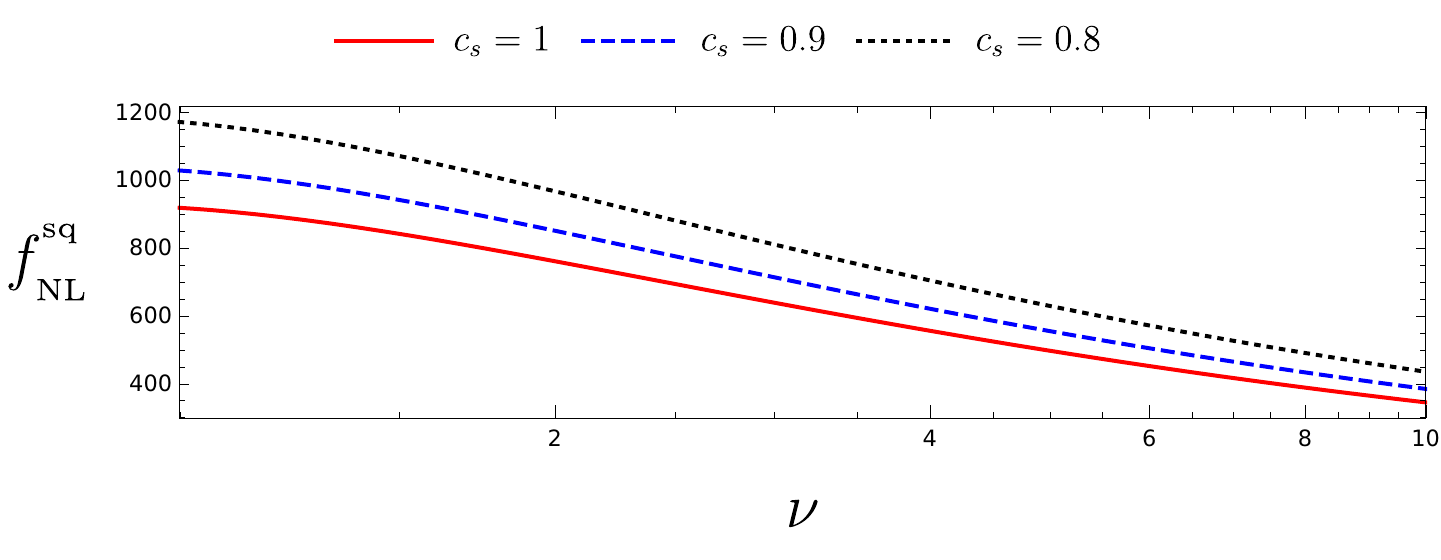}
	\\
	\includegraphics[width=0.49\linewidth]{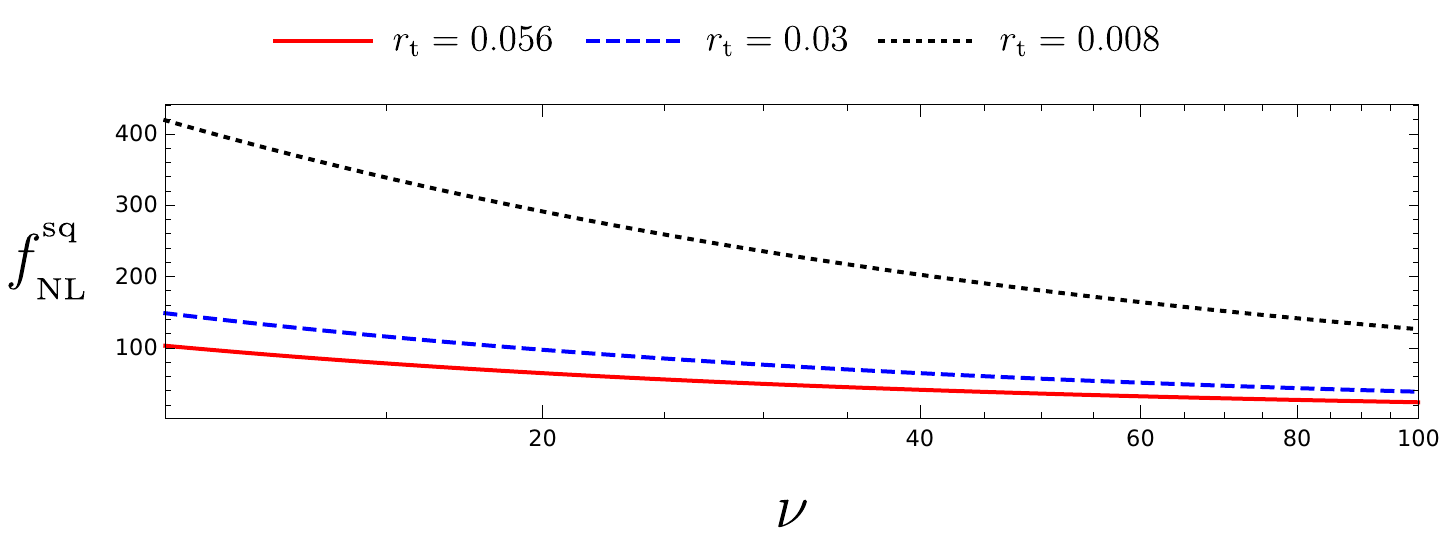}
	\includegraphics[width=0.49\linewidth]{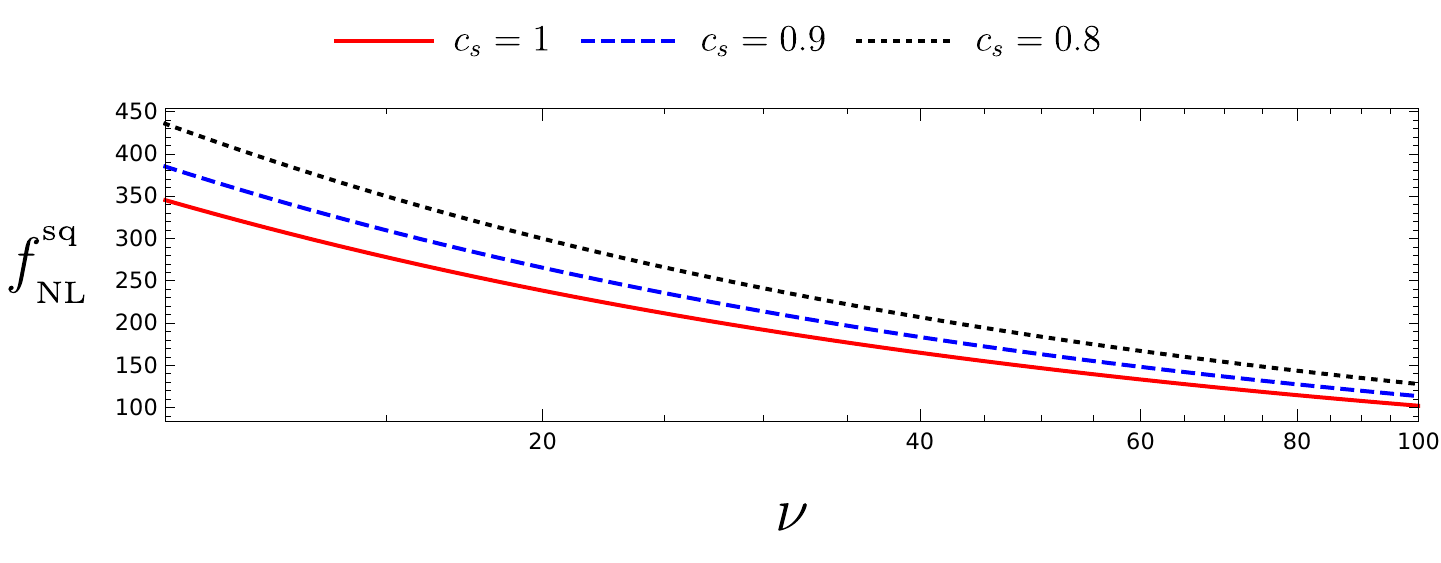}
	\caption{Numerical results for $f_{_{\rm NL}}$ in the range of $\nu=[0.01,100]$ for squeezed configuration. In the left panels we have considered $c_{s}=1$, whereas in the right panels we have set $r_{\rm t}=0.01$.}
	\label{fnv3}
\end{figure}

In Figs. \ref{fnv1}, \ref{fnv2}, and \ref{fnv3}, $f_{_{\rm NL}}$ is presented in the various range of $\nu$ in the  squeezed, equilateral and orthogonal configurations. 
It is worth mentioning that $f_{_{\rm NL}}$ is controlled by three parameters, the sound speed $c_{s}$, the scalar to tensor ratio $r_{\rm t}$ and $\nu$. In the left hand panels of these figures, $f_{_{\rm NL}}$ can take the observationally allowed  values in some  range of $\nu$  by varying $c_s$  while  $r_{\rm t}=0.01$ is held fixed. A similar conclusion holds in the right hand panels where we fix $c_s=1$ and vary $r_{\rm t}$. Generally,  $f_{_{\rm NL}}$  increases by reducing $c_s$ and $r_{\rm t}$. One can find corners of parameter space which yield to acceptable amplitudes for $f_{_{\rm NL}}$ as required by observations in Eq.  \eqref{fnl-Observ}.     


For further studies of bispectrum and its expansion in terms of slow-roll parameters see Appendix \ref{ApendixB1}.

\section{Summaries and Conclusions}
\label{Conclusion}

In this paper we have studied inflationary solution in an extension of mimetic gravity with higher derivative interactions coupled to gravity. It is known that the original mimetic setup is plagued with the ghost and gradient instabilities. These instabilities can be removed with the help of higher derivative interactions coupled to gravity. There are a number of  options to include higher derivative corrections. In this paper we have studied the simplest higher derivative correction in the form $F(\chi) R$ with $\chi \equiv \Box \phi$. It would be interesting to extend the current analysis to include other higher derivative terms 
coupled to gravity such as $F_2(\nabla_{\mu} \nabla_{\nu}\phi R^{\mu \nu})$,   $F_3(\chi \nabla_{\mu} \phi \nabla_{\nu}\phi R^{\mu \nu})$ etc. In addition, in order for the scalar perturbations to become dynamical with a non-zero sound speed $c_s$ we have included the term $P(\chi)$ as well. 

One curious effect in our analysis is that in order to obtain an inflationary solution we have to work with a negative potential. This conclusion is a consequence of the fact that we deal with a constrained theory. More specifically, in oder for the quadratic actions of the scalar and tensor perturbation to be free from instabilities the higher derivative functions $F(\chi)$ and $P(\chi)$ are subject to certain conditions which cause the  potential to be negative. Inflation is achieved while the field rolls up the potential towards $V=0$. While a negative potential may be considered problematic a priori but our analysis show that the setup shows no pathologies either at the background or at the perturbation level.

While the background yields a period of slow-roll inflation the cosmological perturbations in this setup have novel behaviours. Because of the higher derivative interactions the dispersion relation associated with the scalar perturbations 
receives higher order momentum corrections  
as in the model of ghost inflation. Furthermore, the tilt of tensor perturbations can take either signs  in contrast to conventional inflation models \cite{Maldacena:2002vr,Acquaviva:2002ud}. In addition, we obtain a new consistency relation 
between $r_{\rm t}$ and $n_{\rm t}$ which involves $c_s$ and other model parameters encoding the higher derivative interactions.  Despite the presence of
higher derivative corrections  the tensor perturbations propagate with the speed equal to speed of light as strongly implied  by the LIGO observations.

We also studied the predictions of this setup for the amplitudes and shapes of 
non-Gaussianities. Because of higher derivative interactions, various types of interactions are developed in the cubic action endowing the setup with rich non-Gaussianity properties. Depending on model parameters, large amplitudes of non-Gaussianities in various shapes such as equilateral, orthogonal and squeezed  configurations are produced. 

As we mentioned before, it is assumed that the effective Newton constant is stabilized after inflation, corresponding to $F(\chi_e)=1$. We did not provide a dynamical mechanism for this important requirement. In addition, we have not specified the reheating mechanism in this setup. Indeed, it is possible that these two questions are related to each other. While this work was primarily concerned with cosmological perturbations such as the power spectrum and bispectrum, but a concrete picture requires that the questions of reheating and the stabilization of the effective Newton constant to be addressed as well. These are  important questions which are beyond the scope of the current analysis.

\vspace{0.5cm}
\section*{Acknowledgements}
We would like to thank Alireza Vafaei Sadr, Amin Farhang and Mehdi Atashi for helpful discussions on the numerical methods  and  Mohammad Ali Gorji for insightful discussions.

\appendix
\section{Cosmological perturbations}
\label{Ap2}

In this Appendix we present the analysis of cosmological perturbations in comoving gauge in which the calculations are significantly  simpler than other gauges. Moreover, this gauge is especially suitable for investigating the ghost and gradient instabilities. The special property of this gauge is that all perturbation are encoded in the metric sector while the scalar field is kept unperturbed,   $\delta \phi (t, \vec{x})=0$.

Implementing the ADM formalism, the metric is decomposed as
\begin{equation}\label{ADM-metric}
{\rm d}s^2=-N^2{\rm d}t^2+h_{ij}({\rm d}x^i+N^i {\rm d}t)({\rm d}x^j+N^j {\rm d}t) ~\,,
\end{equation}
in which $N$ and $N^i$ are the lapse function and the shift vector respectively while the three-dimensional metric $h_{ij}$ determines the geometry of the spatial hypersurfaces. At the level of background $N=1$, $N_{i}=0$ and $h_{ij}=a(t)^2 \delta_{ij}$. It is then possible to write down perturbations as follows,
\begin{equation}\label{metric per}
N \equiv 1+\alpha \, , \hspace{0.5cm} N_{i} \equiv \partial_{i}\psi \, , \hspace{0.5cm} h_{ij}\equiv  a(t)^2 e^{2 \mathcal{R}} \delta_{ij}+\gamma_{ij} \, ,
\end{equation}
where $\alpha$, $\psi$ and $\mathcal{R}$ are scalar perturbations.  In addition, there is the scalar perturbation $\lambda= \lambda(t)+\delta \lambda(t, \vec{x})$ for the Lagrange multiplier.

Because of the global $O(3)$ symmetry, the scalar, vector and 
tensor perturbations decouple at the linear order of perturbations. Here we have ignored vector perturbations in Eq. (\ref{metric per}) in view of the fact that the vector perturbations decay as usual in an expanding universe. 

At this stage, the above perturbations must be substituted in the action \eqref{action0} to extract the quadratic and cubic actions for the scalar and tensor perturbations. Before doing so, let us first impose the  mimetic constraint Eq. \eqref{mimetric con} at the level of perturbations defined  in  Eq. (\ref{metric per})  which yields 
\ba\label{alpha}
\alpha = 0 \,.
\ea
This result simplifies our following calculations considerably.

\subsection{Linear perturbations: Quadratic action }
Plugging the above perturbations into the action \eqref{action0} and after some integration by parts, the quadratic action in comoving gauge for $\calR$ and $\gamma_{ij}$ is obtained as follows 
\ba
\nonumber  S^{(2)}_{\rm com}  &=&  \int{\rm d}t\, {\rm d}^3\textbf{x}\ \frac{a^3}{2}
\Bigg\{ 
-3\Big(F +\frac{15}{2} H F_{\chi} - 9 H^2 F_{\chi\chi}\Big) \Big(\dot{\calR }^2+\calR  \dot{\calR} \Big)-3{\mathcal{K}}\calR  \dot{\calR} 
 -2F_{\chi}~\partial^2 \psi ~ \partial^2\calR  \nonumber\\
 &&+ \frac{{{\left( \partial \calR \right)}^{2}}}{{a^2}}\big(F + 3 H F_{\chi} -9 \dot{H} F_{\chi\chi}\big) 
-2{\mathcal{K}} \ \partial^2 \psi~\dot{\calR}
+\dfrac{1}{3}\left(\partial^2 \psi\right)^2 \left({\mathcal{K}}+F\right) \nonumber\\
&+&\frac{F}{4} \Big[ (\dot{\gamma}_{ij})^2 - \dfrac{(\partial \gamma_{ij})^2}{a^2} \Big]
\Bigg\} \,
\label{action-R}
\ea
where we have used Eq. \eqref{A} to simplify the result in terms of the function 
${\mathcal{K}}$. Note that in the these analysis there is no assumption on the function ${\mathcal{K}}$ so it is kept general.

It is evident that the $\psi$ mode is a non dynamical degrees of freedom which can be integrated out from the action. Varying Eq. \eqref{action-R} with respect to $\psi$, we find
\begin{align}
\partial^2 \psi = \dfrac{3}{{\mathcal{K}}+F} \Big({\mathcal{K}} \ \dot{\calR} + F_\chi \ \dfrac{\partial^2\calR}{a^2}\Big) \, .
\label{partial_B}
\end{align}
 Substituting the above result into the action \eqref{action-R} and after some integration by parts the quadratic Lagrangian in comoving gauge for $\mathcal{R}$ and $\gamma_{ij}$ is obtained to be
\ba
\label{action-R2}
{\cal L}_2& = & 
\vartheta \frac{a^3}{2} \left\{ \dot{\calR}^2  - \frac{c_s^2}{a^2} {{\big( \partial \calR \big)}^2} - \sigma^2 \Big(\dfrac{\partial^2 \calR}{a^2}\Big)^2
+\frac{F}{4\vartheta} \left[ (\dot{\gamma}_{ij})^2 - \dfrac{(\partial \gamma_{ij})^2}{a^2} \right]  \right\}  \,,
\ea
in which we have defined
\begin{align}
\label{vartheta}
\vartheta \equiv  \dfrac{3{\mathcal{K}}\,F}{{\mathcal{K}}+F}, \hspace{1cm}
\sigma^2 \equiv \dfrac{F^2_\chi}{{\mathcal{K}}\,F} \, ,
\end{align}
and during inflation, the sound speed of the scalar perturbations $c_{s}^2$ as
\ba\label{cs2b}
c_s^2 &\equiv  -\dfrac{1}{3{\mathcal{K}}} \left({\mathcal{K}}+F+3HF_\chi\right)  \, .
\ea

As discussed in the main text, in order to avoid the ghost and gradient instabilities 
we demand that $\vartheta > 0, \ c_s^2 > 0$ and $\sigma^2  > 0$. Applying these conditions on Eqs.~\eqref{vartheta} and \eqref{cs2b}, we obtain the following constraints 
\begin{align}
\label{AFcondition}
{\mathcal{K}}>0 \,, ~~~~~~~~~~ F>0 \,, ~~~~~~~~~~ F_\chi<0\ .
\end{align}


\subsection{Nonlinear scalar perturbations: Cubic action}
\label{cubic}
In this section, we calculate the cubic action for the scalar perturbations which is used to calculate the bispectrum.

Expanding the action \eqref{action0} up to third order, the cubic action is given by
\begin{eqnarray}
\label{cubic action1}
 S^{(3)}_{\rm com}& =&
\int{\rm d}t\, {\rm d}^3x \ a^3 \Big\{
9{\mathcal{K}} ~ \calR \dot{\calR}^2
+
\tilde{f}_1 ~ {\cal R}\dfrac{{\left( \partial \calR \right)}^2}{a^2}
-
9 F_{\chi\chi} ~ \dot{\calR}^2\dfrac{\partial^2 \calR}{a^2} 
-6{\mathcal{K}} \left(\dot{\cal R}\ \partial_i\calR \ \partial_i \psi +  \calR \dot{\calR} \ \partial^2 \psi \right)
\nonumber\\
&-& 6F_\chi \ \partial_i \calR \ \partial_i \psi \ \dfrac{\partial^2 \calR}{a^2} + 2{\mathcal{K}} \ \partial_i\calR \ \partial_i \psi \ \partial^2 \psi - F_\chi \ \dfrac{\left( \partial \calR \right)^2}{a^2} \ \partial^2 \psi +\dfrac{1}{2} \left( \tilde{f}_{2} \ \calR 
- \tilde{f}_{3} \ \dot{\calR} \right) \left(\partial^2 \psi\right)^2
\nonumber\\
&-& 2 \left(F_\chi  \ \calR \  -3F_{\chi\chi} \ \dot{\calR} \right) \dfrac{\partial^2 \calR}{a^2} \ \partial^2 \psi  + \dfrac{3}{2} \left( F \ \calR - F_\chi \ \dot{\calR} \right) \partial_i \partial_j \psi \ \partial_i \partial_j \psi - F_{\chi\chi} \ \dfrac{\partial^2\calR}{a^2} \ \left(\partial^2 \psi\right)^2
\nonumber\\
&&-3\left(2F_{\chi\chi}-3HF_{\chi\chi\chi} \right) \dot{\calR} \ \partial^2 \psi \ \partial^2 \dot{\psi}
\Big\}
\,,
\end{eqnarray}
where,  using the definition of  $\mathcal{K}$  in   Eq. \eqref{A}, the coefficients $\tilde{f}_{i}$   are given by
\begin{align}
\tilde{f}_1 &\equiv F+3HF_\chi+9H^2\epsilon_{_H} F_{\chi\chi}
\\
\tilde{f}_{2} &\equiv 2{\mathcal{K}} - F -9HF_\chi +9H^2F_{\chi\chi}(3-2\epsilon_{_H})+27H^3F_{\chi\chi\chi} \epsilon_{_H}
\\
\tilde{f}_{3} &\equiv 2F_\chi - 9H F_{\chi\chi}
\end{align}

The next step is to eliminate $\psi$ in the above action by utilizing Eq. \eqref{partial_B}. To do this, let us define  $\partial^2\Xi \equiv Q_2 \dot{\calR}$ in which $Q_2 \equiv  3\tilde{\nu}^2/(1+\tilde{\nu}^2) $ with $\tilde{\nu}=\sqrt{{\mathcal{K}}/{F}}$. This definition provides us with the contributions proportional to the linear differential equation of $\mathcal{R}$, i.e.,
\begin{equation}
\dfrac{\delta {\cal L}_2}{\delta{\cal R}}\biggr|_1 =- \left[
\partial_t\left(\vartheta a^3 \dot{\calR}\right)
-\vartheta a c_s^2 \,\,\partial^2 \calR 
+ \dfrac{\vartheta \sigma^2}{a} \,\partial^4 \calR
\right] \,,
\end{equation} 
in the final cubic action. Substituting the relation 
\begin{align}
 \psi = \dfrac{Q_1}{H} \, \dfrac{ \calR}{a^2}+ \Xi  \, , \quad \quad  
Q_1\equiv  -1-c_s^2 Q_2
\label{partial_B2}
\end{align}
into the action \eqref{cubic action1} and performing a lot of integrations by parts and dropping the total derivative terms\footnote{The integration by parts relations presented in Ref. \cite{DeFelice:2011zh} are useful in simplifying the calculations of the cubic action.}, the corresponding cubic Lagrangian is obtained to be   
\begin{align}
\label{cubic action2}
\dfrac{{\cal L}_3}{a^3}&=
f_1 ~ {\cal R}\dfrac{{\big( \partial \calR \big)}^2}{a^2}
+
f_2 ~ \dot{\calR}\dfrac{{\big( \partial \calR \big)}^2}{a^2}
+
f_3 ~ \dot{\calR}^2\dfrac{\partial^2 \calR}{a^2}
+
f_4 ~ \calR \big(\dfrac{\partial^2 \calR}{a^2}\big)^2
+
f_5 ~ \dot{\calR} \big(\dfrac{\partial^2 \calR}{a^2}\big)^2
\nonumber\\
&+
f_6 ~ \big(\dfrac{\partial^2 \calR}{a^2}\big)^3
+
f_7 ~ \calR \dot{\calR}^2
+
f_8 ~ \dot{\calR}^3
+
f_9 ~ \dfrac{\partial^2 \calR}{a^2}\dfrac{{\big( \partial \calR \big)}^2}{a^2}
+
f_{10} ~ \dot{\calR}\, \dfrac{\partial^2 \dot{\calR}}{a^2}  \, \dfrac{\partial^2 \calR}{a^2}
\nonumber\\
&+
f_{11} ~ \dot{\calR}^2 \dfrac{\partial^2 \dot{\calR}}{a^2}
+
f_{12} ~ \big(\dfrac{\partial \calR}{a}\big)^2 \dfrac{\partial^4 \calR}{a^4}
+
f_{13} ~ \big(\dfrac{\partial \calR}{a}\big)^2 \dfrac{\partial^2 \dot{\calR}}{a^2}
+
f_{14} ~ \dfrac{\partial_i \calR}{a} \, \partial^i \partial^j \Xi \, \dfrac{\partial_j \calR}{a}
\nonumber\\
&+
f_{15} ~\dfrac{\partial^2 \calR}{a^2} \partial_i \calR \, \partial^i\Xi
+
f_{16} \dot{\calR} \, \partial_i\calR \, \partial^i\Xi
+
f_{17} \dfrac{\partial^4 \calR}{a^4} \, \partial_i\calR \, \partial^i\Xi
+
f_{18}  \partial^2\calR\big(\partial\Xi\big)^2
\nonumber\\
&+
f_{19} ~ \dfrac{\partial^2 \dot{\calR}}{a^2} \, \partial_i\calR \, \partial^i\Xi
+
f_{20} ~ \dfrac{\partial^2 \calR}{a^2} \, \partial_i\dot{\calR} \, \partial^i\Xi
+
f_{21} ~  \dfrac{\partial^2 \calR}{a^2} \, \partial_i\partial_j\Xi \, \dfrac{\partial^i\partial^j\calR}{a^2}
+
f_{22} ~ \dfrac{\partial^4 \calR}{a^2} \big(\partial\Xi\big)^2
\nonumber\\
&+
{\cal F} ~ \dfrac{\delta {\cal L}_2}{\delta \calR}\biggr|_1
\,.
\end{align}
in which the coefficient in front of $\delta \mathcal{L}_{2}/\delta {R}|_{1}$ is
\newcommand{\epfh}{\dfrac{\epsilon_{_F}^{}}{\epsilon_{_H}}} 
\newcommand{\etfh}{\dfrac{\epsilon_{_F}^{_{(2)}}}{\epsilon_{_H}}} 
\newcommand{\ethh}{\dfrac{\epsilon_{_H}^{_{(2)}}}{\epsilon_{_H}}} 
\newcommand{\dfh}{\dfrac{\epsilon_{_F}^{_{(3)}}}{\epsilon_{_H}}}
\newcommand{\dhh}{\dfrac{\epsilon_{_H}^{_{(3)}}}{\epsilon_{_H}}}
\newcommand{\gfh}{\dfrac{\epsilon_{_F}^{_{(4)}}}{\epsilon_{_H}}}
\newcommand{\ghh}{\dfrac{\epsilon_{_H}^{_{(4)}}}{\epsilon_{_H}}}
\begin{equation}
\nonumber {\cal F} \equiv  \epfh\bigg[\dfrac{Q_2-3}{3H} \big(2Q_2\calR\dot{\calR}- \partial^{-2}\partial^i\partial^j \big(\partial_i\calR \partial_j\Xi \big)+3\partial_{i}\calR \, \partial^{i}\Xi\big)-
\dfrac{Q_1 \kappa }{a^2 H^2}  \partial^{-2}\partial^i\partial^j \big(\partial_i\calR \partial_j\calR \big)
 \bigg] \, .
\end{equation}
Here $\partial^{-2}$ is the inverse Laplacian and $\kappa  \equiv \epsilon_{F}/\epsilon_{H} + \epsilon_{_F}^{^{(2)}}/\epsilon_{H} - \epsilon_{_H}^{^{(2)}}/\epsilon_{H}$. Clearly, all contributions in $\mathcal{F}$ include time and spatial derivatives of $\mathcal{R}$ which vanishes in the large-scale limit ($k \to 0$). When we calculate the bispectrum, we neglect the last term in cubic Lagrangian \eqref{cubic action2} relative to those coming from other terms. The other coefficients are given by
\begin{eqnarray}\label{fi}
	f_1 &=& \dfrac{\mathcal{K}}{\tilde{\nu}^2}
	\Big\{
	1+Q_1Q_2 \big(1-\epfh\big)\big(1+\epsilon_{Q_1}+\epsilon_{Q_2}\big)
	+\dfrac{1}{3}\epsilon_{_F}\big(1+\kappa\big)\big(3-4Q_2+Q_2^2\big(1+Q_1\big)\big)
	\nonumber \\
	&+&
	\dfrac{1}{3}\epfh {\cal D}_1+\epsilon_{_H} Q_1Q_2\big(1-\epfh \kappa\big)+\dfrac{Q_1Q_2}{2}\epsilon_{_F}
	\Big(\big(4+\epsilon_{Q_1}+\epsilon_{Q_2}\big){\cal C}_1+{\cal C}_2 \epsilon_{_H}\Big)
	\Big\}
	\nonumber\\
	&-&
	3{\mathcal{K}}Q_1 \big(1+\epsilon_{_H}+\epsilon_{Q_1} \big)
	+{\mathcal{K}}Q_1 Q_2\big(1+\epsilon_{_H}+\epsilon_{Q_1}+\epsilon_{Q_2} \big) \,,
	\end{eqnarray}
	\begin{eqnarray}
	\nonumber f_2 &=& \dfrac{{\mathcal{K}}}{H}\big(Q_1+\dfrac{Q_2}{6 \tilde{\nu}^2}\epfh\big) \big(Q_2-3\big)+ \dfrac{{\mathcal{K}}Q_1Q_2}{2H \tilde{\nu}^2} \Big\{
	1-\epfh\big(2+3\epfh+3\etfh-3\ethh+\dfrac{Q_2}{3}\big)
	\nonumber\\
\nonumber 	&+&
	{\cal C}_3 \ \epsilon_{_H}
	\Big\} \,,
	\end{eqnarray}
	\begin{eqnarray}
	\nonumber f_3 &=&\dfrac{\mathcal{K}}{36 H^2 \tilde{\nu}^2}\epfh 
	\Big\{ Q_1Q_2 \Big(3\big(Q_2-5\big)+36\kappa+6B\big(-5-\epsilon_{_H}+\epsilon_{Q_1}+\epsilon_{Q_2}\big)\Big)+6 \,{\cal C}_4 \epsilon_{_H}
	\nonumber\\
	\nonumber &-&
	4\big(1+\kappa\big)\big(Q_2-3\big)^2
	\Big\} \,,
	\end{eqnarray}
	\begin{eqnarray}
	\nonumber f_4 &=&\dfrac{{\mathcal{K}}Q_1^2}{H^2}\big(1+\dfrac{1}{\tilde{\nu}^2}\big)+\dfrac{{\mathcal{K}}Q_1}{18H^2 \tilde{\nu}^2} \Big\{3\big(9Q_1\kappa-4\big)\epfh+9Q_1\epsilon_{_F} \big(4\kappa^2+2\kappa\big(1-\etfh\big)-{\cal C}_1\big)
	\\
\nonumber 	&-&2Q_2\big(\epfh\big)^2\big[Q_2\epsilon_{Q_2}+\big(Q_2-3\big)\big(-1+\epsilon_{Q_1}+\epsilon_{Q_2}+\epsilon_{_H}\big(3+2\kappa-\epfh\big)\big)\big]
	\Big\} \,,
	\end{eqnarray}
	\begin{eqnarray}
\nonumber 	f_5 &=&\dfrac{{\mathcal{K}}Q_1}{18H^3\tilde{\nu}^2}\epfh
	\Big\{
	\big(Q_2-3\big)\big(Q_2\epfh-4\big(1+\kappa\big)\big)
	+6Q_1 \big(4\kappa^2+2\kappa\big(1-\etfh\big)-{\cal C}_1\big)
	\nonumber\\
\nonumber 	&-&
	3Q_1\big(2-3\kappa\big)
	\Big\} \,,
	\nonumber
	\end{eqnarray}
	\begin{eqnarray}
	\nonumber f_6 =-\dfrac{{\mathcal{K}}Q_1^2}{9H^4 \tilde{\nu}^2} \ \epfh
	\big(
	1+\kappa
	+\dfrac{3}{2}\big(\epfh\big)\kappa - \dfrac{Q_1}{4}
	\big) \,, 
	\end{eqnarray}
	\begin{eqnarray}
\nonumber 	f_7 &=&\dfrac{{\mathcal{K}}Q_2}{2\tilde{\nu}^2}
	\Big\{
	18\epfh+2Q_2
	+Q_2\big(\epfh\big)^2 \big(\etfh-\dfh\big)\epsilon_{_H}
	+\dfrac{1}{3}\epfh Q_2^2\big(6-\big(\epsilon_{_H}\big(1+\kappa\big)+\epsilon_{Q_2}\big)\big)
	\Big\}
	\nonumber\\
\nonumber	&+&
	{\mathcal{K}}\big(Q_2-3\big)^2+\dfrac{2}{3}\big(\epfh\big)^2Q_2^2\big(Q_2-3\big) \epsilon_{_H} \,,
	\end{eqnarray}
	\begin{eqnarray}
	\nonumber f_8 &=&-\dfrac{{\mathcal{K}}Q_2^2}{9 H \tilde{\nu}^2}\epfh \Big\{
	3-\dfrac{9}{2}
	\kappa-Q_2+\big(3-\epsilon_{Q_2}\big)\big(4\kappa^2+2\kappa\big(1-\etfh\big)-{\cal C}_1\big)
	+ {\cal C}_8 \ \epsilon_{_H}
	\Big\} \,,
	\end{eqnarray}
	\begin{eqnarray}
	\nonumber f_9 &=&-\dfrac{{\mathcal{K}}Q_1}{12H^2\tilde{\nu}^2}\Big\{
	-27Q_1+2\big(\epfh\big)^2Q_2\big(Q_2-3\big)\big(-1+\epsilon_{Q_1}+\epsilon_{Q_2}+\epsilon_{_H} \big(3-\epfh+2\kappa\big)\big)
	\\
\nonumber 	&+& \epfh \big[28+15\kappa+Q_1{\cal D}_2 \big]
	+
	2Q_2^2 \big(\epfh\big)^2 \epsilon_{Q_2}
	\Big\} + \dfrac{2{\mathcal{K}}Q_1^2}{H^2} \,,
	\end{eqnarray}
	\begin{eqnarray}
\nonumber 	f_{10}=\dfrac{{\mathcal{K}}Q_1^2}{3H^4 \tilde{\nu}^2} \ \epfh 
	{\cal C}_9 \,,
	\hspace{0.15cm}
	f_{11}= \dfrac{H Q_2}{2Q_1}f_{10} \,,
	\hspace{0.15cm}
	f_{12}= \dfrac{{\mathcal{K}}Q_1^3}{12H^4 \tilde{\nu}^2} \ \epfh \,,\hspace{0.15cm}	f_{13}=
	-\dfrac{{\mathcal{K}}Q_1Q_2\big(Q_2-3\big)}{6H^3\tilde{\nu}^2} \big(\epfh\big)^2 
\end{eqnarray}
	\begin{eqnarray}
	\nonumber f_{14}= \dfrac{3 {\mathcal{K}}Q_1}{H \tilde{\nu}^2} \ \epfh \,, \hspace{0.25cm}
	f_{15} = \dfrac{Q_1}{H}\big(2{\mathcal{K}}+\kappa \big(\epfh\big)^2\epsilon_{_H}\big) \,, 
	\end{eqnarray}
		\begin{eqnarray}
	\nonumber f_{16}&=& \big(Q_2-3\big)\big(2{\mathcal{K}}+Q_2\big(\epfh\big)^2\epsilon_{_H}\big)+ \dfrac{{\mathcal{K}}}{6\tilde{\nu}^2}\big[\big(54+5Q_2^2\epsilon_{Q_2}\big)\epfh-5Q_2\big(Q_2-3\big)\big(1+\kappa\big)\epsilon_{_F}\big],
	\end{eqnarray}

\begin{eqnarray}
\nonumber 	f_{17}&=&-\dfrac{2}{Q_2}f_{13} \,,
	\hspace{0.5cm}
	f_{18}= \dfrac{1}{6} \big(Q_2-3\big) \big( \epfh \big)^2 \epsilon_{_H} + \dfrac{{\mathcal{K}}}{6\tilde{\nu}^2} \big( 9- \epfh Q_2\epsilon_{Q_2} - \big(Q_2-3\big) \big(1+\kappa\big)\epsilon_{_F} \big),
	\end{eqnarray}	
	\begin{eqnarray}
\nonumber 	 f_{19}= -\dfrac{{\mathcal{K}}Q_1}{H^2 \tilde{\nu}^2} \big(\dfrac{1}{2}+\kappa -\dfrac{Q_2}{6}\big) \epfh ,
	\hspace{0.25cm}
	f_{20}= 
	f_{19} - \dfrac{{\mathcal{K}}Q_1Q_2}{6H^2 \tilde{\nu}^2} \ \epfh
	\end{eqnarray}
	\begin{eqnarray}
\nonumber 	f_{21}= \dfrac{{\mathcal{K}}Q_1}{9 H^3 \tilde{\nu}^2} \big(3Q_1-\big(Q_2-3\big)\dfrac{\epsilon_{_F}}{\epsilon_{_H}}  \big) \dfrac{\epsilon_{_F}}{\epsilon_{_H}},
	\hspace{0.5cm}
	f_{22}= \big(\dfrac{H}{Q_1}\big)^2  f_{12}
\end{eqnarray}
in which  
\begin{align*}
{\cal C}_1 &=
\kappa^2+\kappa \big(1+
\epfh+\dfh-\ethh
\big)+\big(\epfh\big)^2+\epfh\big(\dfh-\ethh\big)+\ethh\big(\dhh-\dfh\big)
\\
{\cal C}_2&= \kappa^3+\big(\epfh\big)^3+2\big(2\epfh+\etfh\big) \kappa^2+\big(3-5\epfh-2\dfh+4\ethh\big)\epfh \kappa+2\big(1-2\ethh\dhh\big)\kappa
\\
&+\big(3\dfh+\big(\dfh\big)^2+\dfh \gfh-\ethh+2\dfh \ethh \big)\big(\kappa-\epfh\big) +\big(3+\dfh-\ethh\big)\big(\kappa^2-\big(\epfh\big)^2\big)
\\
&+\ethh\big(\big(\dfh\big)^2-\big(\dhh\big)^2+\big(3-\ethh\big)\big(\dfh-\dhh\big)+\big(\dfh \gfh - \dhh \ghh
\big)+2\big(1+2\dfh\big)\epfh\big)
\\
{\cal C}_3 &= \big(\epfh+\etfh\big)^2+\big(\epfh+\etfh\big)\big(1-3\ethh\big)+\etfh\big(\epfh+\dfh\big)+\ethh\big(-1+2\ethh-\dhh\big)
\\
{\cal C}_4&= \big(\big(\epfh\big)^2+\big(\etfh\big)^2\big)\big(1-6\ethh\big)+\etfh \big(3\big(\epfh\big)^2+\dfh\big(3\etfh+\dfh+\gfh\big)\big)
\\
&+\etfh\big(3\epfh+\dfh\big)\big(1-6\ethh\big)+\ethh\big(\epfh+\etfh\big)\big(-3+11\ethh-4\dhh\big)
\\
&+2\big(1-3\ethh\big)\big(\ethh\big)^2+\ethh \dhh \big(-1+7\ethh+\dhh-\ghh\big)+\big(\epfh+\etfh\big)^3
\\
&+4\big(\etfh+\dfh\big)\epfh \etfh
\\
{\cal C}_8 &= \big(\epfh\big)^3+2\big(\ethh\big)^3
+\big(3+4\epfh+3\dfh-3\ethh\big)\kappa^2
-\big(3+5\kappa+\dfh-\ethh\big)\big(\epfh\big)^2
\\
&+\dfh \big(2\ethh-\etfh\big)
\big(3+\gfh+
\dfh\big)+\big(2+3\epfh-\epfh\dfh+2\epfh\ethh-4\ethh\dhh\big)\kappa
\\
&-6\big(\ethh\big)^2
+\ethh\big(3\etfh-2\ethh \etfh +4\epfh \dfh -\dhh\big(3-3\ethh+\dhh+\ghh\big)\big)+\kappa^3
\\
{\cal C}_9 &=
\kappa^2+\big(1- \ethh\big)\kappa+\etfh\big(\epfh+\dfh\big)-\ethh \dhh
\\
{\cal D}_1 &= 3-4Q_2\big(1+\epsilon_{Q_2}\big)+Q_2^2\big(1+2\epsilon_{Q_2}\big)+Q_1Q_2^2\big(1+\epsilon_{Q_1}+2\epsilon_{Q_2}\big)+\dfrac{3}{2}Q_1Q_2 \big(1+\epsilon_{Q_1}+\epsilon_{Q_2}\big)
\\
{\cal D}_2 &= 3-6\epsilon_{Q_1}+15\kappa-3\epsilon_{_H}\big(3+\kappa\big)
+
Q_2\big(-1+2\epsilon_{Q_1}+\epsilon_{Q_2}+\epsilon_{_H}\big(3+\kappa\big)\big)
\end{align*}

\section{Explicit expressions for the amplitude of bispectrum}
\label{AppC}

Using Eqs. \eqref{3-point-f} and \eqref{cubic action2} and the definition of $\mathcal{I}_{_{n_{_1},n_{_2}}}^{^{p_{_1},p_{_2},p_{_3}}}$ in Eq. (\ref{I-def}),  it is possible to write down explicitly form of the the non-Gaussian amplitude $\mathcal{A}$ defined in Eq. \eqref{fnl1} for each term of the cubic Lagrangian \eqref{cubic action2} as follows:
\begin{eqnarray}
 \mathcal{A}^{(1)} & =& \frac{ f_{1} H^4}{c_{s}^8  \mathcal{P}_{\mathcal{R}}^2 \vartheta^3 }
k_{1} \mathcal{I}^{0,0,0}_{1,-2} \sum_{i,j,l=1}^{3} |\epsilon_{ijl}|\bfk_{i}.\bfk_{j}\\
\mathcal A^{( 4)} & =& \frac{ f_{4} H^6}{c_{s}^{10}  \mathcal{P}_{\mathcal{R}}^2 \vartheta^3 }
\frac{1}{k_{1}} \mathcal{I}^{0,0,0}_{2,0} \sum_{i,j,l=1}^{3} |\epsilon_{ijl}|k_{i}^2 k_{j}^2\\
\mathcal{A}^{( 6)} & =& \frac{6 f_{6} H^8}{c_{s}^{12} \mathcal{P}_{\cal R}^2 \vartheta^3 }
\frac{k_{2}^2 k_{3}^2}{ k_{1}\sum_{i=1}^{3} k_{i}^3} \mathcal{I}^{0,0,0}_{3,2}\\
\mathcal A^{(9)} & =& \frac{ f_{9} H^8}{c_{s}^{10} \mathcal{P}_{\cal R}^2 \vartheta^3 }
\frac{1}{k_{1} } \mathcal{I}^{0,0,0}_{2,0}\sum_{i,j,l=1}^{3} |\epsilon_{ijl}|k_{i}^2 \ (\bfk_{j}.\bfk_{l})\\
\mathcal A^{(11)} & =& \frac{2 f_{11} H^7}{c_{s}^{8} \mathcal{P}_{\cal R}^2 \vartheta^3 }
\frac{k_{2} k_{3} \sum_{i=1}^{3} k_{i}^2}{k_{1}} \mathcal{I}^{1,1,1}_{1,1} \\
\mathcal A^{(12)}& =& \frac{ f_{12} H^8}{c_{s}^{12} \mathcal{P}_{\cal R}^2 \vartheta^3 }
\frac{1}{k_{1}^3} \mathcal{I}^{0,0,0}_{3,2} \sum_{i,j,l=1}^{3} |\epsilon_{ijl}|k_{i}^4(\bfk_{j}.\bfk_{l})  \, .
\end{eqnarray}
For the case of $p_{1}=1$ and $p_{2}=p_{3}=0$, we have
\begin{eqnarray}
 \mathcal{A}^{(5)}& =& \frac{ f_{5} H^7}{c_{s}^{10} \mathcal{P}_{\cal R}^2 \vartheta^3 }
k_{2}^2 k_{3}^2 \sum_{i,j,l=1}^{3} |\epsilon_{ijl}|\mathcal{I}^{p_{i},p_{j},p_{l}}_{2,1} \ k_{i}^{-1}  \\
\mathcal{A}^{(13)} & =& \frac{ f_{13} H^7}{c_{s}^{10} \mathcal{P}_{\cal R}^2 \vartheta^3 }
\frac{1}{ k_{1}^2} \sum_{i,j,l=1}^{3} |\epsilon_{ijl}|\mathcal{I}^{p_{i},p_{j},p_{l}}_{2,1}k_{i}^3 (\bfk_{j}.\bfk_{l})\\
\mathcal{A}^{(14)} & =& \frac{ Q_{2} f_{14} H^5}{c_{s}^{8} \mathcal{P}_{\cal R}^2 \vartheta^3 }
 \sum_{i,j,l=1}^{3} |\epsilon_{ijl}|\mathcal{I}^{p_{i},p_{j},p_{l}}_{1,-1}\frac{(\bfk_{j}. \bfk_{i})(\bfk_{i}. \bfk_{l})}{k_{i}}\\
\mathcal{A}^{(15)} & =& \frac{ Q_{2} f_{15} H^5}{c_{s}^{8} \mathcal{P}_{\cal R}^2 \vartheta^3 }
 \sum_{i,j,l=1}^{3} |\epsilon_{ijl}|\mathcal{I}^{p_{i},p_{j},p_{l}}_{1,-1}\frac{(\bfk_{i}. \bfk_{j}) k_{l}^{2}}{k_{i}}\\
\mathcal{A}^{(17)} & =& \frac{ Q_{2} f_{17} H^7}{c_{s}^{10} \mathcal{P}_{\cal R}^2 \vartheta^3 }
\frac{1}{k_{1}^3k_{2}k_{3}} \sum_{i,j,l=1}^{3} |\epsilon_{ijl}|\mathcal{I}^{p_{i},p_{j},p_{l}}_{2,1} (\bfk_{i}.\bfk_{j}) k_{j}k_{l}^5\\
\mathcal{A}^{(21)} & =& \frac{ Q_{2} f_{21} H^7}{c_{s}^{10} \mathcal{P}_{\cal R}^2 \vartheta^3 }
\frac{1}{k_{1}^3 k_{2}k_{3}} \sum_{i,j,l=1}^{3} |\epsilon_{ijl}|\mathcal{I}^{p_{i},p_{j},p_{l}}_{2,1} (\bfk_{i}.\bfk_{j})^2 k_{j} k_{l}^3 \, .
\end{eqnarray}
Finally, for $p_{1}=p_{2}=1$ and $p_{3}=0$, the rest of expressions are given by 
\begin{eqnarray}
\mathcal{A}^{ (3)} & =& \frac{ f_{3} H^6}{c_{s}^8 \mathcal{P}_{\cal R}^2 \vartheta^3 }
k_{2}k_{3}\sum_{i,j,l=1}^{3} |\epsilon_{ijl}|\mathcal{I}^{p_{i},p_{j},p_{l}}_{1,0} k_{l} \\
\mathcal{A}^{(7)} & =& \frac{ f_{7} H^4}{c_{s}^{6} \mathcal{P}_{\cal R}^2 \vartheta^3 }
k_{1} \sum_{i,j,l=1}^{3} |\epsilon_{ijl}|\mathcal{I}^{p_{i},p_{j},p_{l}}_{0,-2}k_{i}k_{j}\\
\mathcal{A}^{(10)} & =& \frac{ f_{10} H^8}{c_{s}^{10} \mathcal{P}_{\cal R}^2 \vartheta^3 }
\frac{k_{2}k_{3}}{ k_{1}^{2}} \sum_{i,j,l=1}^{3} |\epsilon_{ijl}|\mathcal{I}^{p_{i},p_{j},p_{l}}_{2,2}(k_{i}^2+k_{j}^2) k_{l}\\
\mathcal{A}^{(16)} & =& \frac{ Q_{2} f_{16} H^4}{c_{s}^{6} \mathcal{P}_{\cal R}^2 \vartheta^3 }
\frac{1}{k_{2}k_{3}} \sum_{i,j,l=1}^{3} |\epsilon_{ijl}|\mathcal{I}^{p_{i},p_{j},p_{l}}_{0,-2} k_{i}^2 (\bfk_{j}.\bfk_{l}) k_{l}\\
\mathcal{A}^{(18)} & =& \frac{ Q_{2}^2 f_{18} H^4}{c_{s}^{6} \mathcal{P}_{\cal R}^2 \vartheta^3 }
\frac{1}{k_{2}k_{3}} \sum_{i,j,l=1}^{3} |\epsilon_{ijl}|\mathcal{I}^{p_{i},p_{j},p_{l}}_{0,-2} (\bfk_{i}.\bfk_{j}) k_{l}^3\\
\mathcal{A}^{(19)} & =& \frac{ Q_{2} f_{19} H^6}{c_{s}^{8} \mathcal{P}_{\cal R}^2 \vartheta^3 }
\frac{1}{k_{1}^2 k_{2}k_{3}} \sum_{i,j,l=1}^{3} |\epsilon_{ijl}|\mathcal{I}^{p_{i},p_{j},p_{l}}_{1,0} (\bfk_{i}.\bfk_{l}) k_{j}^4 k_{l}\\
\mathcal{A}^{(20)} & =& \frac{ Q_{2} f_{20} H^6}{c_{s}^{8} \mathcal{P}_{\cal R}^2 \vartheta^3 }
\frac{1}{k_{1}^2 k_{2}k_{3}} \sum_{i,j,l=1}^{3} |\epsilon_{ijl}|\mathcal{I}^{p_{i},p_{j},p_{l}}_{1,0} (\bfk_{i}.\bfk_{j})(k_{i}^2+k_{j}^2) k_{l}^3\\
\mathcal{A}^{(22)} & =& \frac{ Q_{2}^2 f_{22} H^6}{c_{s}^{8} \mathcal{P}_{\cal R}^2 \vartheta^3 }
\frac{1}{k_{1}^2 k_{2}k_{3}} \sum_{i,j,l=1}^{3} |\epsilon_{ijl}|\mathcal{I}^{p_{i},p_{j},p_{l}}_{1,0} (\bfk_{i}.\bfk_{j}) k_{l}^5 \, .
\end{eqnarray}

\subsection{Expansion in terms of slow-roll parameters}\label{ApendixB1}

\begin{table}[t]
\begin{center}
\begin{tabular}{c ||cccc}
  \hline\hline
  Coefficient &
 $B_{1}$  &  $B_{2}$ & $B_{3}$ & $B_{4}$  \\\hline 
Expansion & $-B_{2} \left(2 c_{s}^2+3\right) $  &  $\frac{\eta}{\varrho^2}$ & $\frac{B_{2}}{2} \left(4 c_{s}^2-7\right)$ & $\frac{11}{9} \frac{\eta}{\varrho^4}$    \\ \hline\hline \hline
 Coefficient   & $B_{5}\approx \frac{30 c_{s}^2 B_{6}}{7}$ & $B_{7}$ &  $B_{8}$  &   $B_{9}$  \\\hline
Expansion  &$\frac{5}{11} B_{4}$& $-12 c_{s}^2 B_{2}$&$5 \eta$& $\frac{6}{7} c_{s}^2 B_{6} $\\ \hline\hline \hline
Coefficient  & $B_{10}$ & $B_{11}$  & $B_{12}$  & $B_{13}$  \\\hline
Expansion  &$-\frac{2}{9} \left(3 +\frac{1}{c_{s}^2}\right) B_{2}$&$\frac{1}{15} (3+\frac{1}{c_{s}^2}) B_{8}$&$\frac{1}{22 c_{s}^2} B_{4} $& $-\frac{1}{c_{s}^2} B_{2}$ \\ \hline\hline \hline
Coefficient  & $B_{14}$& $B_{15}$ &  $B_{16}$  & $B_{17}$   \\\hline
Expansion  &$\frac{18}{11} c_{s}^2 \epsilon_{H} B_{4}$&$\Big(\frac{2}{3 \mathcal{K}} B_{2}+\frac{6}{11} c_{s}^2 B_{4}\Big) \epsilon_{H}^2$&$-\frac{54}{11} c_{s}^{4} B_{4}\epsilon_{H}$& $\frac{6}{11} B_{4}\epsilon_{H}$ \\ \hline\hline \hline
Coefficient  & $B_{18}$& $B_{19}\approx B_{20}$ &  $B_{21}$  & $B_{22}$   \\\hline
Expansion  &$\frac{9}{11} c_{s}^4 B_{4}\epsilon_{H}^2+\frac{1}{3 \mathcal{K}} c_{s}^2 B_{2} \epsilon_{H}^3$&$\frac{3}{11} c_{s}^2 B_{4} \epsilon_{H}$ &$-\frac{4}{11}B_{4}\epsilon_{H}$& $\frac{1}{22} c_{s}^2 B_{4} \epsilon_{H}^2$\\ \hline\hline
\end{tabular}
\caption{The expansion of the shape function coefficient $B_{\rm (i)}$ in terms of the  slow-roll parameter $\epsilon_{_H}$. Here we have defined $\eta \equiv\frac{\pi^2 r_{\rm t}^2}{72 \ c_{s}^6 \ \tilde{\nu}^2}$ and $\varrho \equiv c_{s} \tilde{\nu}$.  }\label{tab1}
\end{center}
\end{table}

In order to derive a simple expression for $f_{_{\rm NL}}$ in the equilateral configuration to the order of the slow roll parameter $\epsilon_{H}$, let us first consider all terms defined in Eq. \eqref{epX} to be much smaller than unity and 
$\tilde{\nu} \lesssim \mathcal{O}(\sqrt{\epsilon_{H}})$.  Then the amplitude of non-Gaussianities can be written as 
\begin{equation}
\mathcal{A}^{(j)}=B_{j}S^{(j)} \, ,
\end{equation} 
where $B_{j}$ are the coefficients coming in front of each shape function $S^{(j)}$
for the amplitude $\mathcal{A}^{(j)}$ listed in previous subsection, for example $B_{1}= f_{1} H^4/(c_{s}^8  \mathcal{P}_{\mathcal{R}}^2 \vartheta^3)$.  With this decomposition,  the shape coefficients can be expanded as shown in Table. \ref{tab1}.   Correspondingly,  the leading contribution to $f_{_{\rm NL}}^{^{\rm equi}}$  is obtained to be
\begin{eqnarray}
 &\dfrac{c_s^6  \ \tilde{\nu}^4}{\pi^4 \ r_{\rm t}^2} \ & f_{_{\rm NL}}^{^{\rm equi,lead}}
\simeq 
-\dfrac{1}{36 }\left(2+\frac{3}{ c_{s}^2}\right)
{\cal I}_{_{1,-2}}^{^{0,0,0}}
-\dfrac{1}{81c_s^2}\left(3+\frac{1}{c_{s}^2}\right)
{\cal I}_{_{2,2}}^{^{1,1,0}}\\
\nonumber &+& \frac{5}{972c_s^6\tilde{\nu}^2}{\cal I}_{_{3,2}}^{^{0,0,0}}
+ \frac{1}{27 c_s^4\tilde{\nu}^2}{\cal I}_{_{2,0}}^{^{0,0,0}}
+
\dfrac{5\tilde{\nu}^2}{36}{\cal I}_{_{0,-1}}^{^{1,1,1}}
+ \dfrac{\tilde{\nu}^2}{108}\left(3+\frac{1}{c_s^2}\right)
{\cal I}_{_{1,1}}^{^{1,1,1}}
\\
\nonumber &+&
\dfrac{1}{324 c_s^4 \tilde{\nu}^2}\left(5-9 \tilde{\nu}^2\right){\cal I}_{_{2,1}}^{^{1,0,0}}
+\dfrac{1}{36c_s^2 } 
 {\cal I}_{_{1,-1}}^{^{1,0,0}}
- \dfrac{1}{3} {\cal I}_{_{0,-2}}^{^{1,1,0}}
+
\dfrac{1}{72} \left(4-\frac{7}{ c_{s}^2} \right){\cal I}_{_{1,0}}^{^{1,1,0}}\\
\nonumber &+& \frac{\epsilon_{H}}{\tilde{\nu}^2}\Big[\dfrac{1}{162 c_s^4}{\cal I}_{_{2,1}}^{^{1,0,0}}+\dfrac{1}{18 c_s^2}{\cal I}_{_{1,-1}}^{^{1,0,0}}+ \dfrac{1}{36 c_s^2}{\cal I}_{_{1,0}}^{^{1,1,0}}-\dfrac{1}{6}{\cal I}_{_{0,-2}}^{^{1,1,0}}\Big]+\frac{\epsilon_{H}^2}{\mathcal{K} c_{s}^2} {\cal I}_{_{1,-1}}^{^{1,0,0}}-\frac{\epsilon_{H}^3}{\mathcal{K}} {\cal I}_{_{0,-2}}^{^{1,1,0}} \, .
\end{eqnarray}
It worth mentioning that one can not discard the sub-leading orders of the slow-roll parameter relative to the leading order, because integral functions $\mathcal{I}^{^{p,q,r}}_{_{n,m}}$ and $\tilde{\nu}$ are running with $\nu$. 

It is interesting that the relative error in this approximation is 
\begin{equation}
\delta= \Big| \frac{f_{_{\rm NL}}^{^{\rm {equi,lead}}}-f_{_{\rm NL}}^{^{\rm equi}}}{f_{_{\rm NL}}^{^{\rm equi}}} \Big| \ll \mathcal{O}(\epsilon_{H}) \, .
\end{equation}
This means that the expansion coefficients presented in Table. \ref{tab1} are near to their exact values  with high accuracies. Having these shape coefficients in hand, one can calculate $f_{_{\rm NL}}$ in other configurations by using  the amplitudes presented in Appendix. \ref{AppC}.

\vspace{1cm}

\bibliographystyle{JHEP}
\bibliography{references}

\providecommand{\href}[2]{#2}\begingroup\raggedright\begin{thebibliography}{10}

\bibitem{Chamseddine:2013kea}
A.~H. Chamseddine and V.~Mukhanov, \emph{{Mimetic Dark Matter}},
  \href{https://doi.org/10.1007/JHEP11(2013)135}{\emph{JHEP} {\bfseries 11}
  (2013) 135}, [\href{https://arxiv.org/abs/1308.5410}{{\ttfamily 1308.5410}}].

\bibitem{Deruelle:2014zza}
N.~Deruelle and J.~Rua, \emph{{Disformal Transformations, Veiled General
  Relativity and Mimetic Gravity}},
  \href{https://doi.org/10.1088/1475-7516/2014/09/002}{\emph{JCAP} {\bfseries
  09} (2014) 002}, [\href{https://arxiv.org/abs/1407.0825}{{\ttfamily
  1407.0825}}].

\bibitem{Yuan:2015tta}
F.-F. Yuan and P.~Huang, \emph{{Induced geometry from disformal
  transformation}},
  \href{https://doi.org/10.1016/j.physletb.2015.03.031}{\emph{Phys. Lett. B}
  {\bfseries 744} (2015) 120--124},
  [\href{https://arxiv.org/abs/1501.06135}{{\ttfamily 1501.06135}}].

\bibitem{Chamseddine:2014vna}
A.~H. Chamseddine, V.~Mukhanov and A.~Vikman, \emph{{Cosmology with Mimetic
  Matter}}, \href{https://doi.org/10.1088/1475-7516/2014/06/017}{\emph{JCAP}
  {\bfseries 1406} (2014) 017},
  [\href{https://arxiv.org/abs/1403.3961}{{\ttfamily 1403.3961}}].

\bibitem{Chamseddine:2016uef}
A.~H. Chamseddine and V.~Mukhanov, \emph{{Resolving Cosmological
  Singularities}},
  \href{https://doi.org/10.1088/1475-7516/2017/03/009}{\emph{JCAP} {\bfseries
  03} (2017) 009}, [\href{https://arxiv.org/abs/1612.05860}{{\ttfamily
  1612.05860}}].

\bibitem{Chamseddine:2016ktu}
A.~H. Chamseddine and V.~Mukhanov, \emph{{Nonsingular Black Hole}},
  \href{https://doi.org/10.1140/epjc/s10052-017-4759-z}{\emph{Eur. Phys. J. C}
  {\bfseries 77} (2017) 183},
  [\href{https://arxiv.org/abs/1612.05861}{{\ttfamily 1612.05861}}].

\bibitem{Mirzagholi:2014ifa}
L.~Mirzagholi and A.~Vikman, \emph{{Imperfect Dark Matter}},
  \href{https://doi.org/10.1088/1475-7516/2015/06/028}{\emph{JCAP} {\bfseries
  06} (2015) 028}, [\href{https://arxiv.org/abs/1412.7136}{{\ttfamily
  1412.7136}}].

\bibitem{Myrzakulov:2015kda}
R.~Myrzakulov, L.~Sebastiani, S.~Vagnozzi and S.~Zerbini, \emph{{Static
  spherically symmetric solutions in mimetic gravity: rotation curves and
  wormholes}},
  \href{https://doi.org/10.1088/0264-9381/33/12/125005}{\emph{Class. Quant.
  Grav.} {\bfseries 33} (2016) 125005},
  [\href{https://arxiv.org/abs/1510.02284}{{\ttfamily 1510.02284}}].

\bibitem{Arroja:2015yvd}
F.~Arroja, N.~Bartolo, P.~Karmakar and S.~Matarrese, \emph{{Cosmological
  perturbations in mimetic Horndeski gravity}},
  \href{https://doi.org/10.1088/1475-7516/2016/04/042}{\emph{JCAP} {\bfseries
  04} (2016) 042}, [\href{https://arxiv.org/abs/1512.09374}{{\ttfamily
  1512.09374}}].

\bibitem{Sebastiani:2016ras}
L.~Sebastiani, S.~Vagnozzi and R.~Myrzakulov, \emph{{Mimetic gravity: a review
  of recent developments and applications to cosmology and astrophysics}},
  \href{https://doi.org/10.1155/2017/3156915}{\emph{Adv. High Energy Phys.}
  {\bfseries 2017} (2017) 3156915},
  [\href{https://arxiv.org/abs/1612.08661}{{\ttfamily 1612.08661}}].

\bibitem{Dutta:2017fjw}
J.~Dutta, W.~Khyllep, E.~N. Saridakis, N.~Tamanini and S.~Vagnozzi,
  \emph{{Cosmological dynamics of mimetic gravity}},
  \href{https://doi.org/10.1088/1475-7516/2018/02/041}{\emph{JCAP} {\bfseries
  02} (2018) 041}, [\href{https://arxiv.org/abs/1711.07290}{{\ttfamily
  1711.07290}}].

\bibitem{Saadi:2014jfa}
H.~Saadi, \emph{{A Cosmological Solution to Mimetic Dark Matter}},
  \href{https://doi.org/10.1140/epjc/s10052-015-3856-0}{\emph{Eur. Phys. J. C}
  {\bfseries 76} (2016) 14}, [\href{https://arxiv.org/abs/1411.4531}{{\ttfamily
  1411.4531}}].

\bibitem{Firouzjahi:2018xob}
H.~Firouzjahi, M.~A. Gorji, S.~A. Hosseini~Mansoori, A.~Karami and T.~Rostami,
  \emph{{Two-field disformal transformation and mimetic cosmology}},
  \href{https://doi.org/10.1088/1475-7516/2018/11/046}{\emph{JCAP} {\bfseries
  11} (2018) 046}, [\href{https://arxiv.org/abs/1806.11472}{{\ttfamily
  1806.11472}}].

\bibitem{Gorji:2019rlm}
M.~A. Gorji, A.~Allahyari, M.~Khodadi and H.~Firouzjahi, \emph{{Mimetic black
  holes}}, \href{https://doi.org/10.1103/PhysRevD.101.124060}{\emph{Phys. Rev.
  D} {\bfseries 101} (2020) 124060},
  [\href{https://arxiv.org/abs/1912.04636}{{\ttfamily 1912.04636}}].

\bibitem{Matsumoto:2015wja}
J.~Matsumoto, S.~D. Odintsov and S.~V. Sushkov, \emph{{Cosmological
  perturbations in a mimetic matter model}},
  \href{https://doi.org/10.1103/PhysRevD.91.064062}{\emph{Phys. Rev. D}
  {\bfseries 91} (2015) 064062},
  [\href{https://arxiv.org/abs/1501.02149}{{\ttfamily 1501.02149}}].

\bibitem{Momeni:2015aea}
D.~Momeni, K.~Myrzakulov, R.~Myrzakulov and M.~Raza, \emph{{Cylindrical
  solutions in Mimetic gravity}},
  \href{https://doi.org/10.1140/epjc/s10052-016-4147-0}{\emph{Eur. Phys. J. C}
  {\bfseries 76} (2016) 301},
  [\href{https://arxiv.org/abs/1505.08034}{{\ttfamily 1505.08034}}].

\bibitem{Astashenok:2015qzw}
A.~V. Astashenok and S.~D. Odintsov, \emph{{From neutron stars to quark stars
  in mimetic gravity}},
  \href{https://doi.org/10.1103/PhysRevD.94.063008}{\emph{Phys. Rev. D}
  {\bfseries 94} (2016) 063008},
  [\href{https://arxiv.org/abs/1512.07279}{{\ttfamily 1512.07279}}].

\bibitem{Sadeghnezhad:2017hmr}
N.~Sadeghnezhad and K.~Nozari, \emph{{Braneworld Mimetic Cosmology}},
  \href{https://doi.org/10.1016/j.physletb.2017.03.039}{\emph{Phys. Lett. B}
  {\bfseries 769} (2017) 134--140},
  [\href{https://arxiv.org/abs/1703.06269}{{\ttfamily 1703.06269}}].

\bibitem{Nozari:2019esz}
K.~Nozari and N.~Rashidi, \emph{{Mimetic DBI Inflation in Confrontation with
  Planck2018 data}},
  \href{https://doi.org/10.3847/1538-4357/ab334b}{\emph{Astrophys. J.}
  {\bfseries 882} (2019) 78},
  [\href{https://arxiv.org/abs/1912.06050}{{\ttfamily 1912.06050}}].

\bibitem{Solomon:2019qgf}
A.~R. Solomon, V.~Vardanyan and Y.~Akrami, \emph{{Massive mimetic cosmology}},
  \href{https://doi.org/10.1016/j.physletb.2019.05.045}{\emph{Phys. Lett. B}
  {\bfseries 794} (2019) 135--142},
  [\href{https://arxiv.org/abs/1902.08533}{{\ttfamily 1902.08533}}].

\bibitem{Shen:2019nyp}
L.~Shen, Y.~Zheng and M.~Li, \emph{{Two-field mimetic gravity revisited and
  Hamiltonian analysis}},
  \href{https://doi.org/10.1088/1475-7516/2019/12/026}{\emph{JCAP} {\bfseries
  12} (2019) 026}, [\href{https://arxiv.org/abs/1909.01248}{{\ttfamily
  1909.01248}}].

\bibitem{Ganz:2019vre}
A.~Ganz, N.~Bartolo and S.~Matarrese, \emph{{Towards a viable effective field
  theory of mimetic gravity}},
  \href{https://doi.org/10.1088/1475-7516/2019/12/037}{\emph{JCAP} {\bfseries
  12} (2019) 037}, [\href{https://arxiv.org/abs/1907.10301}{{\ttfamily
  1907.10301}}].

\bibitem{deCesare:2019pqj}
M.~de~Cesare, \emph{{Reconstruction of Mimetic Gravity in a
  Non-SingularBouncing Universe from Quantum Gravity}},
  \href{https://doi.org/10.3390/universe5050107}{\emph{Universe} {\bfseries 5}
  (2019) 107}, [\href{https://arxiv.org/abs/1904.02622}{{\ttfamily
  1904.02622}}].

\bibitem{Nozari:2019shm}
K.~Nozari and N.~Sadeghnezhad, \emph{{Braneworld mimetic $f(R)$ gravity}},
  \href{https://doi.org/10.1142/S0219887819500427}{\emph{Int. J. Geom. Meth.
  Mod. Phys.} {\bfseries 16} (2019) 1950042}.

\bibitem{deCesare:2018cts}
M.~de~Cesare, \emph{{Limiting curvature mimetic gravity for group field theory
  condensates}}, \href{https://doi.org/10.1103/PhysRevD.99.063505}{\emph{Phys.
  Rev. D} {\bfseries 99} (2019) 063505},
  [\href{https://arxiv.org/abs/1812.06171}{{\ttfamily 1812.06171}}].

\bibitem{Ganz:2018mqi}
A.~Ganz, P.~Karmakar, S.~Matarrese and D.~Sorokin, \emph{{Hamiltonian analysis
  of mimetic scalar gravity revisited}},
  \href{https://doi.org/10.1103/PhysRevD.99.064009}{\emph{Phys. Rev. D}
  {\bfseries 99} (2019) 064009},
  [\href{https://arxiv.org/abs/1812.02667}{{\ttfamily 1812.02667}}].

\bibitem{Ganz:2018vzg}
A.~Ganz, N.~Bartolo, P.~Karmakar and S.~Matarrese, \emph{{Gravity in mimetic
  scalar-tensor theories after GW170817}},
  \href{https://doi.org/10.1088/1475-7516/2019/01/056}{\emph{JCAP} {\bfseries
  01} (2019) 056}, [\href{https://arxiv.org/abs/1809.03496}{{\ttfamily
  1809.03496}}].

\bibitem{Sheykhi:2019gvk}
A.~Sheykhi and S.~Grunau, \emph{{Topological black holes in mimetic gravity}},
  \href{https://arxiv.org/abs/1911.13072}{{\ttfamily 1911.13072}}.

\bibitem{Sheykhi:2020dkm}
A.~Sheykhi, \emph{{Mimetic gravity in $(2+1)$-dimensions}},
  \href{https://arxiv.org/abs/2009.12826}{{\ttfamily 2009.12826}}.

\bibitem{Sheykhi:2020fqf}
A.~Sheykhi, \emph{{Mimetic Black Strings}},
  \href{https://doi.org/10.1007/JHEP07(2020)031}{\emph{JHEP} {\bfseries 07}
  (2020) 031}, [\href{https://arxiv.org/abs/2002.11718}{{\ttfamily
  2002.11718}}].

\bibitem{Nojiri:2014zqa}
S.~Nojiri and S.~D. Odintsov, \emph{{Mimetic $F(R)$ gravity: inflation, dark
  energy and bounce}},  \href{https://arxiv.org/abs/1408.3561}{{\ttfamily
  1408.3561}}.

\bibitem{Astashenok:2015haa}
A.~V. Astashenok, S.~D. Odintsov and V.~Oikonomou, \emph{{Modified
  Gauss\textendash{}Bonnet gravity with the Lagrange multiplier constraint as
  mimetic theory}},
  \href{https://doi.org/10.1088/0264-9381/32/18/185007}{\emph{Class. Quant.
  Grav.} {\bfseries 32} (2015) 185007},
  [\href{https://arxiv.org/abs/1504.04861}{{\ttfamily 1504.04861}}].

\bibitem{Nojiri:2016ppu}
S.~Nojiri, S.~Odintsov and V.~Oikonomou, \emph{{Unimodular-Mimetic Cosmology}},
  \href{https://doi.org/10.1088/0264-9381/33/12/125017}{\emph{Class. Quant.
  Grav.} {\bfseries 33} (2016) 125017},
  [\href{https://arxiv.org/abs/1601.07057}{{\ttfamily 1601.07057}}].

\bibitem{Nojiri:2017ygt}
S.~Nojiri, S.~Odintsov and V.~Oikonomou, \emph{{Ghost-Free $F(R)$ Gravity with
  Lagrange Multiplier Constraint}},
  \href{https://doi.org/10.1016/j.physletb.2017.10.045}{\emph{Phys. Lett. B}
  {\bfseries 775} (2017) 44--49},
  [\href{https://arxiv.org/abs/1710.07838}{{\ttfamily 1710.07838}}].

\bibitem{Nojiri:2016vhu}
S.~Nojiri, S.~Odintsov and V.~Oikonomou, \emph{{Viable Mimetic Completion of
  Unified Inflation-Dark Energy Evolution in Modified Gravity}},
  \href{https://doi.org/10.1103/PhysRevD.94.104050}{\emph{Phys. Rev. D}
  {\bfseries 94} (2016) 104050},
  [\href{https://arxiv.org/abs/1608.07806}{{\ttfamily 1608.07806}}].

\bibitem{Odintsov:2018ggm}
S.~Odintsov and V.~Oikonomou, \emph{{The reconstruction of $f(\phi)R$ and
  mimetic gravity from viable slow-roll inflation}},
  \href{https://doi.org/10.1016/j.nuclphysb.2018.01.027}{\emph{Nucl. Phys. B}
  {\bfseries 929} (2018) 79--112},
  [\href{https://arxiv.org/abs/1801.10529}{{\ttfamily 1801.10529}}].

\bibitem{Casalino:2018wnc}
A.~Casalino, M.~Rinaldi, L.~Sebastiani and S.~Vagnozzi, \emph{{Alive and well:
  mimetic gravity and a higher-order extension in light of GW170817}},
  \href{https://doi.org/10.1088/1361-6382/aaf1fd}{\emph{Class. Quant. Grav.}
  {\bfseries 36} (2019) 017001},
  [\href{https://arxiv.org/abs/1811.06830}{{\ttfamily 1811.06830}}].

\bibitem{Barvinsky:2013mea}
A.~Barvinsky, \emph{{Dark matter as a ghost free conformal extension of
  Einstein theory}},
  \href{https://doi.org/10.1088/1475-7516/2014/01/014}{\emph{JCAP} {\bfseries
  01} (2014) 014}, [\href{https://arxiv.org/abs/1311.3111}{{\ttfamily
  1311.3111}}].

\bibitem{Chaichian:2014qba}
M.~Chaichian, J.~Kluson, M.~Oksanen and A.~Tureanu, \emph{{Mimetic dark matter,
  ghost instability and a mimetic tensor-vector-scalar gravity}},
  \href{https://doi.org/10.1007/JHEP12(2014)102}{\emph{JHEP} {\bfseries 12}
  (2014) 102}, [\href{https://arxiv.org/abs/1404.4008}{{\ttfamily 1404.4008}}].

\bibitem{Ijjas:2016pad}
A.~Ijjas, J.~Ripley and P.~J. Steinhardt, \emph{{NEC violation in mimetic
  cosmology revisited}},
  \href{https://doi.org/10.1016/j.physletb.2016.06.052}{\emph{Phys. Lett. B}
  {\bfseries 760} (2016) 132--138},
  [\href{https://arxiv.org/abs/1604.08586}{{\ttfamily 1604.08586}}].

\bibitem{Firouzjahi:2017txv}
H.~Firouzjahi, M.~A. Gorji and S.~A. Hosseini~Mansoori, \emph{{Instabilities in
  Mimetic Matter Perturbations}},
  \href{https://doi.org/10.1088/1475-7516/2017/07/031}{\emph{JCAP} {\bfseries
  1707} (2017) 031}, [\href{https://arxiv.org/abs/1703.02923}{{\ttfamily
  1703.02923}}].

\bibitem{Ramazanov:2016xhp}
S.~Ramazanov, F.~Arroja, M.~Celoria, S.~Matarrese and L.~Pilo, \emph{{Living
  with ghosts in Ho\v{r}ava-Lifshitz gravity}},
  \href{https://doi.org/10.1007/JHEP06(2016)020}{\emph{JHEP} {\bfseries 06}
  (2016) 020}, [\href{https://arxiv.org/abs/1601.05405}{{\ttfamily
  1601.05405}}].

\bibitem{Capela:2014xta}
F.~Capela and S.~Ramazanov, \emph{{Modified Dust and the Small Scale Crisis in
  CDM}}, \href{https://doi.org/10.1088/1475-7516/2015/04/051}{\emph{JCAP}
  {\bfseries 04} (2015) 051},
  [\href{https://arxiv.org/abs/1412.2051}{{\ttfamily 1412.2051}}].

\bibitem{DeFelice:2015moy}
A.~De~Felice and S.~Mukohyama, \emph{{Phenomenology in minimal theory of
  massive gravity}},
  \href{https://doi.org/10.1088/1475-7516/2016/04/028}{\emph{JCAP} {\bfseries
  04} (2016) 028}, [\href{https://arxiv.org/abs/1512.04008}{{\ttfamily
  1512.04008}}].

\bibitem{Gumrukcuoglu:2016jbh}
A.~E. G\"umr\"uk\c{c}\"uo\u{g}lu, S.~Mukohyama and T.~P. Sotiriou, \emph{{Low
  energy ghosts and the Jeans\textquoteright{} instability}},
  \href{https://doi.org/10.1103/PhysRevD.94.064001}{\emph{Phys. Rev. D}
  {\bfseries 94} (2016) 064001},
  [\href{https://arxiv.org/abs/1606.00618}{{\ttfamily 1606.00618}}].

\bibitem{Babichev:2016jzg}
E.~Babichev and S.~Ramazanov, \emph{{Gravitational focusing of Imperfect Dark
  Matter}}, \href{https://doi.org/10.1103/PhysRevD.95.024025}{\emph{Phys. Rev.
  D} {\bfseries 95} (2017) 024025},
  [\href{https://arxiv.org/abs/1609.08580}{{\ttfamily 1609.08580}}].

\bibitem{Babichev:2017lrx}
E.~Babichev and S.~Ramazanov, \emph{{Caustic free completion of pressureless
  perfect fluid and k-essence}},
  \href{https://doi.org/10.1007/JHEP08(2017)040}{\emph{JHEP} {\bfseries 08}
  (2017) 040}, [\href{https://arxiv.org/abs/1704.03367}{{\ttfamily
  1704.03367}}].

\bibitem{Gorji:2018okn}
M.~A. Gorji, S.~Mukohyama, H.~Firouzjahi and S.~A. Hosseini~Mansoori,
  \emph{{Gauge Field Mimetic Cosmology}},
  \href{https://doi.org/10.1088/1475-7516/2018/08/047}{\emph{JCAP} {\bfseries
  08} (2018) 047}, [\href{https://arxiv.org/abs/1807.06335}{{\ttfamily
  1807.06335}}].

\bibitem{Gorji:2019ttx}
M.~A. Gorji, S.~Mukohyama and H.~Firouzjahi, \emph{{Cosmology in Mimetic SU(2)
  Gauge Theory}},
  \href{https://doi.org/10.1088/1475-7516/2019/05/019}{\emph{JCAP} {\bfseries
  05} (2019) 019}, [\href{https://arxiv.org/abs/1903.04845}{{\ttfamily
  1903.04845}}].

\bibitem{Zheng:2017qfs}
Y.~Zheng, L.~Shen, Y.~Mou and M.~Li, \emph{{On (in)stabilities of perturbations
  in mimetic models with higher derivatives}},
  \href{https://doi.org/10.1088/1475-7516/2017/08/040}{\emph{JCAP} {\bfseries
  08} (2017) 040}, [\href{https://arxiv.org/abs/1704.06834}{{\ttfamily
  1704.06834}}].

\bibitem{Hirano:2017zox}
S.~Hirano, S.~Nishi and T.~Kobayashi, \emph{{Healthy imperfect dark matter from
  effective theory of mimetic cosmological perturbations}},
  \href{https://doi.org/10.1088/1475-7516/2017/07/009}{\emph{JCAP} {\bfseries
  07} (2017) 009}, [\href{https://arxiv.org/abs/1704.06031}{{\ttfamily
  1704.06031}}].

\bibitem{Gorji:2017cai}
M.~A. Gorji, S.~A. Hosseini~Mansoori and H.~Firouzjahi, \emph{{Higher
  Derivative Mimetic Gravity}},
  \href{https://doi.org/10.1088/1475-7516/2018/01/020}{\emph{JCAP} {\bfseries
  01} (2018) 020}, [\href{https://arxiv.org/abs/1709.09988}{{\ttfamily
  1709.09988}}].

\bibitem{ArkaniHamed:2003uz}
N.~Arkani-Hamed, P.~Creminelli, S.~Mukohyama and M.~Zaldarriaga, \emph{{Ghost
  inflation}}, \href{https://doi.org/10.1088/1475-7516/2004/04/001}{\emph{JCAP}
  {\bfseries 0404} (2004) 001},
  [\href{https://arxiv.org/abs/hep-th/0312100}{{\ttfamily hep-th/0312100}}].

\bibitem{Cheung:2007st}
C.~Cheung, P.~Creminelli, A.~Fitzpatrick, J.~Kaplan and L.~Senatore, \emph{{The
  Effective Field Theory of Inflation}},
  \href{https://doi.org/10.1088/1126-6708/2008/03/014}{\emph{JHEP} {\bfseries
  03} (2008) 014}, [\href{https://arxiv.org/abs/0709.0293}{{\ttfamily
  0709.0293}}].

\bibitem{Alishahiha:2004eh}
M.~Alishahiha, E.~Silverstein and D.~Tong, \emph{{DBI in the sky}},
  \href{https://doi.org/10.1103/PhysRevD.70.123505}{\emph{Phys. Rev. D}
  {\bfseries 70} (2004) 123505},
  [\href{https://arxiv.org/abs/hep-th/0404084}{{\ttfamily hep-th/0404084}}].

\bibitem{Akrami:2018odb}
{\scshape Planck} collaboration, Y.~Akrami et~al., \emph{{Planck 2018 results.
  X. Constraints on inflation}},
  \href{https://arxiv.org/abs/1807.06211}{{\ttfamily 1807.06211}}.

\bibitem{Akrami:2019izv}
{\scshape Planck} collaboration, Y.~Akrami et~al., \emph{{Planck 2018 results.
  IX. Constraints on primordial non-Gaussianity}},
  \href{https://arxiv.org/abs/1905.05697}{{\ttfamily 1905.05697}}.

\bibitem{Zheng:2018cuc}
Y.~Zheng, \emph{{Hamiltonian analysis of Mimetic gravity with higher
  derivatives}},  \href{https://arxiv.org/abs/1810.03826}{{\ttfamily
  1810.03826}}.

\bibitem{Golovnev:2013jxa}
A.~Golovnev, \emph{{On the recently proposed Mimetic Dark Matter}},
  \href{https://doi.org/10.1016/j.physletb.2013.11.026}{\emph{Phys. Lett.}
  {\bfseries B728} (2014) 39--40},
  [\href{https://arxiv.org/abs/1310.2790}{{\ttfamily 1310.2790}}].

\bibitem{Khoury:2001wf}
J.~Khoury, B.~A. Ovrut, P.~J. Steinhardt and N.~Turok, \emph{{The Ekpyrotic
  universe: Colliding branes and the origin of the hot big bang}},
  \href{https://doi.org/10.1103/PhysRevD.64.123522}{\emph{Phys. Rev. D}
  {\bfseries 64} (2001) 123522},
  [\href{https://arxiv.org/abs/hep-th/0103239}{{\ttfamily hep-th/0103239}}].

\bibitem{Kallosh:2001ai}
R.~Kallosh, L.~Kofman and A.~D. Linde, \emph{{Pyrotechnic universe}},
  \href{https://doi.org/10.1103/PhysRevD.64.123523}{\emph{Phys. Rev. D}
  {\bfseries 64} (2001) 123523},
  [\href{https://arxiv.org/abs/hep-th/0104073}{{\ttfamily hep-th/0104073}}].

\bibitem{Finelli:2001sr}
F.~Finelli and R.~Brandenberger, \emph{{On the generation of a scale invariant
  spectrum of adiabatic fluctuations in cosmological models with a contracting
  phase}}, \href{https://doi.org/10.1103/PhysRevD.65.103522}{\emph{Phys. Rev.
  D} {\bfseries 65} (2002) 103522},
  [\href{https://arxiv.org/abs/hep-th/0112249}{{\ttfamily hep-th/0112249}}].

\bibitem{Buchbinder:2007ad}
E.~I. Buchbinder, J.~Khoury and B.~A. Ovrut, \emph{{New Ekpyrotic cosmology}},
  \href{https://doi.org/10.1103/PhysRevD.76.123503}{\emph{Phys. Rev. D}
  {\bfseries 76} (2007) 123503},
  [\href{https://arxiv.org/abs/hep-th/0702154}{{\ttfamily hep-th/0702154}}].

\bibitem{Buchbinder:2007tw}
E.~I. Buchbinder, J.~Khoury and B.~A. Ovrut, \emph{{On the initial conditions
  in new ekpyrotic cosmology}},
  \href{https://doi.org/10.1088/1126-6708/2007/11/076}{\emph{JHEP} {\bfseries
  11} (2007) 076}, [\href{https://arxiv.org/abs/0706.3903}{{\ttfamily
  0706.3903}}].

\bibitem{Linde:2001ae}
A.~D. Linde, \emph{{Fast roll inflation}},
  \href{https://doi.org/10.1088/1126-6708/2001/11/052}{\emph{JHEP} {\bfseries
  11} (2001) 052}, [\href{https://arxiv.org/abs/hep-th/0110195}{{\ttfamily
  hep-th/0110195}}].

\bibitem{Hartle:2012qb}
J.~B. Hartle, S.~Hawking and T.~Hertog, \emph{{Accelerated Expansion from
  Negative $\Lambda$}},  \href{https://arxiv.org/abs/1205.3807}{{\ttfamily
  1205.3807}}.

\bibitem{Fujita:2015ymn}
T.~Fujita, X.~Gao and J.~Yokoyama, \emph{{Spatially covariant theories of
  gravity: disformal transformation, cosmological perturbations and the
  Einstein frame}},
  \href{https://doi.org/10.1088/1475-7516/2016/02/014}{\emph{JCAP} {\bfseries
  02} (2016) 014}, [\href{https://arxiv.org/abs/1511.04324}{{\ttfamily
  1511.04324}}].

\bibitem{Corley:1996ar}
S.~Corley and T.~Jacobson, \emph{{Hawking spectrum and high frequency
  dispersion}}, \href{https://doi.org/10.1103/PhysRevD.54.1568}{\emph{Phys.
  Rev.} {\bfseries D54} (1996) 1568--1586},
  [\href{https://arxiv.org/abs/hep-th/9601073}{{\ttfamily hep-th/9601073}}].

\bibitem{Corley:1997pr}
S.~Corley, \emph{{Computing the spectrum of black hole radiation in the
  presence of high frequency dispersion: An Analytical approach}},
  \href{https://doi.org/10.1103/PhysRevD.57.6280}{\emph{Phys. Rev.} {\bfseries
  D57} (1998) 6280--6291},
  [\href{https://arxiv.org/abs/hep-th/9710075}{{\ttfamily hep-th/9710075}}].

\bibitem{Martin:2000xs}
J.~Martin and R.~H. Brandenberger, \emph{{The TransPlanckian problem of
  inflationary cosmology}},
  \href{https://doi.org/10.1103/PhysRevD.63.123501}{\emph{Phys. Rev.}
  {\bfseries D63} (2001) 123501},
  [\href{https://arxiv.org/abs/hep-th/0005209}{{\ttfamily hep-th/0005209}}].

\bibitem{Martin:2002kt}
J.~Martin and R.~H. Brandenberger, \emph{{The Corley-Jacobson dispersion
  relation and transPlanckian inflation}},
  \href{https://doi.org/10.1103/PhysRevD.65.103514}{\emph{Phys. Rev.}
  {\bfseries D65} (2002) 103514},
  [\href{https://arxiv.org/abs/hep-th/0201189}{{\ttfamily hep-th/0201189}}].

\bibitem{Ashoorioon:2011eg}
A.~Ashoorioon, D.~Chialva and U.~Danielsson, \emph{{Effects of Nonlinear
  Dispersion Relations on Non-Gaussianities}},
  \href{https://doi.org/10.1088/1475-7516/2011/06/034}{\emph{JCAP} {\bfseries
  06} (2011) 034}, [\href{https://arxiv.org/abs/1104.2338}{{\ttfamily
  1104.2338}}].

\bibitem{Ashoorioon:2018uey}
A.~Ashoorioon, R.~Casadio, M.~Cicoli, G.~Geshnizjani and H.~J. Kim,
  \emph{{Extended Effective Field Theory of Inflation}},
  \href{https://doi.org/10.1007/JHEP02(2018)172}{\emph{JHEP} {\bfseries 02}
  (2018) 172}, [\href{https://arxiv.org/abs/1802.03040}{{\ttfamily
  1802.03040}}].

\bibitem{Ashoorioon:2018ocr}
A.~Ashoorioon, \emph{{Non-Unitary Evolution in the General Extended EFT of
  Inflation \textbackslash{}\& Excited Initial States}},
  \href{https://doi.org/10.1007/JHEP12(2018)012}{\emph{JHEP} {\bfseries 12}
  (2018) 012}, [\href{https://arxiv.org/abs/1807.06511}{{\ttfamily
  1807.06511}}].

\bibitem{abramowitz1948handbook}
M.~Abramowitz and I.~A. Stegun, \emph{Handbook of mathematical functions with
  formulas, graphs, and mathematical tables}, vol.~55.
\newblock US Government printing office, 1948.

\bibitem{Chen:2006nt}
X.~Chen, M.-x. Huang, S.~Kachru and G.~Shiu, \emph{{Observational signatures
  and non-Gaussianities of general single field inflation}},
  \href{https://doi.org/10.1088/1475-7516/2007/01/002}{\emph{JCAP} {\bfseries
  0701} (2007) 002}, [\href{https://arxiv.org/abs/hep-th/0605045}{{\ttfamily
  hep-th/0605045}}].

\bibitem{Monitor:2017mdv}
{\scshape LIGO Scientific, Virgo, Fermi-GBM, INTEGRAL} collaboration, B.~Abbott
  et~al., \emph{{Gravitational Waves and Gamma-rays from a Binary Neutron Star
  Merger: GW170817 and GRB 170817A}},
  \href{https://doi.org/10.3847/2041-8213/aa920c}{\emph{Astrophys. J. Lett.}
  {\bfseries 848} (2017) L13},
  [\href{https://arxiv.org/abs/1710.05834}{{\ttfamily 1710.05834}}].

\bibitem{PhysRevLett.119.251303}
J.~Sakstein and B.~Jain, \emph{Implications of the neutron star merger gw170817
  for cosmological scalar-tensor theories},
  \href{https://doi.org/10.1103/PhysRevLett.119.251303}{\emph{Phys. Rev. Lett.}
  {\bfseries 119} (Dec, 2017) 251303}.

\bibitem{PhysRevLett.119.251301}
T.~Baker, E.~Bellini, P.~G. Ferreira, M.~Lagos, J.~Noller and I.~Sawicki,
  \emph{Strong constraints on cosmological gravity from gw170817 and grb
  170817a}, \href{https://doi.org/10.1103/PhysRevLett.119.251301}{\emph{Phys.
  Rev. Lett.} {\bfseries 119} (Dec, 2017) 251301}.

\bibitem{Khoury:2006fg}
J.~Khoury, \emph{{Fading gravity and self-inflation}},
  \href{https://doi.org/10.1103/PhysRevD.76.123513}{\emph{Phys. Rev. D}
  {\bfseries 76} (2007) 123513},
  [\href{https://arxiv.org/abs/hep-th/0612052}{{\ttfamily hep-th/0612052}}].

\bibitem{Koshelev:2020foq}
A.~S. Koshelev, K.~Sravan~Kumar, A.~Mazumdar and A.~A. Starobinsky,
  \emph{{Non-Gaussianities and tensor-to-scalar ratio in non-local R$^{2}$-like
  inflation}}, \href{https://doi.org/10.1007/JHEP06(2020)152}{\emph{JHEP}
  {\bfseries 06} (2020) 152},
  [\href{https://arxiv.org/abs/2003.00629}{{\ttfamily 2003.00629}}].

\bibitem{Babichev:2007dw}
E.~Babichev, V.~Mukhanov and A.~Vikman, \emph{{k-Essence, superluminal
  propagation, causality and emergent geometry}},
  \href{https://doi.org/10.1088/1126-6708/2008/02/101}{\emph{JHEP} {\bfseries
  02} (2008) 101}, [\href{https://arxiv.org/abs/0708.0561}{{\ttfamily
  0708.0561}}].

\bibitem{PhysRev.182.1400}
Y.~Aharonov, A.~Komar and L.~Susskind, \emph{Superluminal behavior, causality,
  and instability}, \href{https://doi.org/10.1103/PhysRev.182.1400}{\emph{Phys.
  Rev.} {\bfseries 182} (Jun, 1969) 1400--1403}.

\bibitem{GARRIGA1999219}
J.~Garriga and V.~F. Mukhanov, \emph{Perturbations in k-inflation},
  \href{https://doi.org/https://doi.org/10.1016/S0370-2693(99)00602-4}{\emph{Physics
  Letters B} {\bfseries 458} (1999) 219 -- 225}.

\bibitem{Maldacena:2002vr}
J.~M. Maldacena, \emph{{Non-Gaussian features of primordial fluctuations in
  single field inflationary models}},
  \href{https://doi.org/10.1088/1126-6708/2003/05/013}{\emph{JHEP} {\bfseries
  05} (2003) 013}, [\href{https://arxiv.org/abs/astro-ph/0210603}{{\ttfamily
  astro-ph/0210603}}].

\bibitem{Acquaviva:2002ud}
V.~Acquaviva, N.~Bartolo, S.~Matarrese and A.~Riotto, \emph{{Second order
  cosmological perturbations from inflation}},
  \href{https://doi.org/10.1016/S0550-3213(03)00550-9}{\emph{Nucl. Phys. B}
  {\bfseries 667} (2003) 119--148},
  [\href{https://arxiv.org/abs/astro-ph/0209156}{{\ttfamily
  astro-ph/0209156}}].

\bibitem{DeFelice:2011zh}
A.~De~Felice and S.~Tsujikawa, \emph{{Primordial non-Gaussianities in general
  modified gravitational models of inflation}},
  \href{https://doi.org/10.1088/1475-7516/2011/04/029}{\emph{JCAP} {\bfseries
  04} (2011) 029}, [\href{https://arxiv.org/abs/1103.1172}{{\ttfamily
  1103.1172}}].

\end{thebibliography}\endgroup
\bibliographystyle{JHEP}

\end{document}